\documentclass[12pt,english,prd,preprint,nofootinbib,floatfix,tightenlines,showpacs]{revtex4}
\usepackage[T1]{fontenc}
\usepackage[latin1]{inputenc}
\usepackage{amsmath}
\usepackage{graphicx}
\usepackage{color}

\makeatletter
\@ifundefined{textcolor}{}
{%
 \definecolor{BLACK}{gray}{0}
 \definecolor{WHITE}{gray}{1}
 \definecolor{RED}{rgb}{1,0,0}
 \definecolor{GREEN}{rgb}{0,1,0}
 \definecolor{BLUE}{rgb}{0,0,1}
 \definecolor{CYAN}{cmyk}{1,0,0,0}
 \definecolor{MAGENTA}{cmyk}{0,1,0,0}
 \definecolor{YELLOW}{cmyk}{0,0,1,0}
 }

\usepackage{hyperref}

\makeatother

\usepackage{babel}
\begin{document}
\setcounter{footnote}{0}

\preprint{MSUHEP-100707, SMU-HEP-10-10, arXiv:1007.2241{[}hep-ph{]}}

\title{New parton distributions for collider physics}

\author{Hung-Liang Lai,$^{1,2}$ Marco Guzzi,$^{3}$ Joey Huston,$^{1}$ Zhao
Li,$^{1}$ Pavel M. Nadolsky,$^{3}$ Jon Pumplin,$^{1}$ and C.-P.
Yuan$^{1}$}

\affiliation{$^{1}$Department of Physics and Astronomy, Michigan State University,
East Lansing, MI 48824-1116, U.S.A.\\
 $^{2}$Taipei Municipal University of Education, Taipei, Taiwan\\
 $^{3}$Department of Physics, Southern Methodist
University, Dallas, TX 75275-0175, U.S.A. }

\date{\today}
\begin{abstract}
We extract new parton distribution functions (PDFs) of the proton by 
global analysis of hard scattering data in the general-mass framework 
of perturbative quantum chromodynamics. Our analysis includes new theoretical 
developments together with the most recent collider data from deep-inelastic 
scattering, vector boson production, and single-inclusive jet production.
Due to the difficulty in fitting both the D\O~ Run-II 
$W$ lepton asymmetry data and some fixed-target DIS data, 
we present two families of PDFs, CT10 and CT10W, without and with these 
high-luminosity $W$ lepton asymmetry data included in the 
global analysis. 
With both sets of PDFs, we study theoretical 
predictions and uncertainties for a diverse selection of processes 
at the Fermilab Tevatron and the CERN Large Hadron Collider.

\end{abstract}



\maketitle
\newpage
\tableofcontents{}

\newpage

\section{Introduction \label{sec:Introduction}}

Parton distribution functions (PDFs) of the proton are essential 
for making theoretical predictions, and potentially obtaining breakthrough
physics results, from experiments at high-energy hadron colliders
such as the Fermilab Tevatron and the CERN Large Hadron Collider (LHC).
An accurate determination of PDFs, and their corresponding uncertainties, 
from the global
analysis is therefore crucial. 
There have been continuous efforts on this front by several groups
\cite{Alekhin:2009ni,Ball:2010de,Pumplin:2009nk,2009wt,Martin:2009iq}.
In this paper, we describe several theoretical advancements in the
global QCD analysis that was used to produce the previous CTEQ6.6 \cite{Nadolsky:2008zw}
and CT09 \cite{Pumplin:2009nk} PDFs, and also present a study 
of the impact on the PDFs by 
new precision collider data. 
We begin by summarizing the principal changes in the theoretical treatment.

First, we now treat the systematic uncertainty associated with the
overall normalization factor in each of the data sets in the same 
manner that all other systematic error parameters are handled. Since 
the log-likelihood is an approximately 
quadratic function of the normalization parameters, their 
best-fit values can be computed algebraically for any values 
of the other fitting parameters.  This development 
simplifies the fitting procedure,
since explicit numerical minimization of  
the experimental normalizations is no longer required. 
It improves the estimate of 
uncertainties, expanding them slightly, by allowing the 
estimated normalizations to vary 
during the process of finding the most extreme acceptable fits.

Second, we now compute $\chi^{2}$, which measures the consistency 
between a given set of PDFs and the data, using
weight 1 for all experiments (with just one exception to be discussed below).
In the previous CTEQ fits, weights larger than 1 were applied to some data sets
to disallow bad fits to these sets, especially in the course of defining
eigenvector PDF sets that delimit the uncertainty. That goal is now handled
by adding an extra contribution to the total $\chi^{2}$, to guarantee  
the quality of fit to each individual data set and halt 
the displacement along any eigenvector early, if necessary,
to prevent one or more individual data sets from being badly described.

Third, we use more flexible PDF parametrizations for some parton flavors
($d$, $s$, and $g$) at  the initial scale $\mu = 1.3 \, \mathrm{GeV}$ 
in order to reduce 
parametrization dependence. This increases the uncertainty in
the strange quark and gluon distributions in kinematical regions 
where the constraints from the data are still limited. In
total, the CT10 PDF parametrizations include 26 free parameters, 
expanded from 22 used in the CTEQ6.6 analysis. 

Besides these theoretical advancements, the CT10 analysis includes 
new precise experimental data in every major category of scattering processes: 
deep-inelastic scattering (DIS), vector boson production (VBP), 
and single-inclusive jet production. In Ref.~\cite{Pumplin:2009nk},
we compared the Tevatron Run-II single-inclusive jet production data \cite{Aaltonen:2008eq,:2008hua} with the Run-I jet data sets \cite{Affolder:2001fa,Abbott:2000kp} 
and examined their impact.
In addition 
to these Run-I and Run-II jet data sets, the CT10 analysis 
includes other recent data from HERA and Tevatron experiments. 
The HERA-1 {}``combined'' data set on $e^\pm p$ DIS
\cite{2009wt}, developed by a collaboration between the H1
and ZEUS experiments, has replaced eleven original independent HERA-1 data sets. 
We also include data on the rapidity distribution
of $Z^{0}$ production, which has been measured at the Tevatron by both the
CDF \cite{Aaltonen:2010zza} and D\O~ \cite{Abazov:2007jy} collaborations.
Finally, we consider data on the measurement of the Tevatron Run-II 
$W$ lepton asymmetry, $A_\ell(y_\ell)$: the asymmetry
in the rapidity distribution $y_\ell$ of the charged
lepton $\ell$ from $W$ boson decay \cite{Acosta:2005ud,d0_e_asy,d0_mu_asy}. 
These data are sensitive
to the flavor content of the proton, especially to the ratio of 
down- and up-quark PDFs, $d(x)/u(x)$.

The high-luminosity Run-II lepton asymmetry data by the D\O~ Collaboration \cite{d0_e_asy,d0_mu_asy}
play a special role in this study. While being precise, they 
run into disagreement with some previous data sets;
and in addition, they exhibit some tension among themselves.  
Because of these disagreements, 
we present results from two different PDF fits: CT10, in which 
the D\O~ data on $A_\ell$ are 
ignored; and CT10W, in which these data are emphasized 
by moderately increasing their $\chi^2$ weights, 
which suffices for getting an acceptable fit to these data sets.

Another aspect of this paper consists of a study of the 
quality of the fit to the various data sets. 
This study aims to quantify the degree of consistency of constraints
imposed on the PDFs by different sets of experiments, in order to
establish the extent of the PDF uncertainty
allowed by the experimental measurements. Similar questions 
have been recently addressed by an examination of $\chi^2$
contributions provided by the individual experiments
\cite{Pumplin:2009nk,Pumplin:2009sc}, using techniques discussed 
in Refs.~\cite{Collins:2001es,Pumplin:2009nm}. 
Here, we explore the quality of fit issues with the help of a function 
$S$ defined in Eq.~\ref{eq:S_n},
which is convenient for comparing the goodness-of-fit 
among data sets containing different numbers of data points $N$.
Using this function, we demonstrate 
that non-negligible tensions between the fitted data
sets (also noticed in Ref.~~\cite{Pumplin:2009nm}) persist regardless
of the number of PDF parameters introduced in the global fit.

The organization of the paper is as follows. Sec.\ \ref{sec:NewTheory} 
discusses the new features in CT10 theoretical treatment in more detail.
Sec.\ \ref{sec:NewData} overviews the newly included data sets. 
Sec.\ \ref{sec:HERA-1} discusses the impact of the combined HERA-1 data.
Sec.\ \ref{sec:WLasy} examines the D\O~ Run-II lepton asymmetry data.
Sec.\ \ref{sec:CT10CT10W} compares the PDFs obtained from the CT10 and CT10W 
global fits.
Sec.\ \ref{sec:quality} examines the quality of the fits to each data set 
in terms of the statistical variable $S$ defined in this Section.
Sec.\ \ref{sec:applications} presents typical applications 
of the new PDFs to collider physics, such as jet pair production 
at the Tevatron,  electroweak and Higgs boson production 
at the Tevatron and LHC, and various processes beyond the Standard Model. 
Sec.\ \ref{sec:Conclusions} presents our conclusions. 
Finally, the appendix contains a detailed comparison 
of the CT10 fits with the HERA-1 DIS data in various $x,Q$ regions.
We also comment on the agreement of the combined HERA-1 data set
with the next-to-leading order (NLO) DGLAP evolution of CT10 distributions in the probed region
of $x$ and $Q$.

\section{Theoretical developments \label{sec:NewTheory}}
We implemented several new features in the global analysis procedures, as  
compared to the CTEQ6.6 \cite{Nadolsky:2008zw} and CT09 \cite{Pumplin:2009nk} studies.  

In the new fits, the normalization 
uncertainty in each experiment is handled just like any other 
systematic error parameter. Under a reasonable assumption that 
the normalization errors obey quasi-Gaussian statistics, 
the normalization 
choice that minimizes $\chi^2$ can be determined algebraically, by following
the approach in Refs.~\cite{LM,Pumplin:2002vw}.
This revision simplifies the fitting procedure, by eliminating the 
need to assign an explicit search parameter 
(up to 30-40 extra MINUIT \cite{MINUIT} parameters in total)
to each normalization factor during the numerical minimization.  
At the same time, it improves the 
estimate of PDF uncertainties, by correctly allowing 
the normalization factors to vary, as 
the total log-likelihood $\chi^2$ is explored along each eigenvector 
direction to determine uncertainty limits.  In previous 
CTEQ analyses, the normalizations were frozen during that 
exploration, so that this upgrade results in a small increase 
in the final estimated uncertainty range.  (We have checked that the 
normalization shifts found in the fits, both for the central 
fit and the eigenvector uncertainty sets, lie within a reasonable range,   
when compared to the published normalization uncertainty of each data set.)

At the initial scale $\mu_0 = 1.3 \, \mathrm{GeV}$ for DGLAP evolution \cite{Gribov:1972ri,Altarelli:1977zs,Dokshitzer:1977sg}, both CT10 and CTEQ6.6 sets 
assume the same functional form for valence quark PDFs: 
\begin{equation}
q_v(x,\mu_0) = q(x,\mu_0) - \bar{q}(x,\mu_0) = a_0 \, x^{a_1} \, (1 - x)^{a_2} \, 
\exp(a_3 x \, + \, a_4 x^2 \, + \, a_5 \sqrt{x}),
\label{eq:uvparam}
\end{equation}
where $q=u$ or $d$. While all parameters $a_1$,...,$a_5$ are varied freely in CT10, 
the coefficient $a_5$ for $d(x)$ was set to zero in CTEQ6.6; 
consequently, the CT10 down-quark PDF is more flexible at large $x$ 
than that of CTEQ6.6. 
(The coefficients $a_2$,\dots,$a_5$ for $u_v(x)$ and $d_v(x)$ are 
taken to be independent. The $a_1$ values, expected to be close to 0.5 based
on Regge theory,  are set equal to each other.)

For the gluon, CTEQ6.6 also used the form (\ref{eq:uvparam}) with 
$a_5 = 0$.  The same form is employed in CT10, multiplied by 
an additional factor $\exp(-a_6 x^{-a_7})$ 
to allow for extra freedom of the gluon 
at small $x$.  This extra term is not required for getting
the best fit to the current data, since it reduces the minimum $\chi^2$ 
by only 6 units. Rather, it allows 
us to better explore the uncertainty in the small-$x$ region, where 
the current data provide little constraint on $g(x,\mu)$.  

For strangeness PDF, CTEQ6.6 used an \emph{ad hoc} prescription designed to
avoid fits in which the ratio of strange to non-strange sea quark PDFs,
$R_s = (s(x) + \bar{s}(x))/(\bar{u}(x) + \bar{d}(x))$,
was counterintuitively large at $x \lesssim 10^{-2}$, where this ratio is not constrained by the current data \cite{Nadolsky:2008zw}.  
In CT10, $s(x,\mu_0)$ is given 
by a more flexible form (\ref{eq:uvparam}) with $a_4 = 0$.  
The desire to impose
reasonable expectations on $R_s$ in the $x\rightarrow 0$ 
is handled in CT10 by adding 
a soft constraint (a $\chi^2$ penalty term) 
such that solutions with $R_s$ outside of the range 0.4-1 
are disfavored 
at $x$ below $10^{-3}$.
(The same power-law behavior was assumed for 
$\bar{u}(x)$, $\bar{d}(x)$, and $\bar{s}(x)$ in the limit $x \to 0$, 
based on Regge theory; with the same coefficient for $\bar{u}$ and $\bar{d}$,
so that $\bar{u}(x) / \bar{d}(x) \to 1$ and 
$\bar{s}(x) / \bar{d}(x) \to R_s = \mathrm {const}$ as $x \to 0$.)
For simplicity, an assumption of symmetry between the strangeness 
and anti-strangeness PDFs was made, $s(x,\mu)=\bar{s}(x,\mu)$, 
similarly to CTEQ6.6.

When computing the $\chi^2$ measure of consistency between the PDFs and 
the data in CT10, we follow the usual CTEQ analysis 
approach \cite{Hessian,LM} of requiring agreement at the confidence
level (CL) of about 90\%  with each experiment included in the
fit, for each final PDF eigenvector set provided to compute the PDF
uncertainty. This is achieved, on average, by defining an upper
bound on the excursion of the global $\chi^2$ from its minimum value, 
chosen so as
to
keep the $\chi^2$ function of each individual experiment within the 
90\% CL computed (for the number of data points 
in this experiment \cite{Pumplin:2002vw}). In addition to this
overall
tolerance condition, CTEQ6.6 and the earlier fits 
assigned weights greater than 1 to some data sets---particularly those 
with a small number of points---to ensure that the fits to those data sets 
remained acceptable for all of the eigenvector sets that define the uncertainty 
range.
The procedure for the choice of weights was time-consuming and 
varied depending on the selection of experiments and 
``tensions'' between them. It might also give some experiments with extra weights 
an 
undue
influence on the best fit.
 
In the CT10 fit, we introduce a different approach, which reaches the same
objective of enforcing the 90\% CL agreement with all experiments in a
more efficient way. Each data set is assigned weight $1$ in CT10, 
with the exception of the D\O~ Run-II lepton asymmetry data.  
We define a variable 
\begin{equation}
S_n \, = \, \sqrt{2\chi^{2}(N_n)}-\sqrt{2N_{n}-1} 
\label{eq:S_n}
\end{equation}
for each data set $n$ with $N_n$ data points.
On statistical grounds explained in Sec.\ \ref{sec:quality}, 
$S_n$ is expected to 
be well approximated by a standard normal distribution (with a mean of
zero, variance of 1, and negligible skewness), independently of the 
number of points $N_n$ for $N_n > 10$. Thus, in an ideal situation,
it is easy to assign a confidence level to each excursion of
$S_n$ from its central value, for all practical $N_n$. 
For example, a 90\% CL excursion in the $n$-th experiment 
would correspond to $S_n \approx 1.3$, cf. Fig.~\ref{figs:pma}.

In reality, the
distribution of $S_n$ values is broader than a Gaussian of unit variance
even in the best fit (cf. Sec.\ \ref{sec:quality}), due to 
some incompatibility between the different data sets.
For this reason, in the experiments that have $\chi^2_n > N_n$ already in the
best fit, we compute $S_n$ by dividing the $\chi^2$ value 
by its best-fit value, to bring 
the $S_n$ distribution in close agreement with the standard normal 
distribution. We then add a penalty term 
to the log-likelihood function $\chi^2$ 
(which also includes the usual $\chi^2$ contributions from the
individual data points, of the type shown in Eq.~(\ref{Chi2sys})) to
exclude solutions with improbable positive $S_n$ values. 

The specific penalty term we chose is 
\begin{equation}
P \, = \, \sum_{n}\, S_n^{\, k} \, \theta(S_n).
\label{eq:penalty}
\end{equation}
It applies only to experiments with $S_n > 0$, 
as indicated by the theta function
$\theta(S_n)$. Individual $S_n$ values are raised to power
$k=16$, selected so that $P$ is negligible 
in most of the allowed parameter region, but grows rapidly when a
90\% CL boundary for some experiment 
(corresponding to $S_n = 1.3$ for this experiment) 
is reached. 

The final PDF uncertainty shows little dependence 
on the exact form of $P$, provided that it is small in the bulk 
of the allowed region and grows rapidly near the 90\% CL boundaries.
The penalty term warrants that none of the alternative eigenvector
PDFs disagrees strongly with any individual data set within the
estimated PDF uncertainty range. Because of the large power law $k$, 
it can quickly halt the displacement along any eigenvector direction, 
owing to the $90\%$ CL criterion.  

The procedure described captures the idea of preserving the 90\% CL
agreement among the data sets \cite{Hessian,LM,Martin:2009iq} 
explicitly and automatically, while still retaining 
most of the original importance of the criterion based on the 
global $\chi^2$. In particular, 
we find that the $S_n$ penalties are important for 
about half of the final eigenvector sets. They guarantee that data sets 
with small numbers of data points are not ignored in a large global
fit, even in situations when a significant increase in 
$\chi^2$ of a specific small data set is misconstrued as 
a harmless minor change in the global $\chi^2$.
(The two-part structure of $\chi^2$ 
loosely resembles a bicameral legislature such 
as the US Congress, where votes in the House are proportional 
to population---data points in our case---while votes 
in the Senate represent specific entities---experiments 
or data sets in our case.)

The CT10 and CT10W central fits and their eigenvector uncertainty
sets were computed using QCD parameters
$\alpha_s(M_Z) = 0.118$ (evolved 
by numerically solving the RG differential equation at two loops 
with the HOPPET program \cite{Salam:2008qg}),
$m_c = 1.3 \, \mathrm{GeV}$ and 
$m_b = 4.75 \, \mathrm{GeV}$.
The value chosen for $\alpha_s(m_Z)$ is close to the world average
value, which is constrained most strongly 
by electroweak precision experiments that 
are not directly included in the PDF fitting. In addition to the
eigenvector PDF sets for this central $\alpha_s(M_Z)$ value, the CT10(W)
distributions \cite{ct10website} provide several PDFs 
for alternative $\alpha_s(M_Z)$ values in the interval 0.113-0.123. Those
can be used to evaluate the combined uncertainty due to the PDFs and
$\alpha_s(M_Z)$ in any physical process of interest, by following 
a convenient procedure that is spelled out and derived 
in Ref.~\cite{Lai:2010nw}.
The procedure is to add the PDF and $\alpha_s$ uncertainties in
quadrature, which is sufficient for evaluating the combined uncertainty,
including the full correlation between the PDFs and $\alpha_s$.

Our choice of the input charm mass $m_c=1.3$ GeV
is based on a mild preference for that value
in $\chi^2$ for the global fit.  (The charm mass behaves as 
phenomenological parameter in the PDF fit at NLO --- in part because
it plays a role in approximating phase space effects.) 
A systematic study of the allowed range for $m_c$ and $m_b$ will be
undertaken in a future publication.
 
The calculations at NLO accuracy in various processes in this
and previous CTEQ analyses~\cite{Nadolsky:2008zw,Pumplin:2009nk,Lai:2010nw}
are summarized as follows.  
The NLO terms are 
included directly for DIS and VBP processes.  To speed up the 
calculations, inclusive jets and the W lepton asymmetry are calculated 
using a lookup table which gives the ratio NLO/LO~\footnote{
The contribution from next-to-next-leading-logarithm(NNLL) resummation 
at small transverse momenta of $W$ bosons is added into the
NLO term for the case of $A_\ell$. } 
separately for each 
data point. This table depends only very weakly on the parameters for the 
input PDFs. The table is updated in the course of the fitting, and we 
check its agreement with the final run of the fitting to be sure that the 
calculation is accurate at NLO. The method is therefore just a 
calculational convenience, not an approximation.  That is, there is an 
effective NLO calculation for every inclusive jet and $A_\ell$ data point 
used in the fitting. To be clear, the same PDFs and 2-loop $\alpha_s$ are used 
in both the numerator and denominator for the table. The lookup 
table just summarizes the effects of the NLO corrections to the matrix 
elements for each data point.

\section{Overview of new data sets \label{sec:NewData}}

In the past two years, several new precise data sets became available,
expanding the scope of the earlier data used in the previous 
CTEQ6.6 and CT09 analyses. 

The H1 and ZEUS collaborations at the HERA $ep$ collider released a 
joint data set \cite{2009wt} that
combines results from eleven measurements in neutral-current (NC) and
charged-current (CC)
deep inelastic scattering (DIS) processes at HERA-1. In our previous analyses, 
which included the HERA results as separate data sets, 
each one was handled independently from the other ten sets, 
and the correlations between systematic errors in the distinct data sets were
neglected. Since many systematic factors 
are common to both experiments and affect 
all results in a correlated way, 
Ref.~\cite{2009wt} presents the HERA-1 DIS
results as a single data set, with all 114 correlated systematic
effects shared by each data point. The combined data set has a reduced
total systematic uncertainty, as a result of cross calibration 
between H1 and ZEUS measurements. When the combined
HERA-1 data set is used, we observe a reduction in the PDF uncertainty,
compared to a counterpart fit based on the separate HERA-1 data sets.
We shall discuss the impact of the combined HERA-1 data 
in Sec.~\ref{sec:HERA-1}.

New data on the asymmetry in the rapidity distribution of the charged lepton 
from $W$ boson decay, measured in $p\bar{p}$ collisions 
at $\sqrt{s}=1.96 \, \mathrm{TeV}$, have been published by both the CDF and 
D\O~ Collaborations.
The lower luminosity CDF Run-I \cite{Abe:1994rj}
and Run-II \cite{Acosta:2005ud} data agree well
with the other data sets used in the global analysis. 
The high-luminosity D\O~ Run-II data \cite{d0_e_asy, d0_mu_asy}
conflict with some of the fixed-target DIS experiments.
Since the D\O~ Run-II $W$ lepton asymmetry data  
show significant tension with respect to the other data ---
and to some extent with themselves---
we produce  two separate fits: CT10, from which the D\O~ $W$ lepton
asymmetry sets are excluded; and CT10W, in which they are
included. The fits to the $A_\ell$ data are presented 
in Sec.\ \ref{sec:WLasy}, and the resulting PDFs are compared in Sec.\
\ref{sec:CT10CT10W}. 

Measurements by the CDF \cite{Aaltonen:2010zza} and D\O~
\cite{Abazov:2007jy} collaborations of the rapidity distribution for $Z^0$ 
bosons at the Tevatron are also included in this analysis. The D\O~ measurement
with integrated luminosity of $0.4\mbox{ fb}^{-1}$ 
agrees very well with the theory prediction,
with $\chi^2 = 16\ (15)$ for 28 data points in the CT10 (CT10W) fit. 
The agreement with the (more precise) CDF data at  
$2.1\mbox{ fb}^{-1}$ is slightly worse, with $\chi^2= 41 (34)$ for 28 data points.  
The CDF data show a slight preference for CT10W over CT10. Comparisons
of these data sets with the NLO theoretical predictions based on CT10
and CT10W PDFs are shown in Fig.~\ref{figs:DISTZTev}.
Overall, the impact of the $Z$ rapidity data sets on the best fit 
is quite mild.

The analyses presented here also include Run-II inclusive
jet data from CDF and D\O~  \cite{Aaltonen:2008eq,:2008hua}, 
present in the CT09 analysis \cite{Pumplin:2009nk}, but not in CTEQ6.6. 
In total, the CT10 (CT10W) fit is based on 29 (31) data sets 
with a total of 2753 (2798) data points.

\begin{figure}[b]
\includegraphics[width=0.48\textwidth,height=0.8\textheight,keepaspectratio]{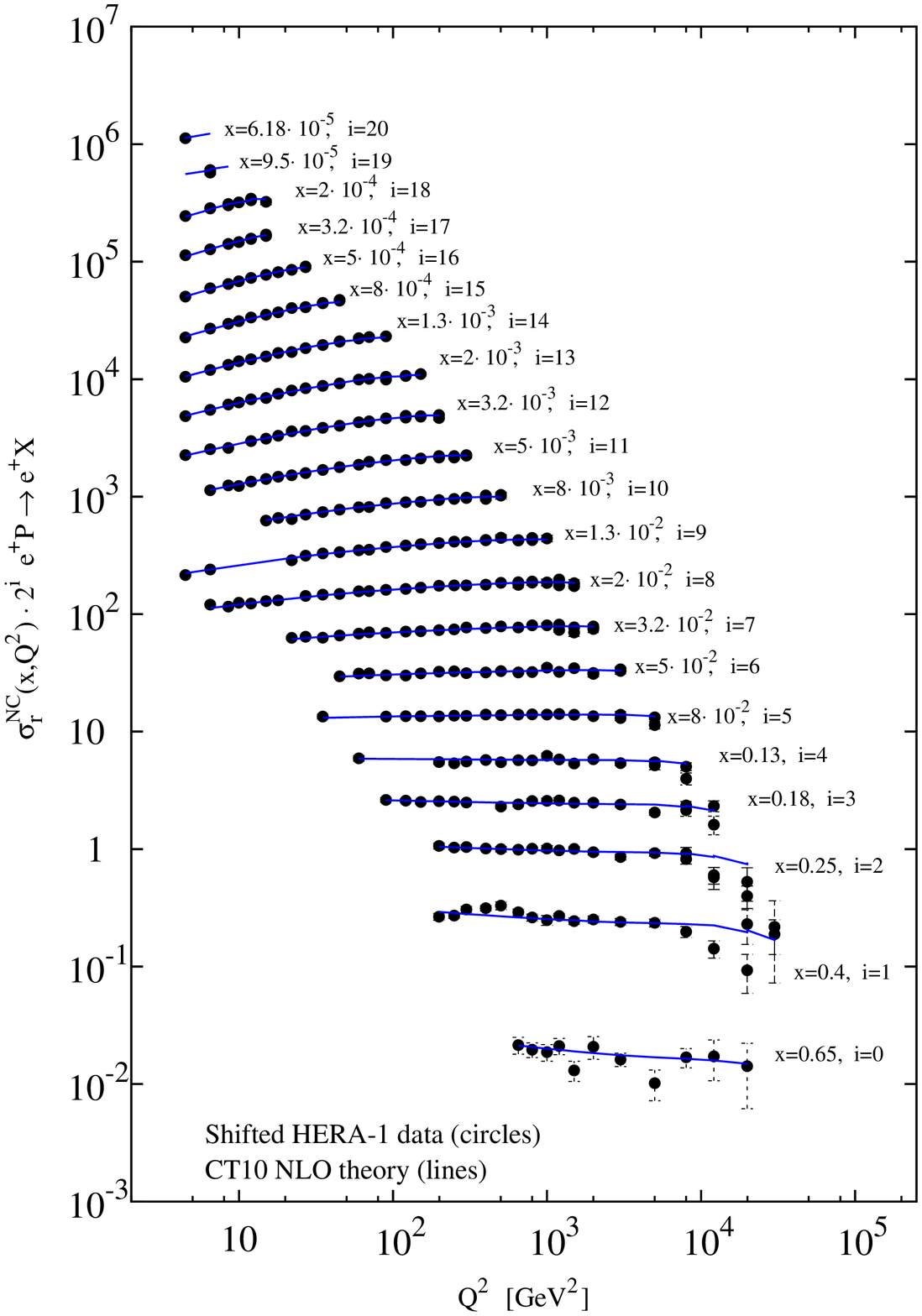}
\includegraphics[width=0.48\textwidth,height=0.8\textheight,keepaspectratio]{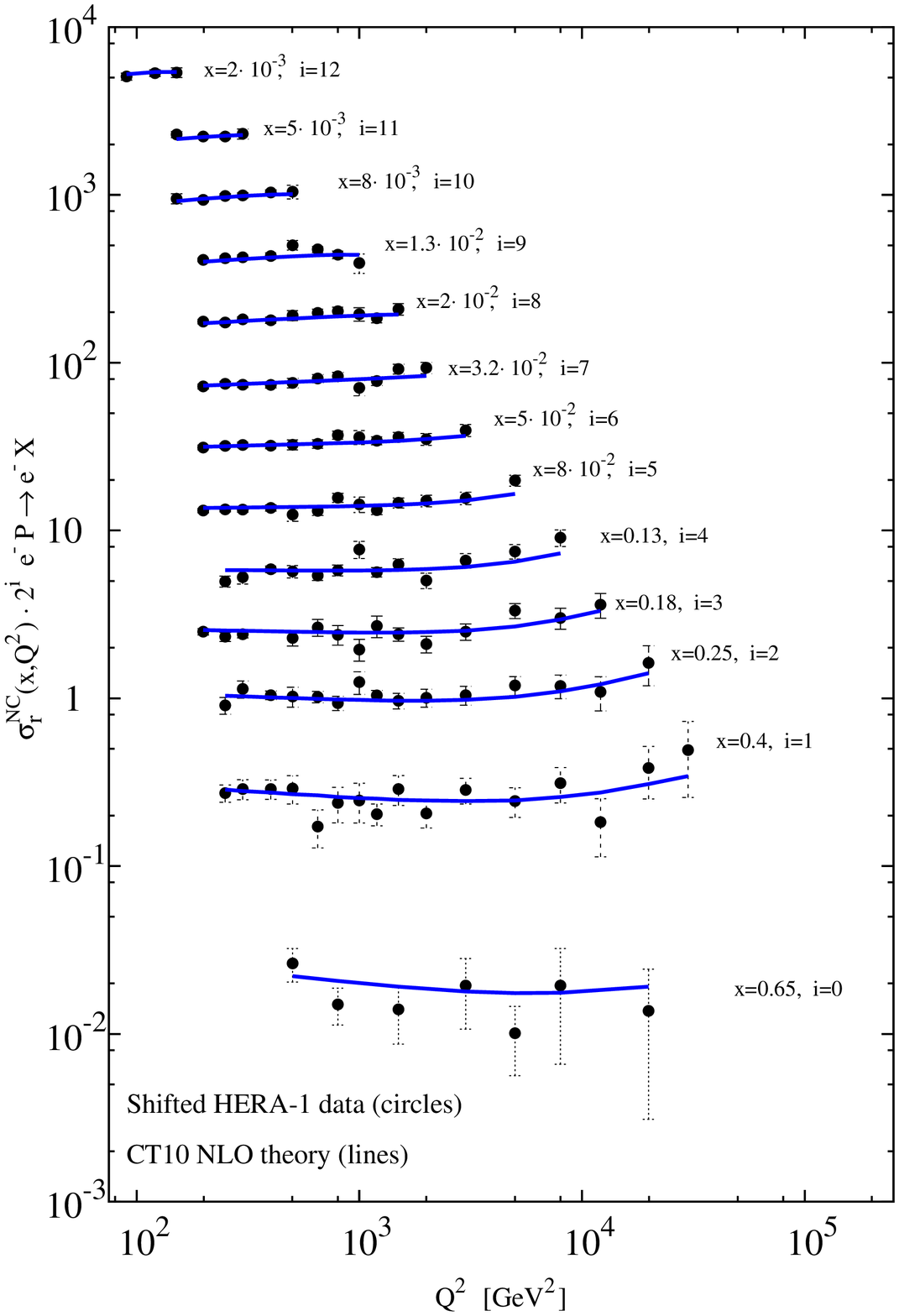}
\caption{Comparison of CT10 NLO predictions for reduced cross sections
  in $e^+p$ (left) and $e^-p$ (right) 
neutral-current DIS with the combined HERA-1 data \cite{2009wt}, with
  correlated systematic shifts included.\label{fig:CT10redNC}}
\end{figure}

\section{Impact of the combined HERA-1 data \label{sec:HERA-1}}

The combined H1/ZEUS data set for DIS at HERA-1 \cite{2009wt} is included in
our analysis, together with the estimates of the correlated
experimental uncertainties 
provided by the HERA experiments \cite{HERAcorrmat}. 
When comparing each experimental value
$D_{k}$ with the respective theory value $T_{k}(\{a\})$ (dependent
on PDF parameters $\{a\}$), we account for the possible systematic shifts
in the data, as estimated by the correlation matrix $\beta_{k\alpha}$.
There are $N_{\lambda}=$114
independent sources of experimental systematic uncertainties, quantified
by the parameters $\lambda_{\alpha}$ that should obey the standard normal
distribution. The contribution of the combined HERA-1 set to the log-likelihood
function $\chi^{2}$ is given by

\begin{equation}
\chi^{2}(\{a\},\{\lambda\}) \, = \, 
\sum_{k=1}^{N}\frac{1}{s_{k}^{2}}\left(D_{k} - 
T_{k}(\{a\}) - \sum_{\alpha=1}^{N_{\lambda}}
\lambda_{\alpha}\beta_{k\alpha}\right)^{2} \, + \, 
\sum_{\alpha=1}^{N_{\lambda}}\lambda_{\alpha}^{2},
\label{Chi2sys}
\end{equation}
where $N$ is the total number of points, and 
$s_{k}=\sqrt{s_{k,\mbox{\small stat}}^{2}+s_{k,\mbox{\small uncor sys}}^{2}}$
is the total uncorrelated error on the measurement $D_{k}$, equal 
to the statistical and uncorrelated systematic errors on $D_{k}$ added
in quadrature. Minimization of $\chi^2$ with respect to the systematic
parameters $\lambda_\alpha$ is realized algebraically \cite{LM,Pumplin:2002vw}.

\begin{figure}[tb]
\begin{centering}
\includegraphics[width=0.6\columnwidth]{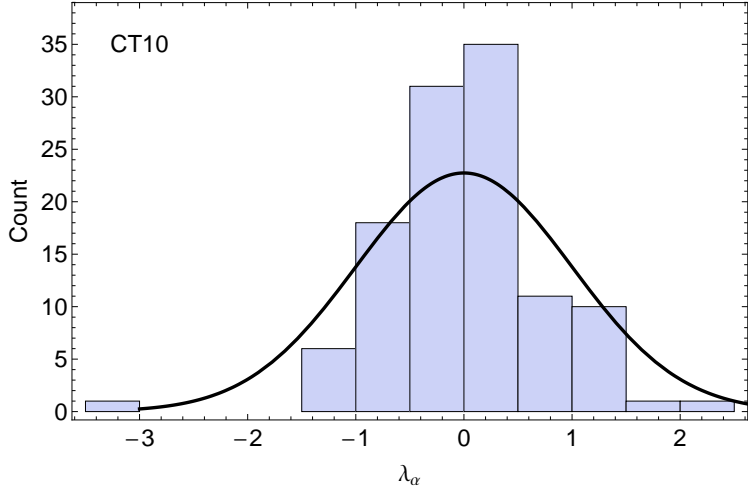}
\end{centering}
\caption{Distribution of systematic parameters $\lambda_{\alpha}$ 
of the combined HERA-1 data set in the CT10 best fit (CT10.00).
\label{fig:lambdas}}
\end{figure}

Both the CT10 and CT10W central fits, designated as CT10.00 and CT10W.00
respectively, show acceptable agreement 
with the combined H1/ZEUS set of reduced DIS cross sections. For the
rest of this section, we discuss the CT10 fit. The outcome of the
CT10W fit is very similar; figures comparing 
the CT10W fit to the combined HERA data are available at \cite{ct10website}.
For the HERA-1 sample, we obtain $\chi^2\approx 680$ 
for the $N=579$ points that pass our kinematical cuts for the DIS data: 
$Q > 2\mbox{ GeV}$ and $W>3.5\mbox{ GeV}$.  
A comparison of theory predictions with the NC $e^+p$ and $e^-p$ data
is shown in Fig.~\ref{fig:CT10redNC}.  
Apart from some excessive scatter of the NC $e^\pm p$ data around theory
predictions, which results in a slightly higher-than-ideal value of 
$\chi^2/N=1.18$, 
NLO theory describes the overall data well, 
without obvious systematic discrepancies.

The data points shown in Fig.~\ref{fig:CT10redNC} include systematic
shifts bringing the theoretical and experimental values in closer agreement, by allowing the systematic parameters $\lambda_\alpha$
to take their most optimal values 
within the bounds allowed 
by the correlation matrix $\beta_{k\alpha}$.  As expected, the 
best-fit values of $\lambda_{\alpha}$ are distributed consistently with
the standard normal distribution. Their contribution $\sum_\alpha
\lambda_\alpha^2\approx 65$ to $\chi^2$ 
in Eq.~(\ref{Chi2sys}) 
is better than the expected value of $114$.

The histogram of $\lambda_\alpha$ values obtained in the best CT10 fit (CT10.00) 
is shown in Fig.~\ref{fig:lambdas}, with 
an overlaid standard normal distribution. 
The histogram is clearly compatible with its
stated Gaussian behavior. 
With many eigenvector sets, one observes 1-2 values at $(\pm)$2-3$\sigma$, 
but such large displacements are not persistent.

The overall agreement with the combined HERA-1 data is slightly worse than
with the separate HERA-1 data sets, as a consequence of some increase
in $\chi^2/N$ for the NC data 
at $x<0.001$ and $x>0.1$. To investigate the origin of this increase,
we compare the CT10 fit to an alternative
fit, in which the combined HERA-1 set is replaced by the eleven separate
HERA DIS data sets, and with the
rest of the inputs kept identical to those in the CT10 fit. 
In this ``alternative CT10 fit'', 
each HERA-1 data set contributes a $\chi^{2}$ term of the same
form as in Eq.~(\ref{Chi2sys}), but with independent correlation
matrices $\beta_{k\alpha}$ 
and systematic parameters $\{\lambda_{\alpha}\}$ in each measurement.

The Appendix examines the contributions of the individual data
points to $\chi^2$ in the CT10 and alternative CT10 fits 
and finds them to be consistent with random 
point-to-point fluctuations of the combined data in the small-$x$ and large-$x$
ranges.
The fluctuations are somewhat irregular and larger than normally expected. 
Their spread widens upon the
combination of the data sets. Thus, this analysis
does not reveal significant
systematic differences between the NLO QCD theory 
and the full sample of the HERA-1 DIS data. In the same spirit, we
demonstrate in the appendix 
that the HERA-1 set is compatible with the NLO DGLAP evolution of
CT10 PDFs, whether those PDFs are fitted to the whole DIS sample, 
or only to a specially selected subsample of it with points at large
$x$ and $Q$.

\begin{figure}[tbp]
\includegraphics[width=0.49\columnwidth,height=0.49\columnwidth,keepaspectratio,angle=-90]{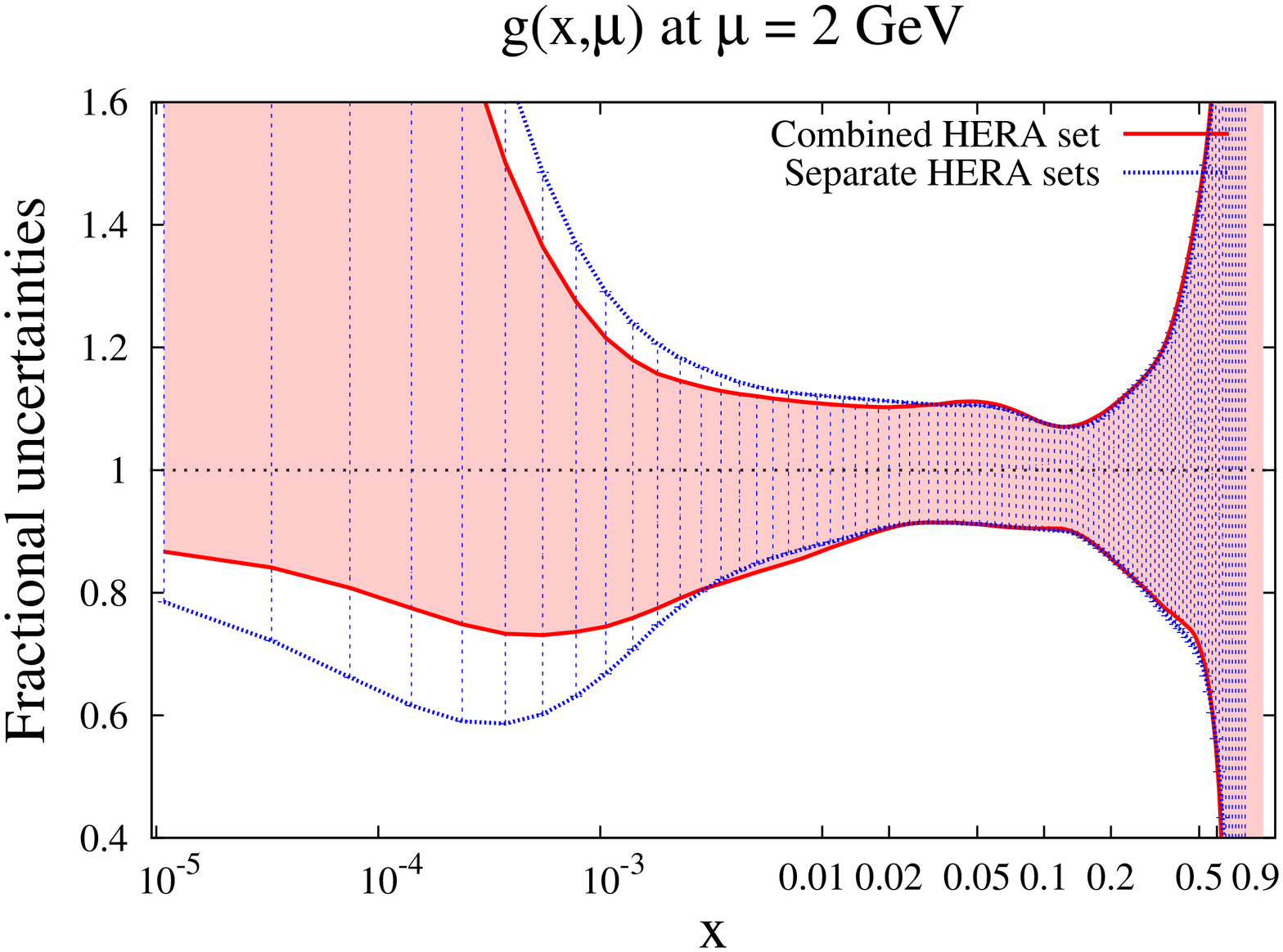}\includegraphics[width=0.49\columnwidth,height=0.49\columnwidth,keepaspectratio,angle=-90]{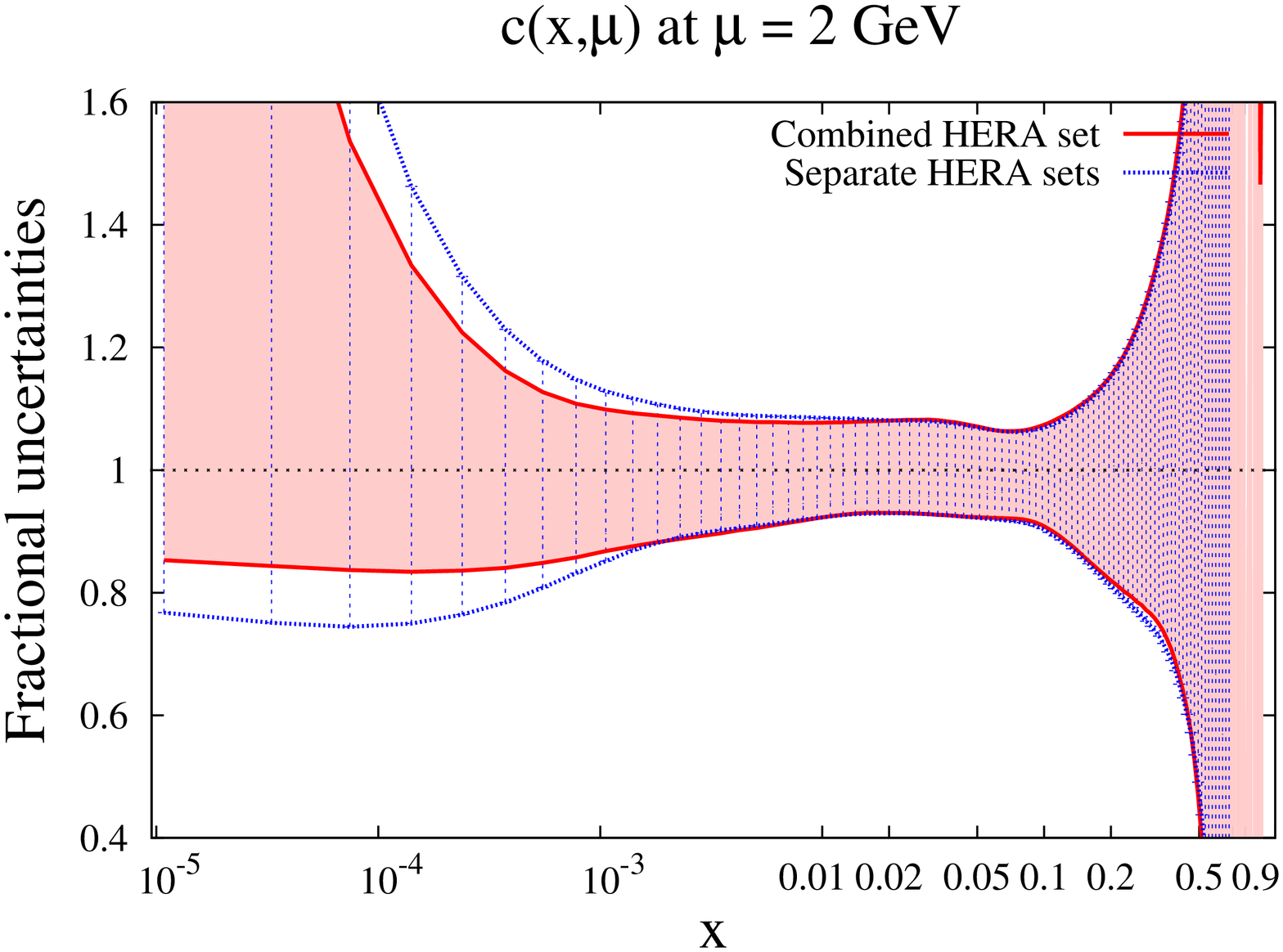}
\hspace{1cm}(a)
\hspace{8.3cm}(b)
\caption{{\small Impact of the combination of HERA-1 data sets on the PDFs uncertainties: $g(x,\mu),c(x,\mu)$,
$\mu=2$ GeV.}}
\centering{}\label{err-no_norm} 
\end{figure}

Modifications induced by the combination of HERA-1 sets 
are illustrated by figures comparing the PDFs in the CT10 and
alternative CT10 fits.  
Figs.~\ref{err-no_norm}(a,b) show error bands for (a) the
gluon and (b) the charm quark, as a function
of Bjorken $x$ at $\mu=2$ GeV. 
These PDFs are chosen because they
exhibit the largest changes upon the combination of the HERA-1 sets. The
modifications in the bottom quark are comparable to those in gluon and
charm, while the changes for other flavors are smaller. 

The error bands in Figs.~\ref{err-no_norm}(a,b) represent
the asymmetric positive and negative uncertainties 
of the PDFs $f_a(x,\mu)\equiv f$, 
computed as \cite{Nadolsky:2001yg}
\begin{eqnarray}
 &  & \delta^{+}f=\sqrt{\sum_{i=1}^{N_a}\left[\textrm{max}\left(f_{i}^{(+)}-f_{0},f_{i}^{(-)}-f_{0},0\right)\right]^{2}},\nonumber \\
 &  & \delta^{-}f=\sqrt{\sum_{i=1}^{N_a}\left[\textrm{max}\left(f_0 - f_{i}^{(+)},f_0 - f_{i}^{(-)},0\right)\right]^{2}}, 
\end{eqnarray}
in terms of $f_{0}$, the best-fit (central) PDF value, 
and $f^{\pm}_i$, the PDFs for positive and
 negative variations of the PDF parameters along the $i$-th
 eigenvector direction in the $N_a$-dimensional PDF parameter space.
The red solid band corresponds to the combined HERA set,
and the blue hatched band corresponds to the separate sets.
The uncertainties are shown as ratios to the central
PDFs in their respective fits,
\begin{eqnarray}
\Delta^{(+)}f(x)=\frac{f_0+\delta^{+}f}{f_0}; &  & \Delta^{(-)}f(x)=\frac{f_0-\delta^{-}f}{f_0}.
\end{eqnarray}
The impact of the HERA-1 data on the uncertainties
of the gluon and charm PDFs is quite clear in the small-$x$ region,
starting from $x=10^{-3}$ and going down to $x=10^{-5}$, where we observe
contraction of the error bands. In the large $x$ region, the error
bands for the combined and separate HERA data sets are
almost coincident.

\begin{figure}

\begin{centering}
\includegraphics[width=0.49\columnwidth,height=0.49\columnwidth,keepaspectratio,angle=-90]{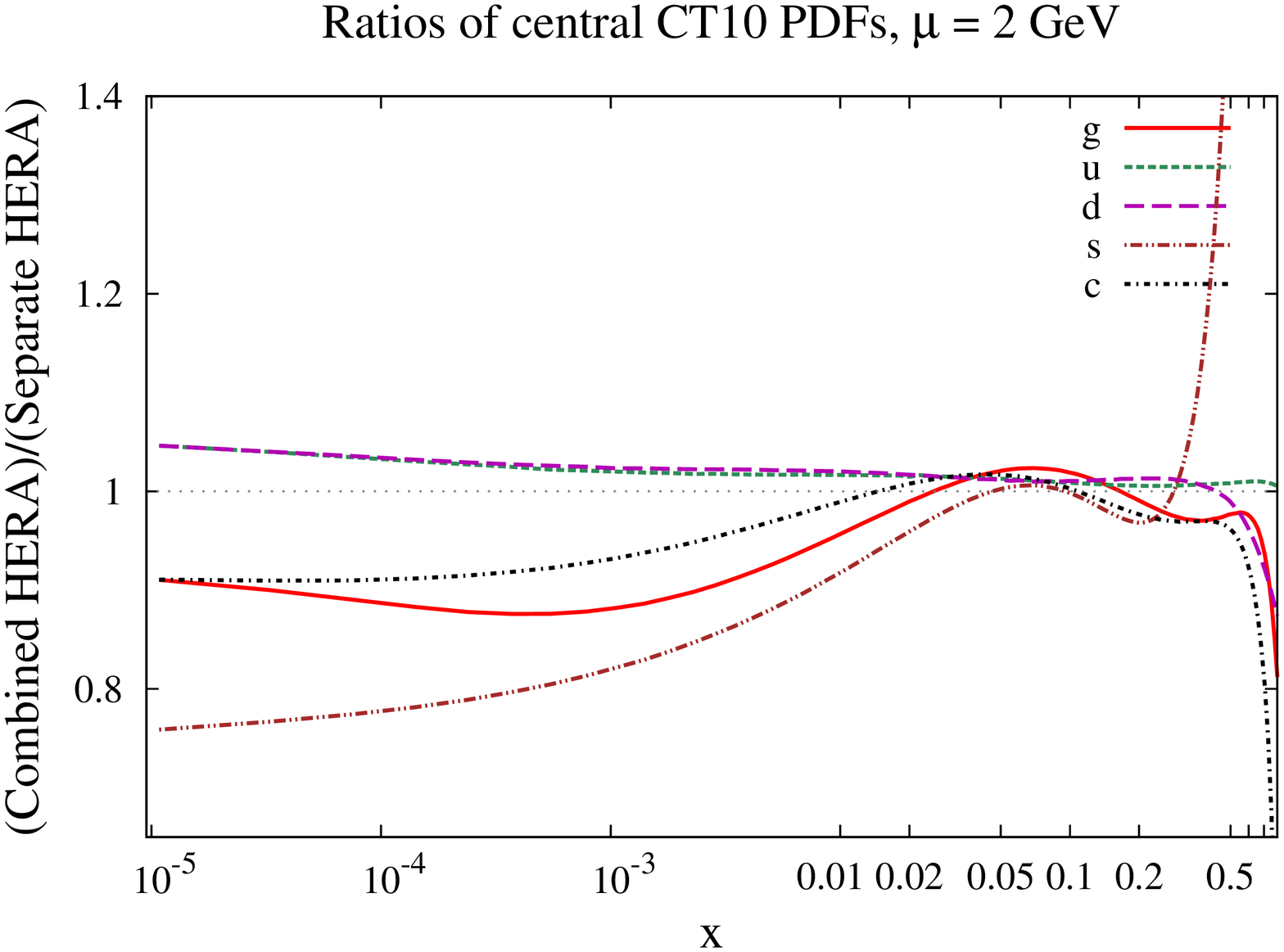}\includegraphics[width=0.49\columnwidth,height=0.49\columnwidth,keepaspectratio,angle=-90]{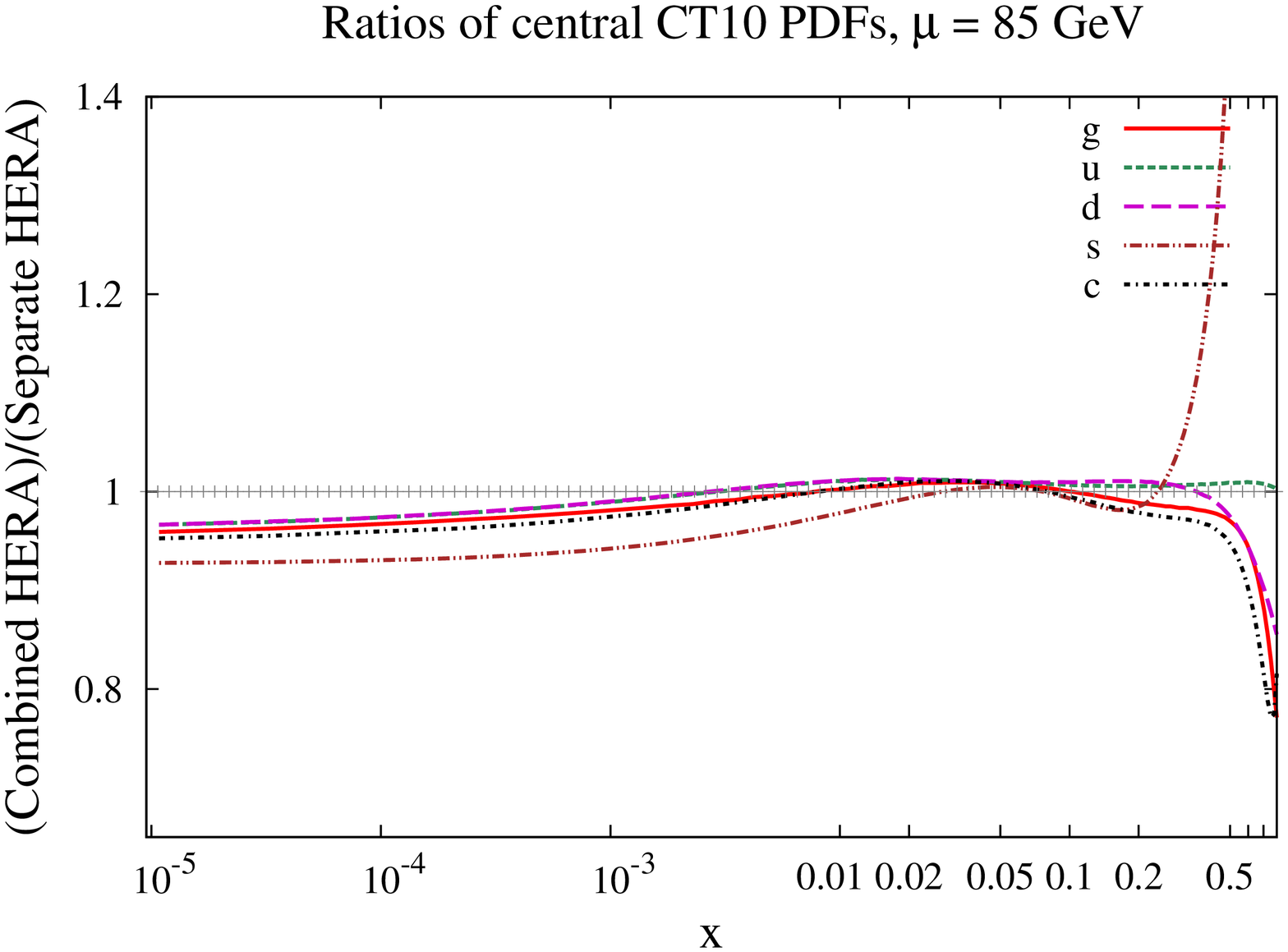}
\par\end{centering}

\hspace{1cm}(a)
\hspace{8.3cm}(b)

\caption{{\small Ratios of CT10 PDFs (fitted to the combined HERA data
    set)
to the alternative CT10 PDFs (fitted to the separate HERA data sets),
for $\mu=2$ and $85$ GeV.}}

\centering{}\label{ratio_BA} 
\end{figure}

 Ratios of the PDFs in the central PDF sets 
of the CT10 and alternative CT10 fits 
are shown in Figs.~\ref{ratio_BA}(a) and (b), 
 at $\mu=2$ GeV and $85$ GeV. At $\mu=2$ GeV (Fig.~\ref{ratio_BA}(a)),
the effect of the new data is again most
 evident in the behavior of the gluon and charm PDFs at $x$ below $10^{-2}$. 
 These PDFs are suppressed by up to 10\% 
 upon the combination of the HERA sets.
In addition, one observes a
 suppression of the strange (anti-)quark PDF, 
which, however, is small compared to the large PDF uncertainty
 associated with this flavor. The light-quark PDFs are slightly
 enhanced at small $x$
, while at medium to large $x$ region, down-quark PDF becomes smaller and
up-quark PDF remains about the same. 

Fig.~\ref{ratio_BA}(b) shows how these ratios are impacted 
by the DGLAP evolution to $\mu=85$ GeV. Some suppression persists 
in the gluon PDF at $x<0.01$, but this is diminished by the singlet
evolution, which also suppresses 
the ratios for all quarks in the same $x$ region.
At medium to large $x$, the features of PDFs are similar to 
those at $\mu=2$ GeV described above.

All the differences observed between the PDFs using the combined and the 
separate HERA data sets are fully contained in the respective error bands,
so no tension between the best-fit solutions of the two
fits is evident. The resulting changes in predictions for collider
observables, with the exception of those sensitive to gluon or
heavy-quark scattering at $x<0.01$ and small momentum scales, are thus
expected to be mild. 

\begin{table}
\begin{tabular}{|c|c|c|c|}
\hline 
Collider/observable & $\sigma\left(pp\!\!\!\!{}^{{}^{(-)}}\rightarrow (W^{\pm}\rightarrow\ell\nu_{\ell})X\right)$ & $\sigma\left(pp\!\!\!\!{}^{{}^{(-)}}\rightarrow(Z^{0}\rightarrow\ell\bar{\ell})X\right)$ & $\frac{\sigma\left(pp\!\!\!\!{}^{{}^{(-)}}\rightarrow (W^{\pm}\rightarrow\ell\nu_{\ell})X\right)}{\sigma\left(pp\!\!\!\!{}^{{}^{(-)}}(Z^{0}\rightarrow\ell\bar{\ell})X\right)}$\tabularnewline
\hline
\hline 
Tevatron, $\sqrt{s}=1.96$ TeV & $+2.6\%$ & $+2.5\%$ & $-0.05\%$\tabularnewline
\hline 
LHC, $\sqrt{s}=7$ TeV  & $+0.7\%$ & $+0.7\%$ & $-0.02\%$\tabularnewline
\hline 
LHC, $\sqrt{s}=14$ TeV  & $-0.5\%$ & $-0.5\%$ & $-0.07\%$\tabularnewline
\hline
\end{tabular}

\caption{Percent changes in CT10 total cross sections for inclusive $W^{\pm}$
and $Z^{0}$ boson production at the Tevatron Run-II and LHC, caused
by the replacement of separate HERA-1 cross sections by the combined
HERA data set.\label{tab:CombHERAWZ}}

\end{table}

As an illustration,
Table~\ref{tab:CombHERAWZ} shows the changes, due to the
combination of the HERA sets, in inclusive $W$ and
$Z$ boson production cross sections at the Tevatron and LHC, as well
as in their ratios, computed at NLO in $\alpha_s$ in accordance with
the settings discussed in Sec.~\ref{sec:applications}. 
The largest observed change is an increase of
2.5\% in the $W$ and $Z$ cross sections in the Tevatron Run-II.
Changes in the LHC cross sections are about 0.7\% at most. 
These changes are well correlated in the $W$ and
$Z$ scattering processes, so that the ratio of the $W$ and $Z$ cross
sections, shown in the last column of the table, changes (decreases)
marginally by 0.02-0.07\%.

\section{$W$ lepton asymmetry in the global PDF analysis \label{sec:WLasy}}

The interest in the Tevatron $W$ boson charge asymmetry $A_\ell$
originated in the late 1980's \cite{Berger:1988tu,Martin:1988aj}, 
when its measurement
was proposed in order to resolve a controversy between
constraints on the ratio of up and down quark PDFs, $d(x,\mu)/u(x,\mu)$, 
obtained from DIS on hydrogen and deuterium
targets. At the time, a discrepancy between the $d/u$ values derived from 
DIS data by BCDMS \cite{Benvenuti:1989rh,Benvenuti:1989fm}, 
EMC \cite{Aubert:1987da}, and, to some extent, 
SLAC \cite{Whitlow:1990dr} limited the accuracy of predictions
of $W$ and $Z$ boson observables in the early Tevatron runs, 
notably $\sigma(W)/\sigma(Z)$, $\Gamma_W/\Gamma_Z$, and $M_W$.
A more precise measurement of proton and deuteron DIS cross
sections by NMC \cite{Amaudruz:1992bf} was found 
to be in better agreement with
BCDMS than with EMC. Several 
theoretical \cite{Badelek:1991qa,Virchaux:1991jc} and
experimental \cite{Bazizi:1991mq,Milsztajn:1990cc} 
factors were also identified that could
cause the discrepancy and, in the long run, limit the accuracy of
determination of the $d/u$ ratio from the DIS cross sections. 
So, when CDF measured $A_\ell$ 
\cite{Abe:1994rj} and found it to agree with the PDFs
fitted to the BCDMS+NMC data and disagree with the PDFs 
fitted to the EMC data, the controversy was generally assumed to 
be resolved in favor of BCDMS and NMC. The combination of the BCDMS,
NMC, and CDF $A_\ell$ data sets has been used since then 
as a self-consistent input by MRSA \cite{Martin:1994kn},
CTEQ3 \cite{Lai:1994bb}, and subsequent global analyses.  

This status quo has been challenged recently by high-luminosity
measurements of $W$ charge asymmetry in electron and muon channels by
D\O~ \cite{d0_e_asy, d0_mu_asy}. The D\O~ data disagree
significantly with NLO theoretical predictions based on CTEQ6.1 and 6.6 PDFs
\cite{d0_e_asy, d0_mu_asy}. They disagree even more with
the PDFs produced by the other groups \cite{Catani:2010en}. 
When the D\O~ $A_\ell$ data are included in our global fit, 
they show significant tension with the NMC ratio 
$F_2^d(x,Q)/F_2^p(x,Q)$, BCDMS $F_2^d(x,Q)$, and CDF Run I $A_\ell$,
but are generally
compatible with the other data sets -- not unexpectedly, since it is
mostly the above three sets that probe the same PDF ratio $d/u$. In
addition, there appears to be some disagreement among 
the subsets of the D\O~ $A_\ell$ data themselves, as will be discussed below. 

To understand how the $W$ charge asymmetry data can seriously contradict
some PDF sets in spite of the agreement of these PDFs 
with other precise measurements, note that
$A_\ell(y)$ is very sensitive to the average slope of $d(x,M_W)/u(x,M_W)$ 
in the relevant kinematic region
\cite{Berger:1988tu,Martin:1988aj}. 
Small differences between the slopes of distinct PDF sets can significantly
change the behavior of $A_\ell$; see, for instance, Figs. 2 and
19 of Ref.~\cite{Lai:1994bb}. It is therefore not 
surprising that the existing PDF sets, while being 
compatible with the available fixed-target DIS cross sections, can
vary drastically in their predictions for $A_\ell$. 

The emerged discord in $W$ asymmetry measurements poses a
dilemma for our global analysis. On one hand, $W$ boson production is not
affected by hard-to-control uncertainties typical for DIS on a deuterium
target. Several factors beyond the leading-power perturbative QCD
affect deuterium DIS cross sections at $x > 0.1$,
including target-mass, dynamical higher-twist, and 
nuclear binding  effects \cite{OwensLargeX}. 
(No nuclear corrections to the deuteron DIS data are included 
in this analysis.)
In principle, these factors themselves 
need to be determined from the DIS data, increasing 
the uncertainty in the resulting PDFs. In practice, 
their impact is minimized by the selection cuts imposed on the DIS data 
included in the global analysis. Even with 
the safeguards, the large-$x$ quark PDFs may have residual 
sensitivity to these uncertainties 
beyond the leading-twist QCD \cite{OwensLargeX, Thorne:2010kj}.

On the other hand,
the fixed-target DIS experiments continue to provide significant
  constraints on the PDFs both at intermediate and large $x$
 \cite{Pumplin:2009sc} and cannot be
  discarded without increasing the PDF uncertainties; nor are the tensions 
between the subsets of the $A_\ell$ data fully
understood yet. Until these issues are clarified, our
  provisional solution is to present two separate sets
  of PDFs, CT10 without the D\O~ Run-II $A_\ell$ data, and CT10W with them,
  in order to explore possible implications for collider experiments sensitive 
  to the $d/u$ ratio.

\subsection{Detailed comparison to D\O~ lepton asymmetry data}
The Tevatron charge asymmetry studied here is constructed from
rapidity distributions, $d\sigma/dy_\ell$, of the charged
lepton $\ell = e$ or $\mu$ from the decay of the $W$ boson:
\begin{equation}
A_\ell(y_\ell)=\frac{d\sigma(p\bar p\rightarrow (W^+\rightarrow
  \ell^+\nu_\ell) X)/dy_\ell \ -\ d\sigma(p\bar
  p\rightarrow (W^-\rightarrow \ell^-\bar \nu_\ell) X)/dy_\ell}
  {d\sigma(p\bar p\rightarrow (W^+\rightarrow
  \ell^+\nu_\ell )X)/dy_\ell \ + \ 
d\sigma(p\bar p\rightarrow (W^-\rightarrow \ell^-\bar
  \nu_\ell)X)/dy_\ell}.
\end{equation}
These distributions are observed directly; selection cuts
are usually imposed on transverse momentum $p_T^\ell$ of $\ell$ in
various bins to emphasize the sensitivity of this distribution to
$d(x,\mu)/u(x,\mu)$ in different ranges of $x$ \cite{Acosta:2005ud}.

We compute $A_\ell(y_\ell)$ using the program ResBos
\cite{Balazs:1995nz,Balazs:1997xd,Landry:2002ix}, which returns fully
differential cross sections for both decay leptons at NLO and, in
addition, performs  next-to-next-to-leading-logarithm (NNLL) 
resummation at small transverse momenta of $W$ bosons. The $A_\ell(y_\ell)$
distributions with cuts on $p_T^\ell$ have some sensitivity to
the resummed and NNLO corrections \cite{Balazs:1997xd,Catani:2010en},
which can reach a few percent at the largest values of $y_\ell$ accessible
at the Tevatron. We examined the magnitude 
of these corrections and found them to be unimportant
in comparison to the current experimental errors.\footnote{The
  sensitivity to NNLO effects was examined by redoing the calculation 
  for $A_\ell(y_\ell)$ after adding
  the exact $\alpha^2_s$ correction for $W$ bosons produced 
with non-zero transverse momentum. This correction captures a large
part of the full NNLO effect. The changes in the results were
 found to be small and 
comparable to the difference between the exact NLO and NNLO $A_\ell$
values found in Ref.~\cite{Catani:2010en}.}

\begin{table}

\begin{tabular}{|c|c|c|c|c|c|c|}
\hline 
Bin & $\ell$ & Cut & Points & $\chi^2$ (CT10) & $\chi^2$ ($w$=1) & $\chi^2$ (CT10W)\tabularnewline
\hline
\hline 
1 & $e$ & $p_T^\ell>25$ GeV & 12 & 79.5 & 37.2 & 25.3\tabularnewline
\hline 
2 & $e$ & $25<p_T^\ell<35$ GeV & 12 & 20.7 & 20.3 & 25.5\tabularnewline
\hline 
3 & $e$ & $p_T^\ell>35$ GeV & 12 & 91.4 & 41.7 & 26.5\tabularnewline
\hline 
4 & $\mu$ & $p_T^\ell>20$ GeV & 9 & 8.3 & 10.8 & 13.5\tabularnewline
\hline
\end{tabular}

\caption{$\chi^2$ of D\O~ Run-II $W$ lepton asymmetry data in
  representative PDF fits.\label{tab:wlasy}}

\end{table}

Any fit that agrees with the Run-II $A_\ell$ 
must sacrifice some of the agreement with the  
Run-I $A_\ell$ data and some DIS experiments, as both are
probing similar PDF kinematics. To obtain a reasonable $\chi^2$ in the
CT10W fit, we find it necessary
to increase the $\chi^2$ weight 
of the D\O~ Run-II $A_\ell$ data, as we did,
{\it e.g.}, for a special PDF set (CTEQ4HJ) for high $E_T$ jets 
from the Tevatron in 1995 \cite{cteq4}. From the sample of D\O~ muon $A_\ell$, 
only one bin, with $p_T^\mu > 20 $ GeV, has reasonable $\chi^2$ when
fitted together with the electron $A_\ell$ data; the $\chi^2$ values
in the muon bins with $20 < p_T^\mu < 35 $ GeV and $p_T^\mu > 35 $ GeV
stay above 15 for 9 data points for all combinations of the weights tried.
The CT10W fit
therefore includes three electron $p_T^e$ bins and the compatible 
muon $p_T^\mu$ bin, as shown in Table~\ref{tab:wlasy}.\footnote{The missing transverse energy 
$\not\!\!E_{T}$ is required to be larger than 25 GeV 
in the electron asymmetry data, and 20 GeV in the muon data.}
The impact of the weights on the $\chi^2$ values 
for the D\O~ Run-II $A_\ell$ data is also shown in this table. 

The table demonstrates that the CT10 PDFs, 
obtained without the D\O~ $A_\ell$ data,
disagree strongly with bins 1 and 3 of  $A_\ell$.
In the next column, taken from a fit that includes 
the D\O~ $A_\ell$ data with weight $w=1$,  
the $\chi^2$ values in bins 1 and 3 are still rather poor.
Because the number of D\O~ $A_\ell$ data points is small, 
this fit tends to ignore them when they conflict with 
the other high-statistics data sets. To emphasize 
the four most compatible $A_\ell$ data sets,
the $\chi^2$ function of the CT10W fit, shown in the rightmost column,
includes their contributions 
with weights (5,2,5,2). The weights make 
$\chi^2$ values in this column more
acceptable, even though 
still not entirely perfect.

A measure of the tension between D\O~ $A_\ell$ and the other data 
sets can be obtained by examining 
the increase in the total $\chi^2$ for the other data sets, after 
the D\O~ $A_\ell$ data are included. 
The resulting increase is 67, so the CT10W fit 
can be considered acceptable within the CT10 analysis based on the
90\% global tolerance criterion. 
Of the total increase in $\chi^2$ of 67 units, 33  units 
are contributed by the NMC $F_{2}^{d}/F_{2}^{p}$ ratio data
\cite{NMC}. 
The other major source of conflict comes from the BCDMS
deuterium data \cite{Benvenuti:1989fm}, with an increase in $\chi^2$
of 19. 
Also significantly worse is the fit to the CDF Run-I W lepton
asymmetry data \cite{Abe:1994rj}, with an increase of $\chi^2$ 
by 5, for only 11 
data points.  
Aside from those three sets, 
all other sets accommodate CT10 and CT10W equally well.

The D\O~ Run-II W lepton asymmetry data sets 
also appear to have considerable tension among themselves.
For example,  the fit to $p_T^\ell$ bin 2 is worse in CT10W than in
CT10. 

\begin{figure}[p]
\begin{centering}
\includegraphics[clip,scale=0.72]{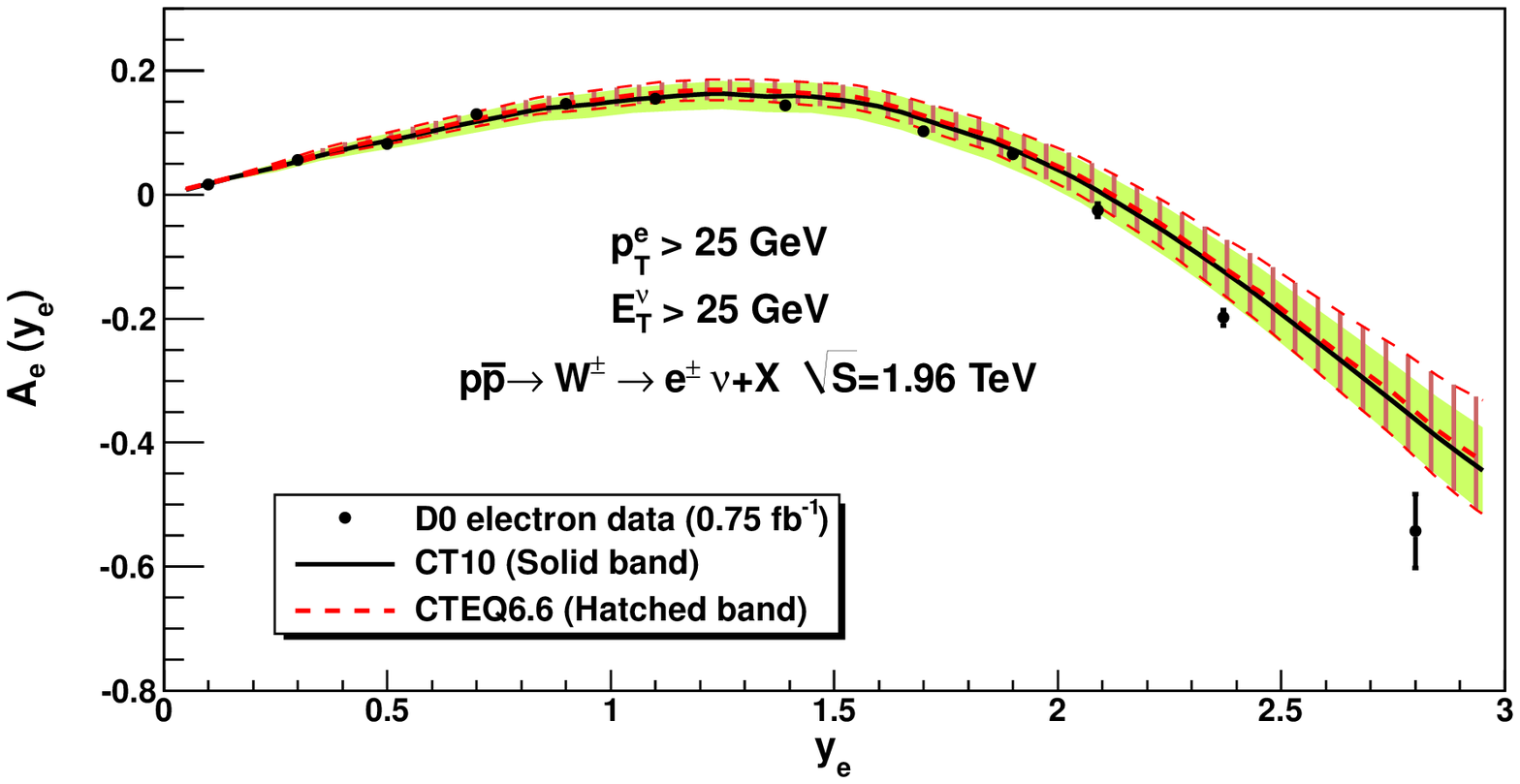} 
\includegraphics[clip,scale=0.72]{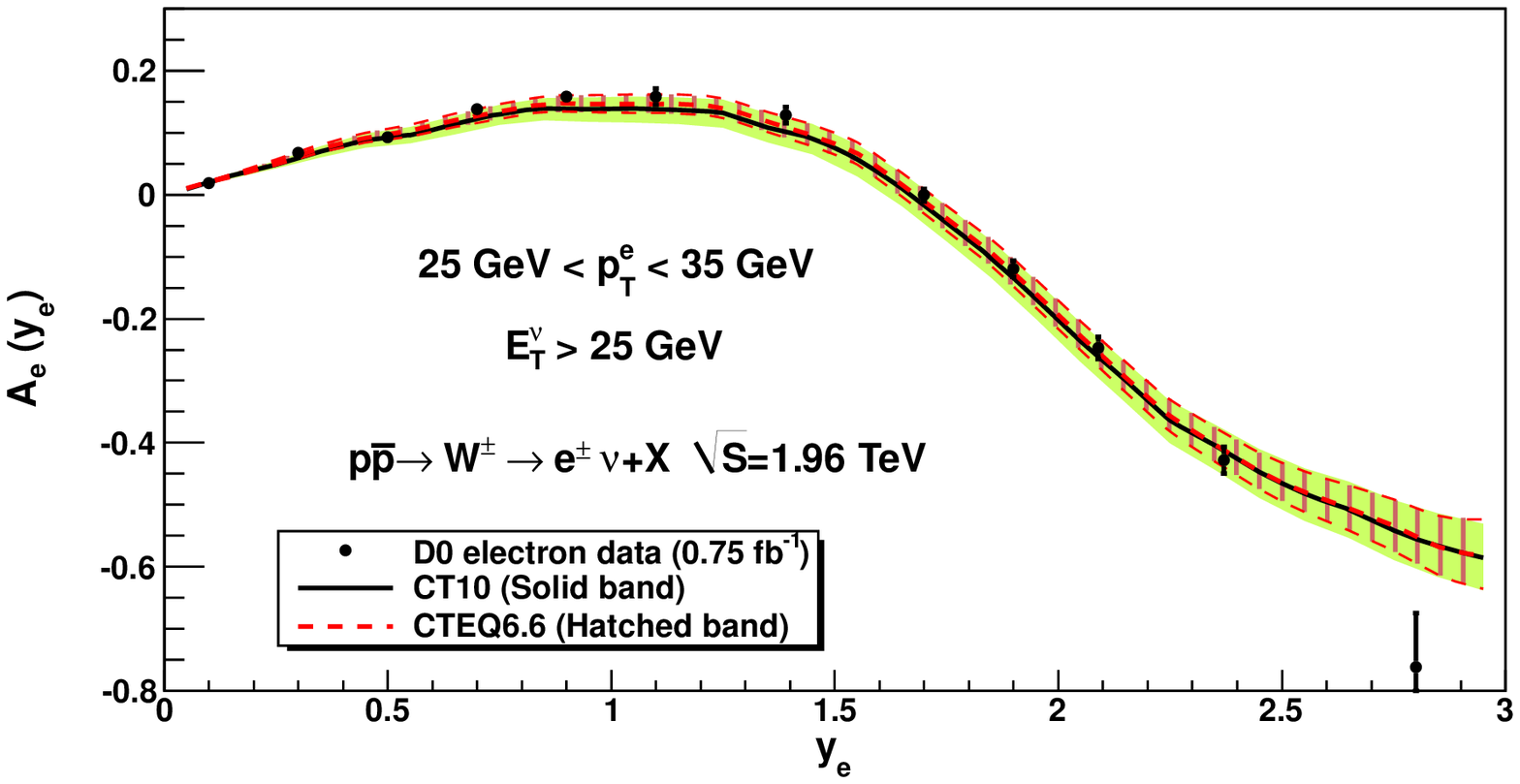}
\includegraphics[clip,scale=0.72]{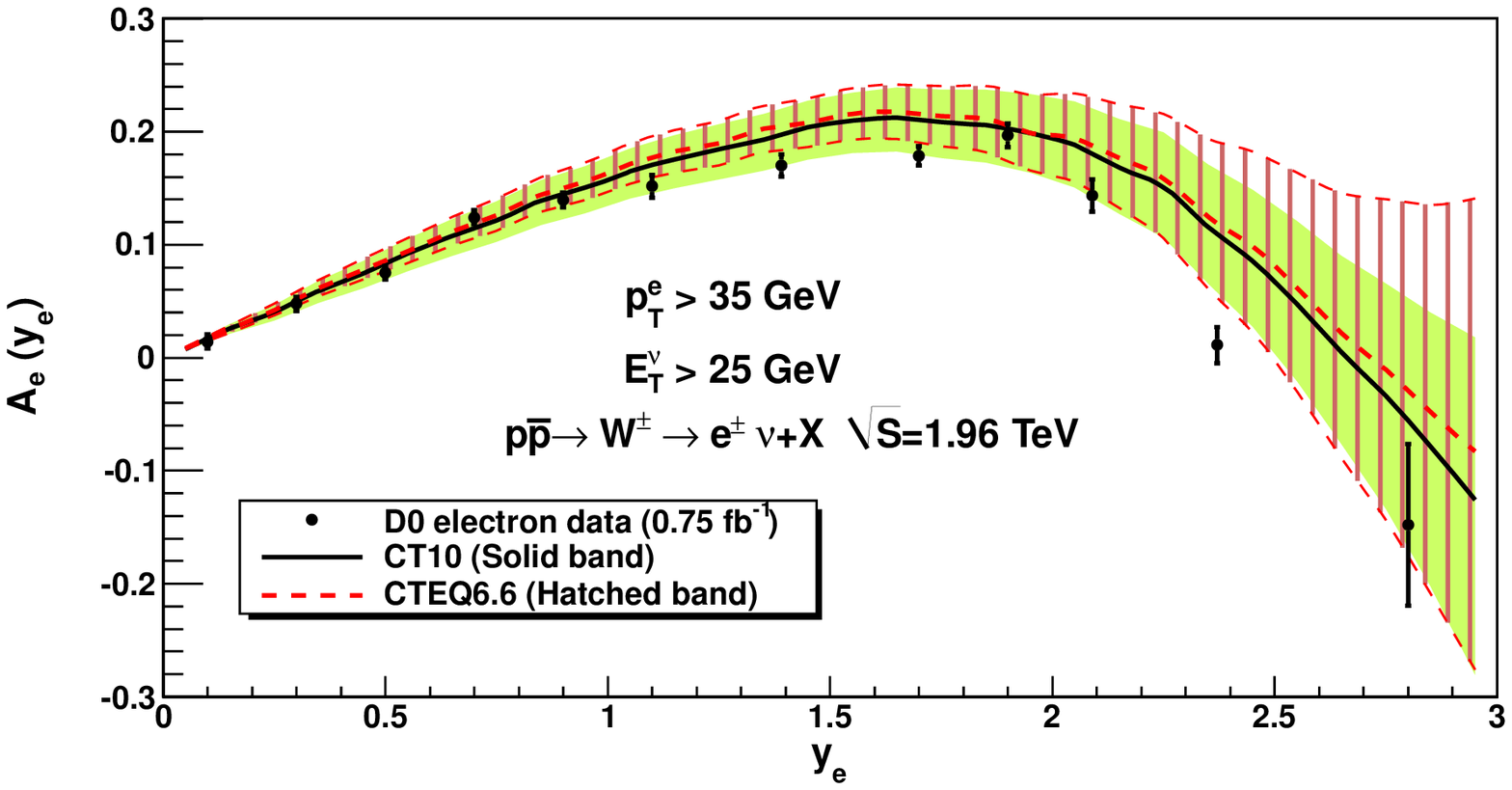} 
\par\end{centering}
\caption{
Comparison of the CT10 and CTEQ6.6 predictions with the 
D\O~ Run-II data for the electron charge asymmetry $A_e(y_e)$  for 
an integrated luminosity of 0.75 ${\rm fb}^{-1}$ \cite{d0_e_asy} .
}
\label{figs:eleAct10} 
\end{figure}

\begin{figure}[p]
\begin{centering}
\includegraphics[clip,scale=0.7]{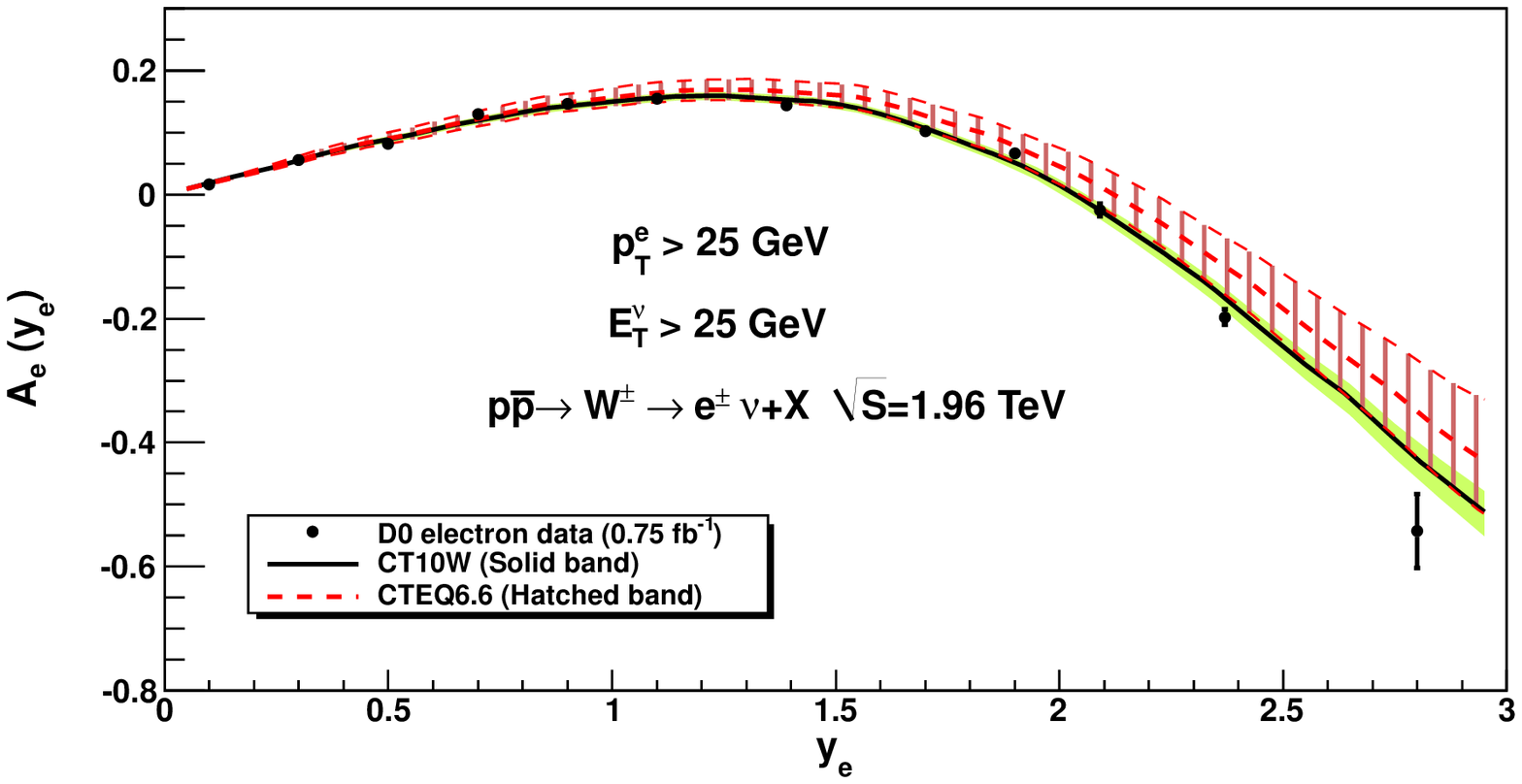} 
\includegraphics[clip,scale=0.7]{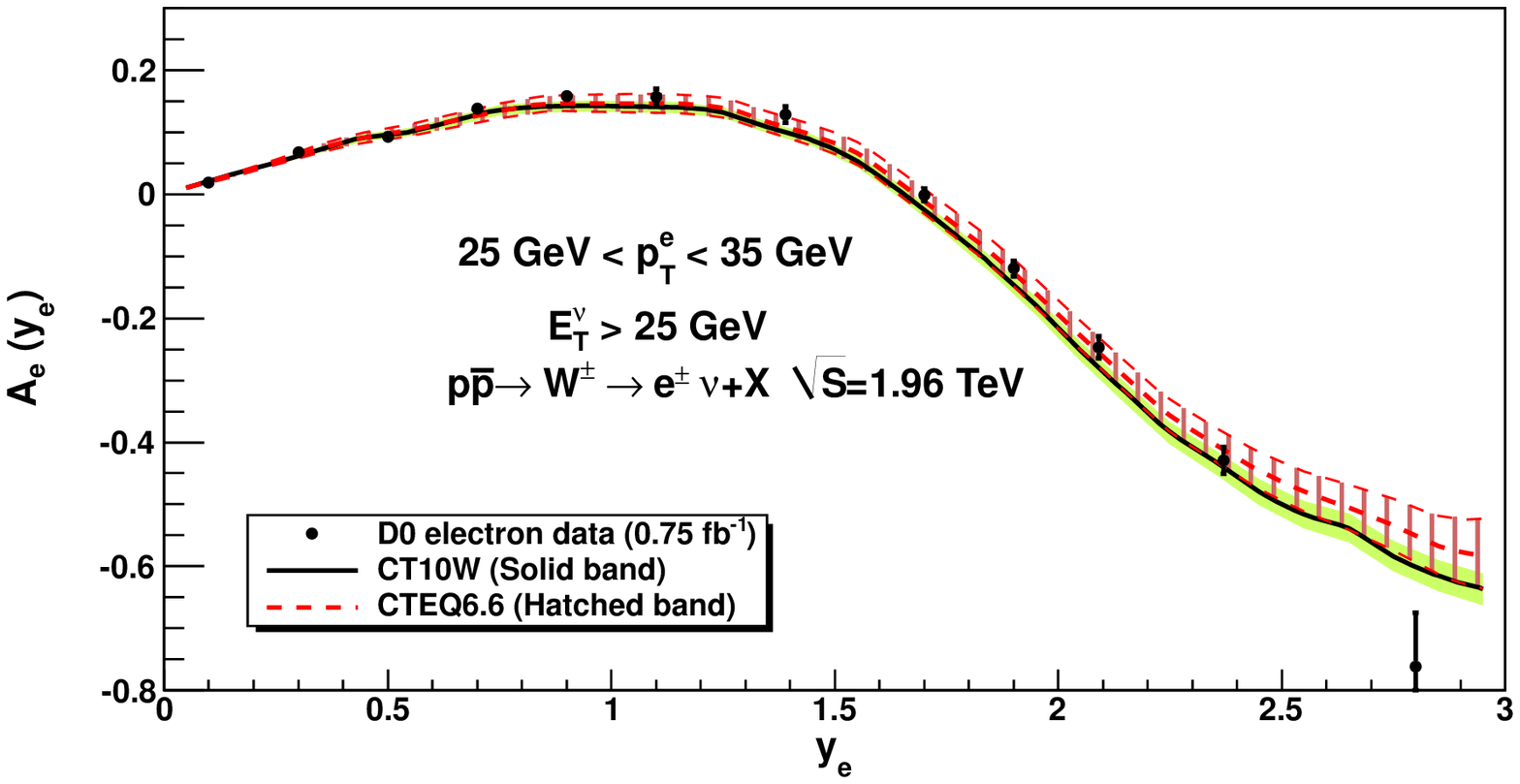}
\includegraphics[clip,scale=0.7]{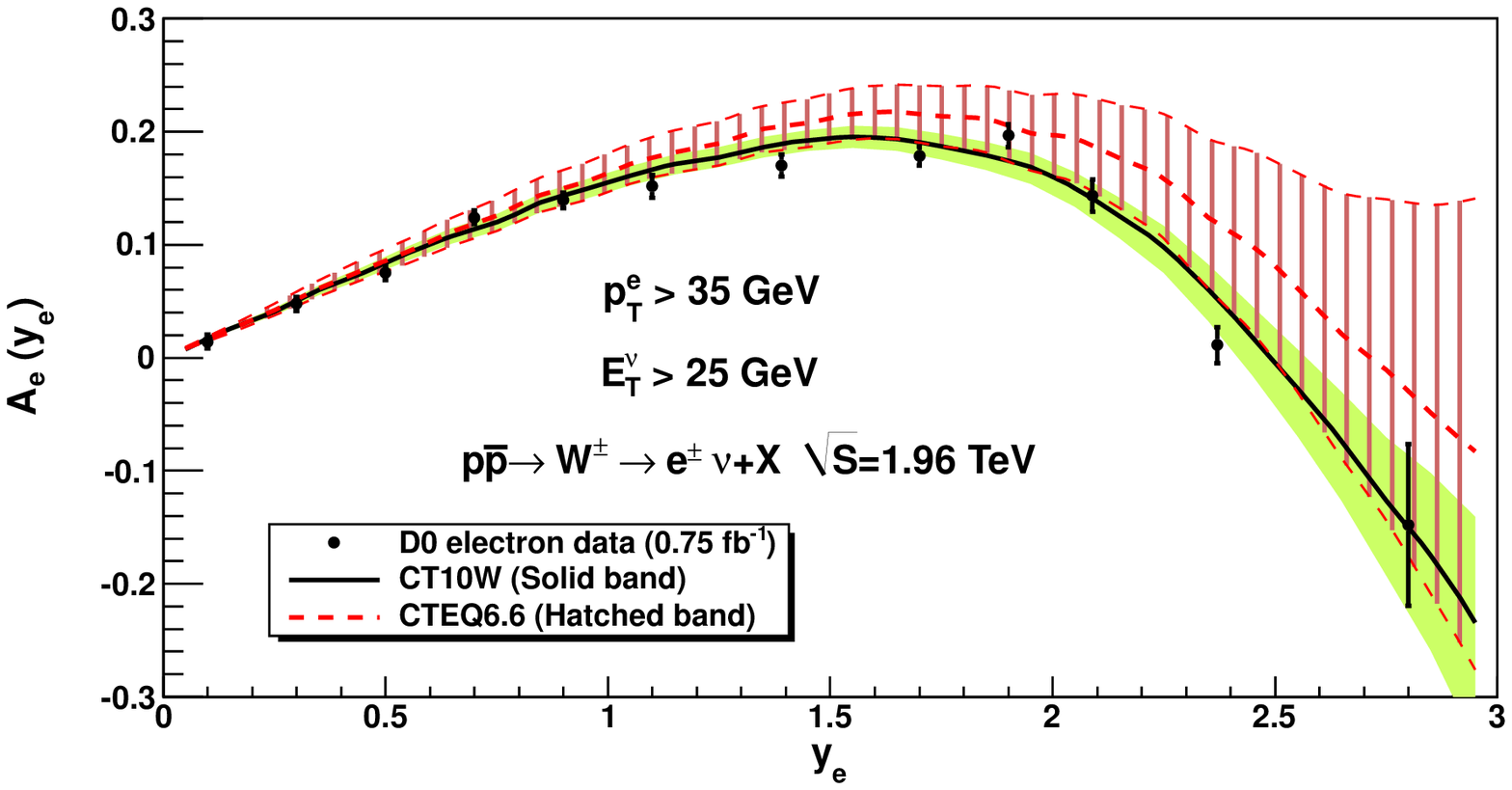} 
\par\end{centering}
\caption{Same as Fig.~\ref{figs:eleAct10}, for the CT10W PDFs.}
\label{figs:eleAct10w} 
\end{figure}

\begin{figure}[p]
\begin{centering}
\includegraphics[clip,scale=0.7]{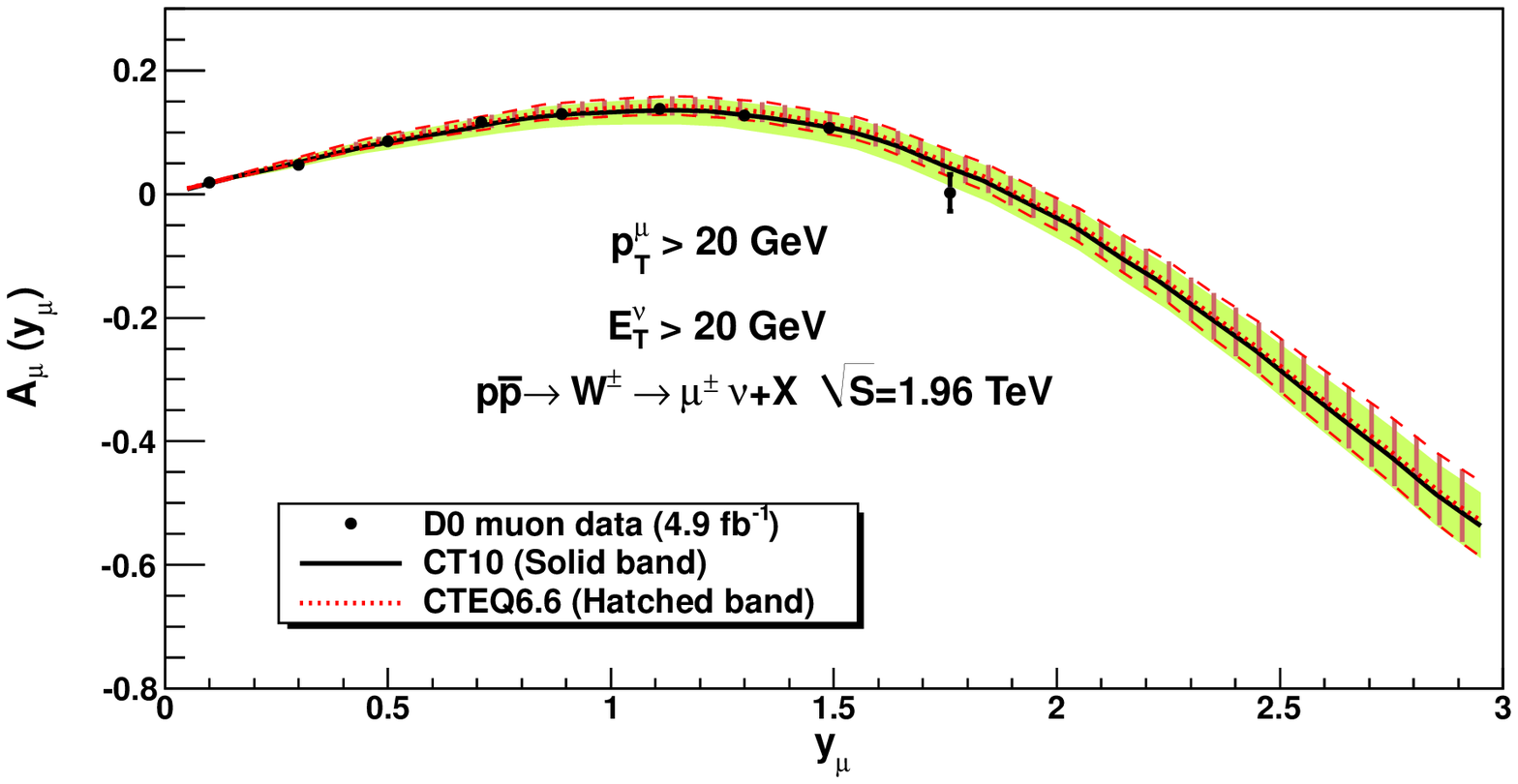} 
\includegraphics[clip,scale=0.7]{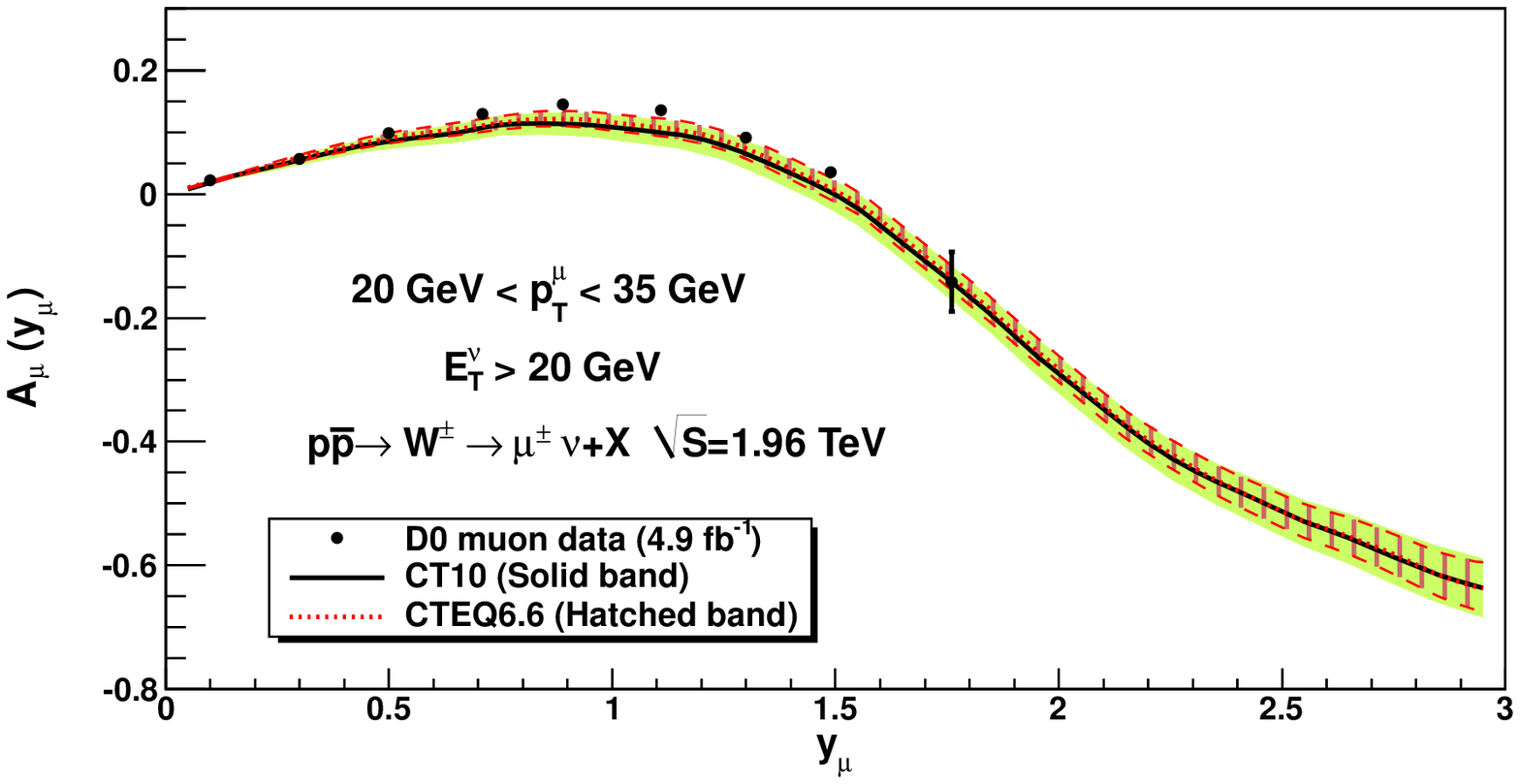}
\includegraphics[clip,scale=0.7]{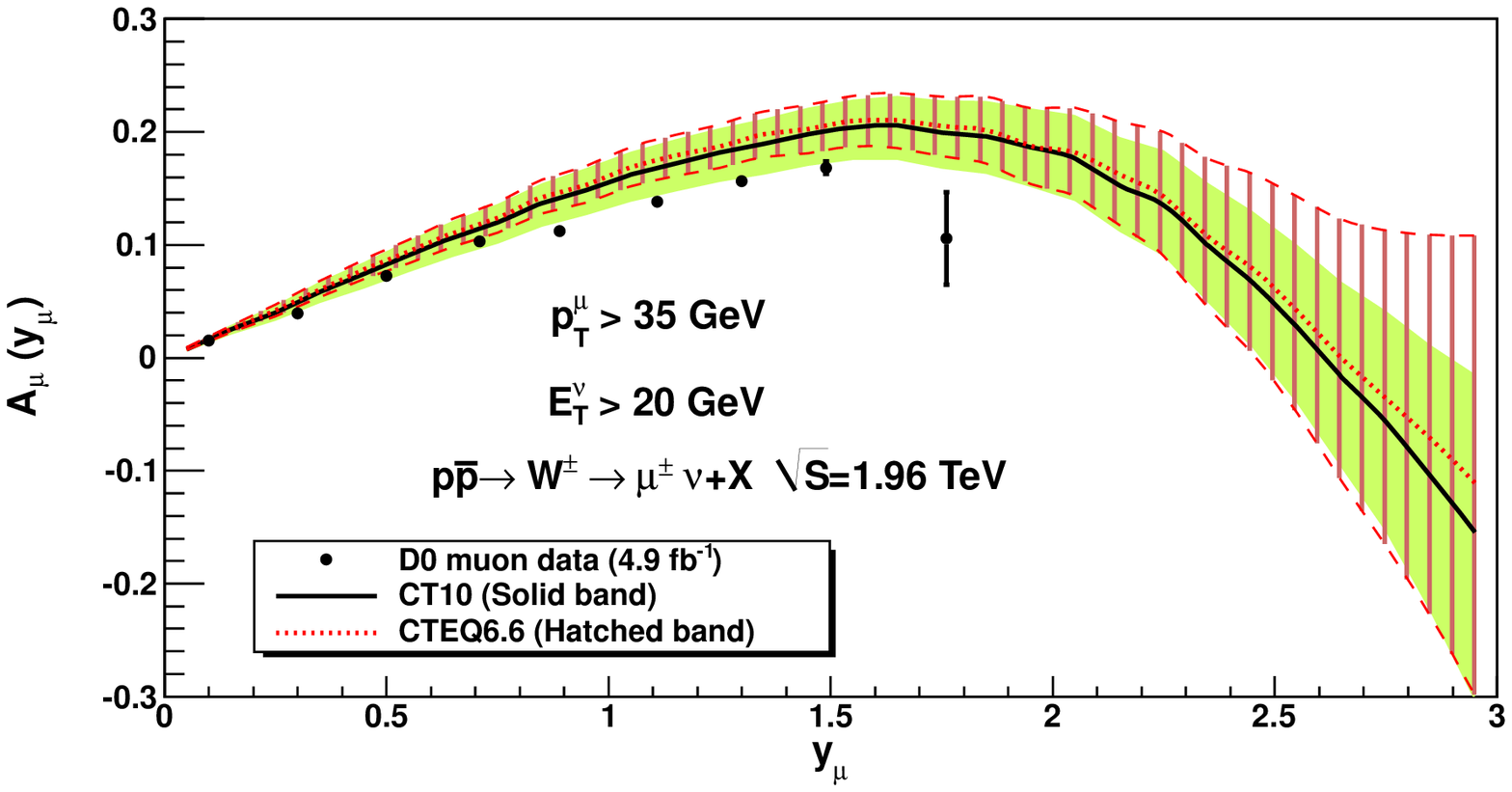} 
\par\end{centering}
\caption{
Comparison of the CT10 and CTEQ6.6 predictions with the 
D\O~ Run-II data for the muon charge asymmetry $A_\mu(y_\mu)$ 
for an integrated
luminosity of 4.9 ${\rm fb}^{-1}$ \cite{d0_mu_asy}. 
}
\label{figs:muonAct10} 
\end{figure}

\begin{figure}[p]
\begin{centering}
\includegraphics[clip,scale=0.7]{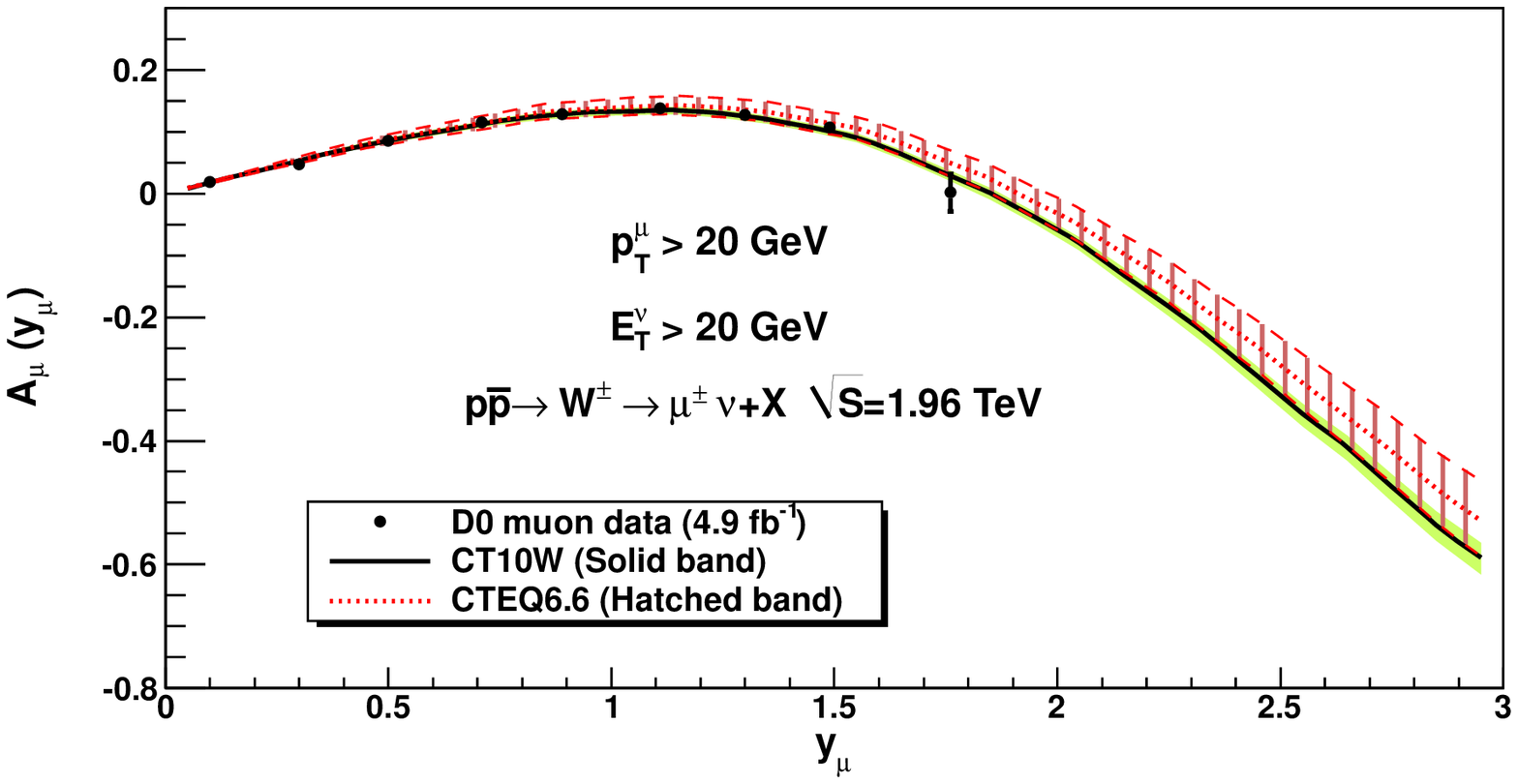} 
\includegraphics[clip,scale=0.7]{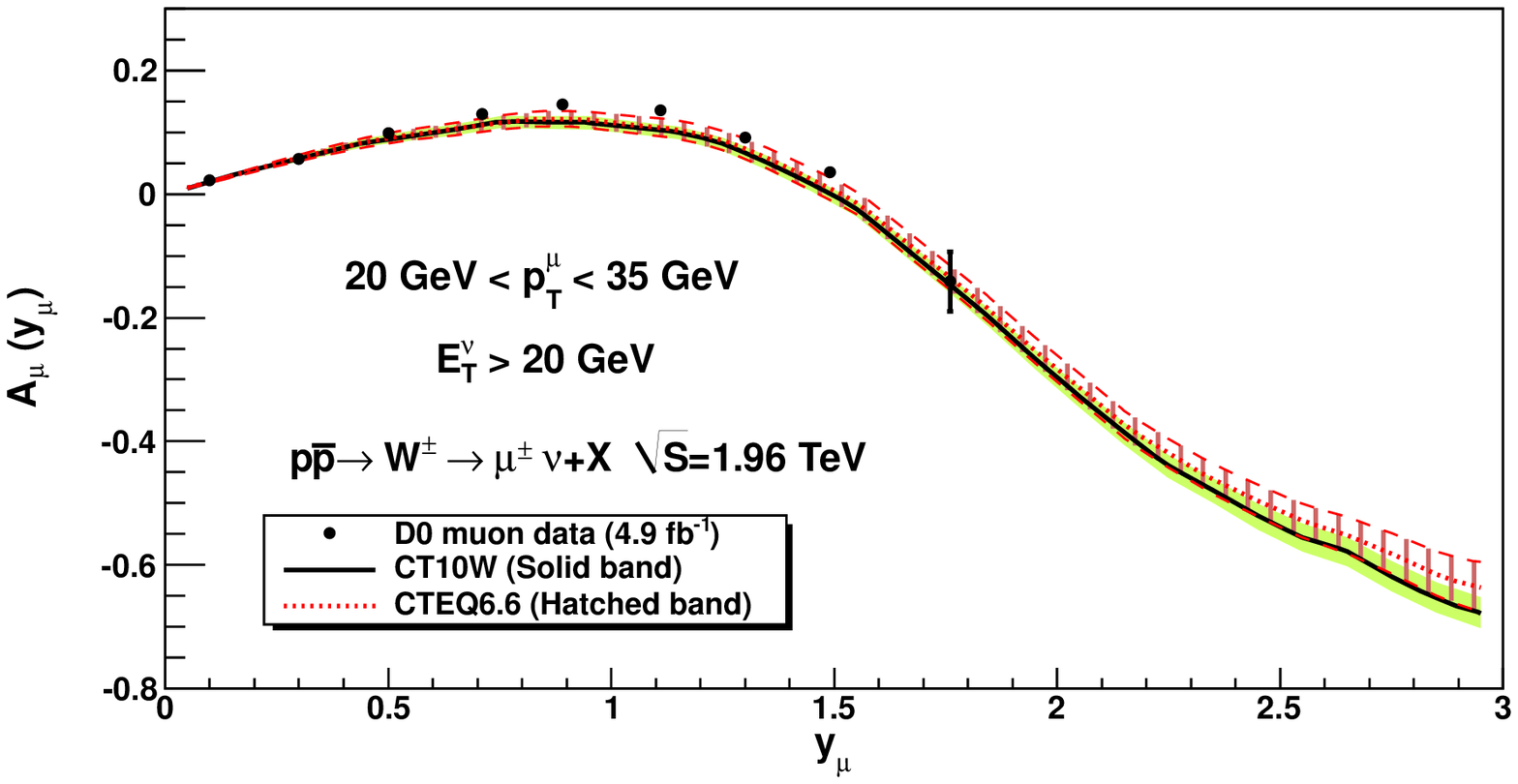}
\includegraphics[clip,scale=0.7]{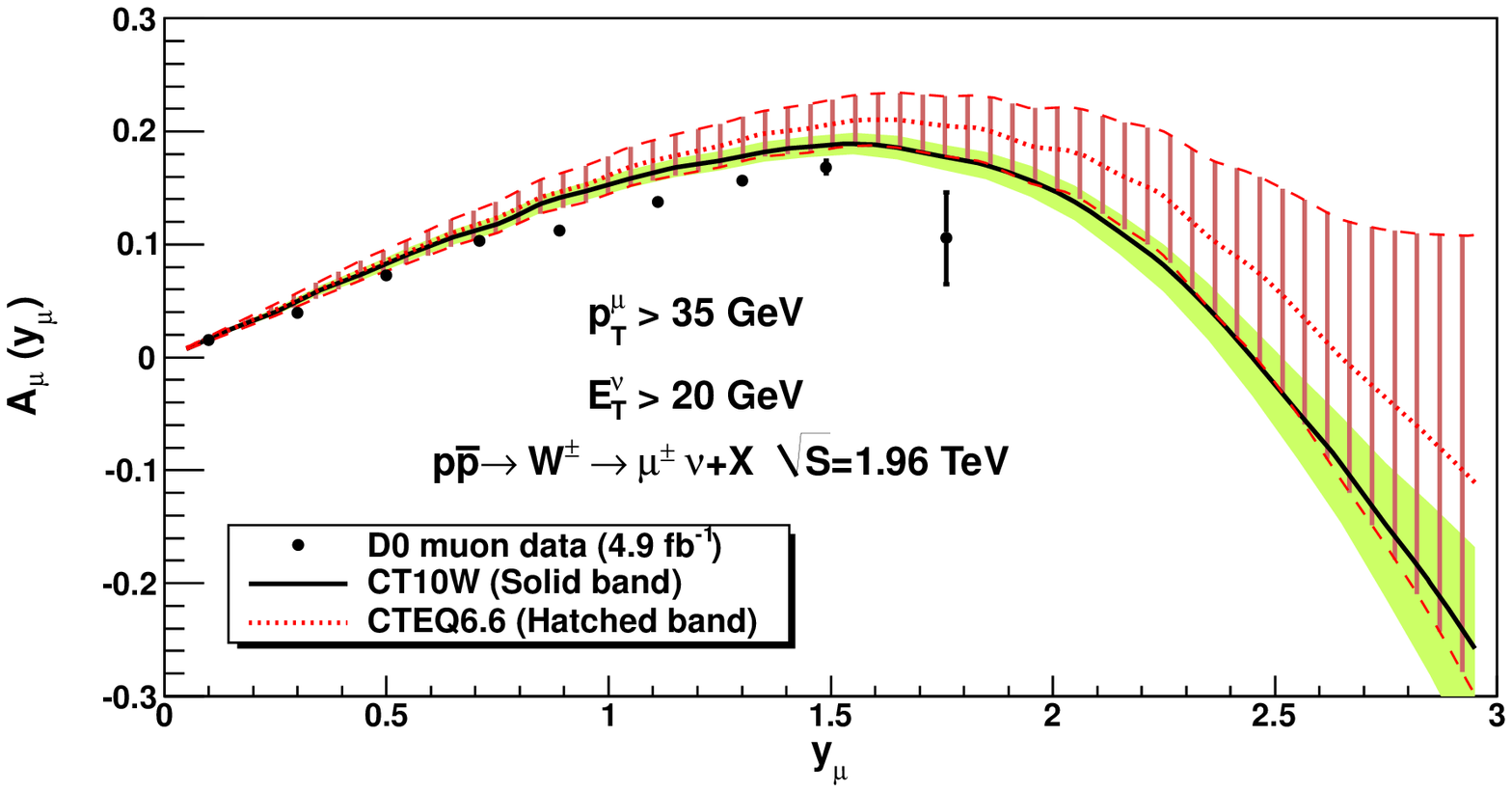} 
\par\end{centering}
\caption{
Same as Fig.\,\protect{\ref{figs:muonAct10}}, for the CT10W PDFs.
}
\label{figs:muonAct10w} 
\end{figure}

Agreement of the individual
D\O~ $A_\ell$ data points with NLO theoretical predictions based on 
CTEQ6.6, CT10, and CT10W PDF's is illustrated by Figs.\
\ref{figs:eleAct10}-\ref{figs:muonAct10w}, for the cuts on 
$p_{T}^{\ell}$ and $\not\!\!{E}_{T}$ specified in the figures.
In the case of the electron asymmetry shown in Fig.\ \ref{figs:eleAct10}, 
CT10 central values and PDF
uncertainties are similar to those obtained with CTEQ6.6,
except for the large-rapidity region 
($|y|>2$) in the bin $p_{T}^{e}>35\mbox{ GeV}$,
where CT10 predicts a somewhat smaller PDF uncertainty. 
It is obvious that the CT10 prediction  does not describe the $A_e$ data 
better than CTEQ6.6. (Note again that these data
are not included in the CT10 fit.).

In contrast, the CT10W prediction in Fig.~\ref{figs:eleAct10w},
obtained upon including the  D\O~ Run-II $A_\ell$ data, agrees with
these data much better. Most noticeably, the PDF uncertainty band of 
the CT10W set is narrower than that of CTEQ6.6 or CT10. 
As we will see in the next section, 
this reflects significant reduction in the
uncertainty of the (slope of the) $d/u$ ratio, once  
the $A_\ell$ data are included to constrain it.

Figs.\ \ref{figs:muonAct10} and \ref{figs:muonAct10w} are similar to 
Figs.\ \ref{figs:eleAct10} and \ref{figs:eleAct10w}, but 
show the D\O~ Run-II muon charge asymmetry. 
In the $p_T^\mu > 20\mbox{ GeV}$ bin,
the agreement of the CT10W.00 prediction
with the data is actually slightly worse than that of the best-fit CTEQ6.6
set (CTEQ6.6M) 
and CT10.00 predictions. All three theoretical predictions (CTEQ6.6, CT10, and
CT10W) disagree with the data in the other two $p_T^\mu$ bins.
Taken together, Figs.\ \ref{figs:muonAct10} and \ref{figs:muonAct10w}
suggest that only one $p_T^\mu$ bin of
the muon asymmetry data can be accommodated in the fit.

\begin{figure}
\begin{centering}
\includegraphics[clip,scale=0.8]{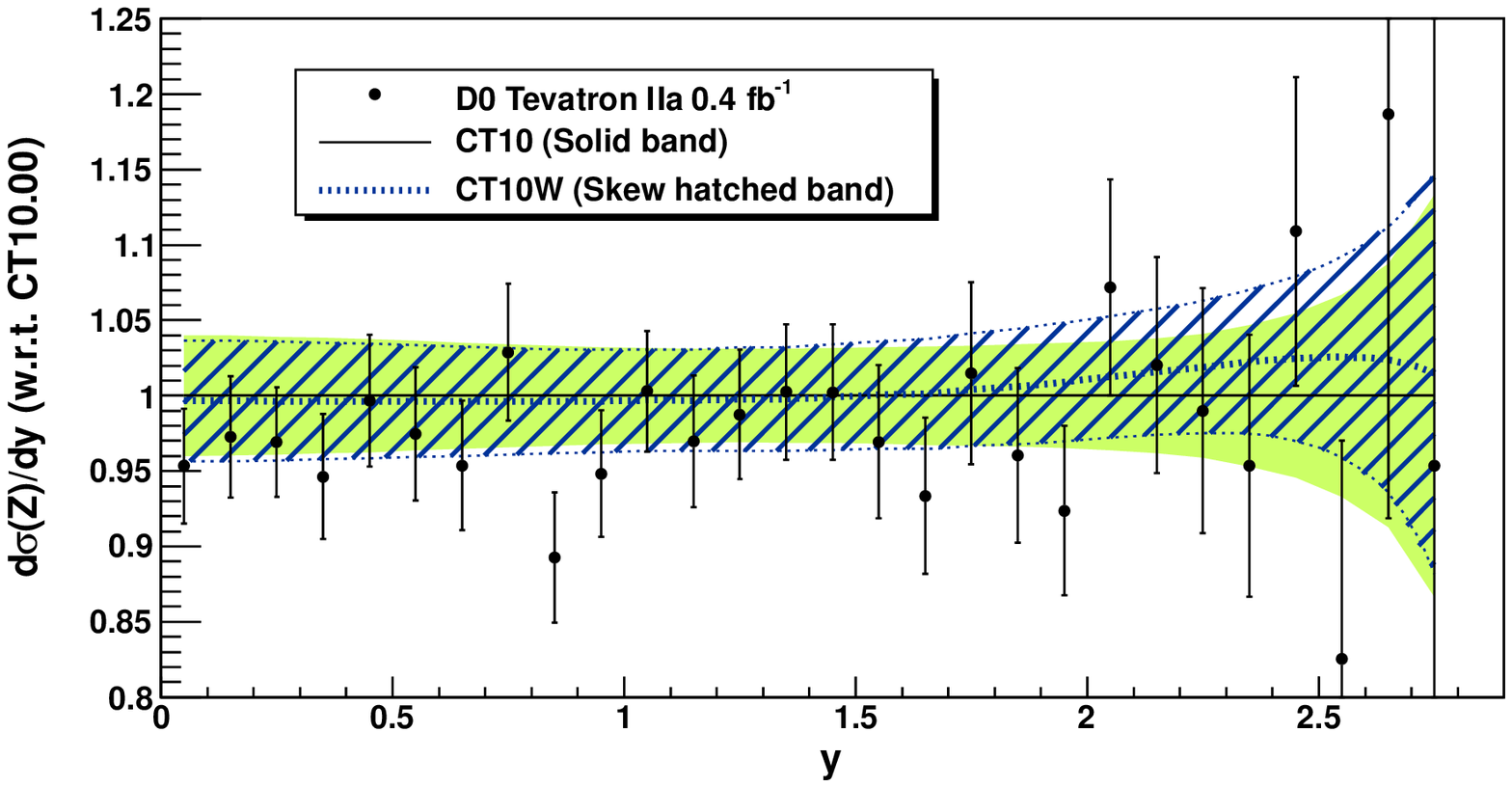} 
\includegraphics[clip,scale=0.8]{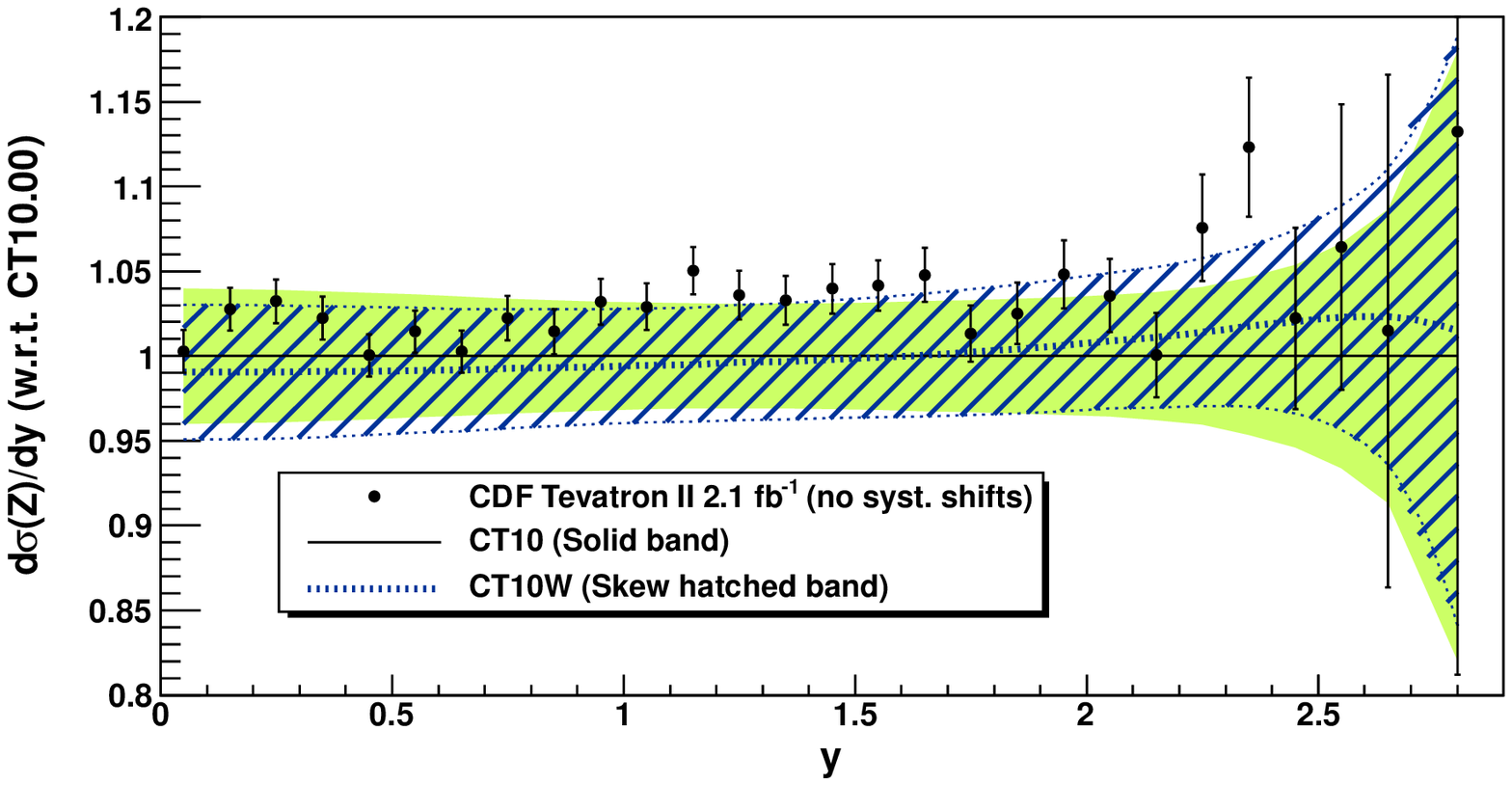}
\par\end{centering}
\caption{ Ratios of the NLO rapidity distributions for $Z$ boson production, 
relative to the CTEQ10.00 prediction, at 
the Tevatron Run-II. }
\label{figs:DISTZTev} 

\end{figure}

Figure \ref{figs:DISTZTev} 
compares NLO theoretical predictions for rapidity ($y$)
distributions of $Z$ bosons with the experimental data by CDF Run-II
\cite{Aaltonen:2010zza} and D\O~ Run-II \cite{Abazov:2007jy}. 
Both CT10 and CT10W sets give similar predictions and are in good agreement
with the data. Among the two experimental measurements, 
the more precise CDF Run-II $Z$ rapidity data (in the lower inset) 
appear to be closer 
to the CT10W prediction at $|y|>2$ than to the CT10 prediction, 
{\it i.e.,} to mildly favor the trend suggested by the latest $A_e$ data.  
CDF has published the systematic uncertainties of their measurement.
Those are included in our fit and produce
additional correlated shifts of the data toward the theoretical
values; however, the lower inset of Fig. \ref{figs:DISTZTev} 
shows these data without such shifts. 
With the systematic shifts included, the agreement
between NLO theory and CDF Run-II $Z$ $y$ data is even better than
is seen in Fig.~\ref{figs:DISTZTev}.

\clearpage

\section{Comparison of CTEQ6.6, CT10, and CT10W PDF sets \label{sec:CT10CT10W}}

\begin{figure}[h]
\begin{centering}
\resizebox*{0.49\textwidth}{!}{ \includegraphics[clip]{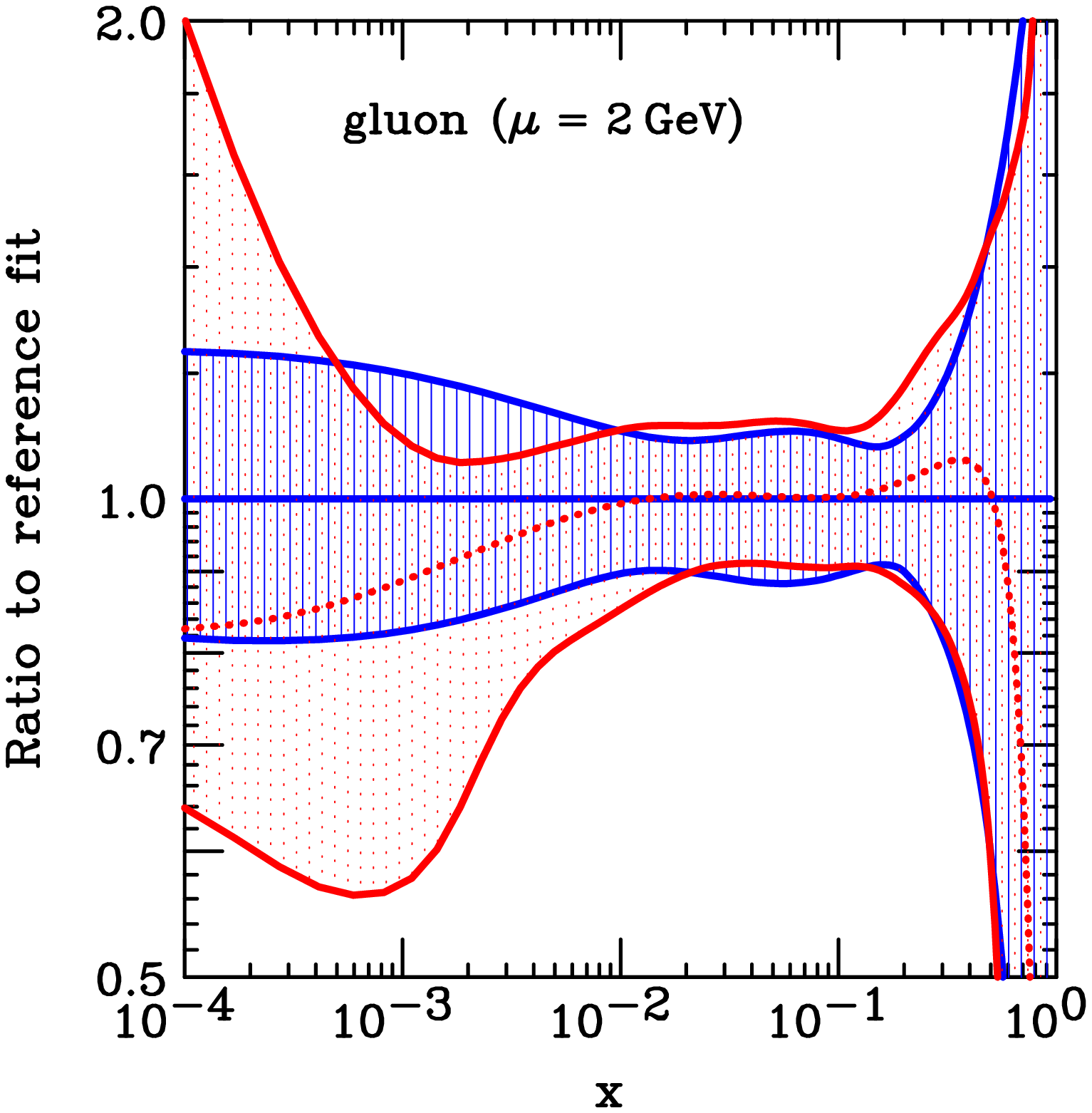}}
\hfill{}\resizebox{0.49\textwidth}{!}{ \includegraphics[clip]{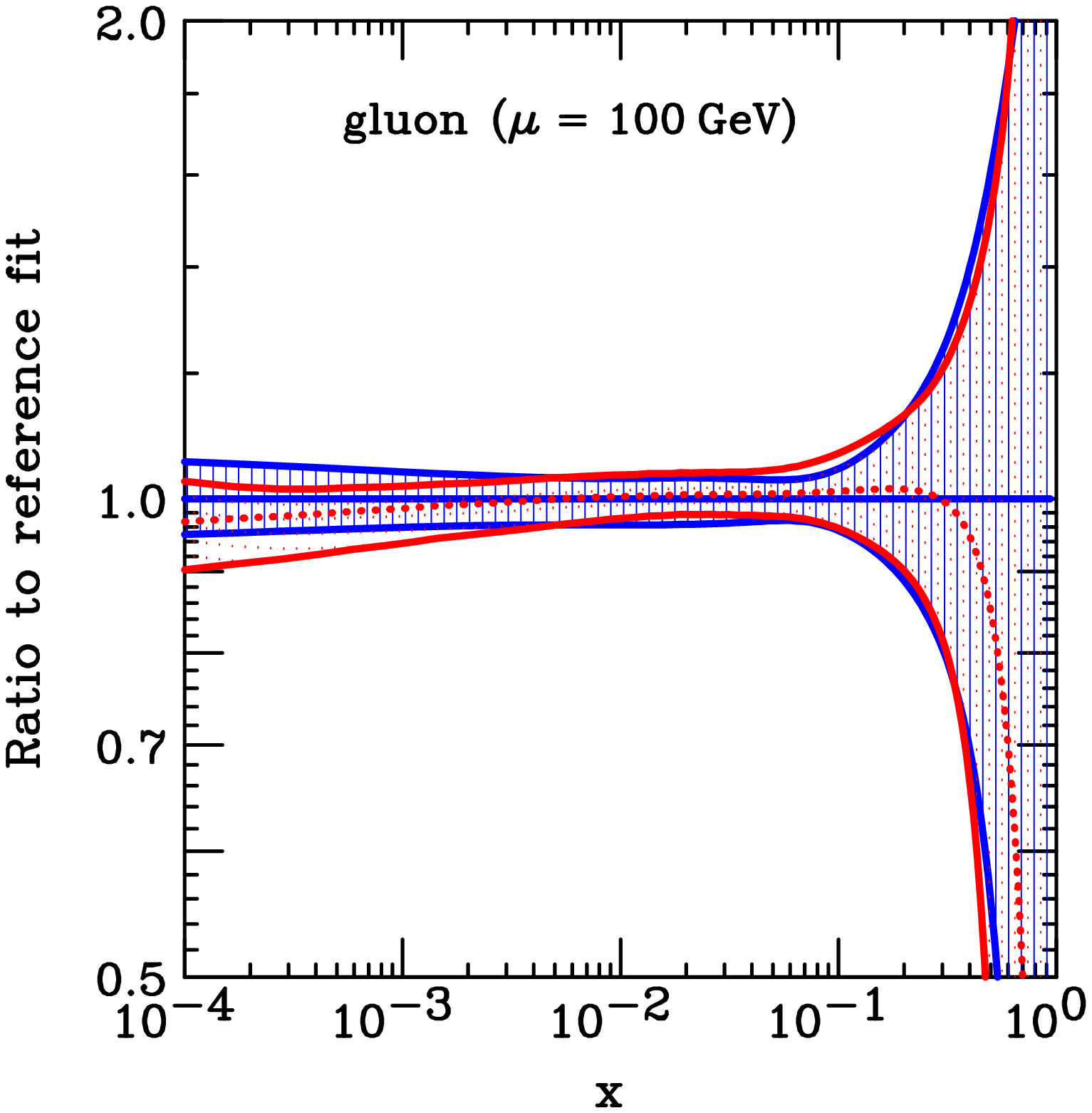}} 
\par\end{centering}
\caption{ Comparisons of the CTEQ6.6 and CT10 best-fit gluon PDFs and their
  uncertainties at $\mu=2$ GeV (left) and 100 GeV (right). The
  best-fit CTEQ6.6 gluon distribution is used as a reference. The
  CTEQ6.6 (CT10) best-fit PDFs and uncertainties are indicated by
  solid curves and hatched bands, while those of CT10 are indicated 
  by dashed curves and dotted bands.}
\label{figs:ct10gluon} 
\end{figure}

Figure \ref{figs:ct10gluon} shows the best-fit PDFs
and uncertainty ranges of 
the gluon distribution  
in CTEQ6.6 and CT10 eigenvector PDF sets, relatively to the CTEQ6.6 
best-fit PDF, CTEQ6.6M.  The two error bands are similar, 
except at small $x$, where the more flexible parametrization of the CT10
gluon PDF allows for a wider uncertainty.
The CT10 uncertainty range can be \emph{larger} than that of CTEQ6.6, because
the additional 
constraints from new experimental data are offset by
the combined effect of
allowing the experimental normalization 
factors to vary during eigenvector set 
searches, the increased freedom in the 
parametrizations, and the change to
weight 1 for every data set,
as discussed in Sec.\ \ref{sec:NewTheory}.

\begin{figure}[tbh]
\begin{centering}
\resizebox*{0.49\textwidth}{!}{ \includegraphics[clip]{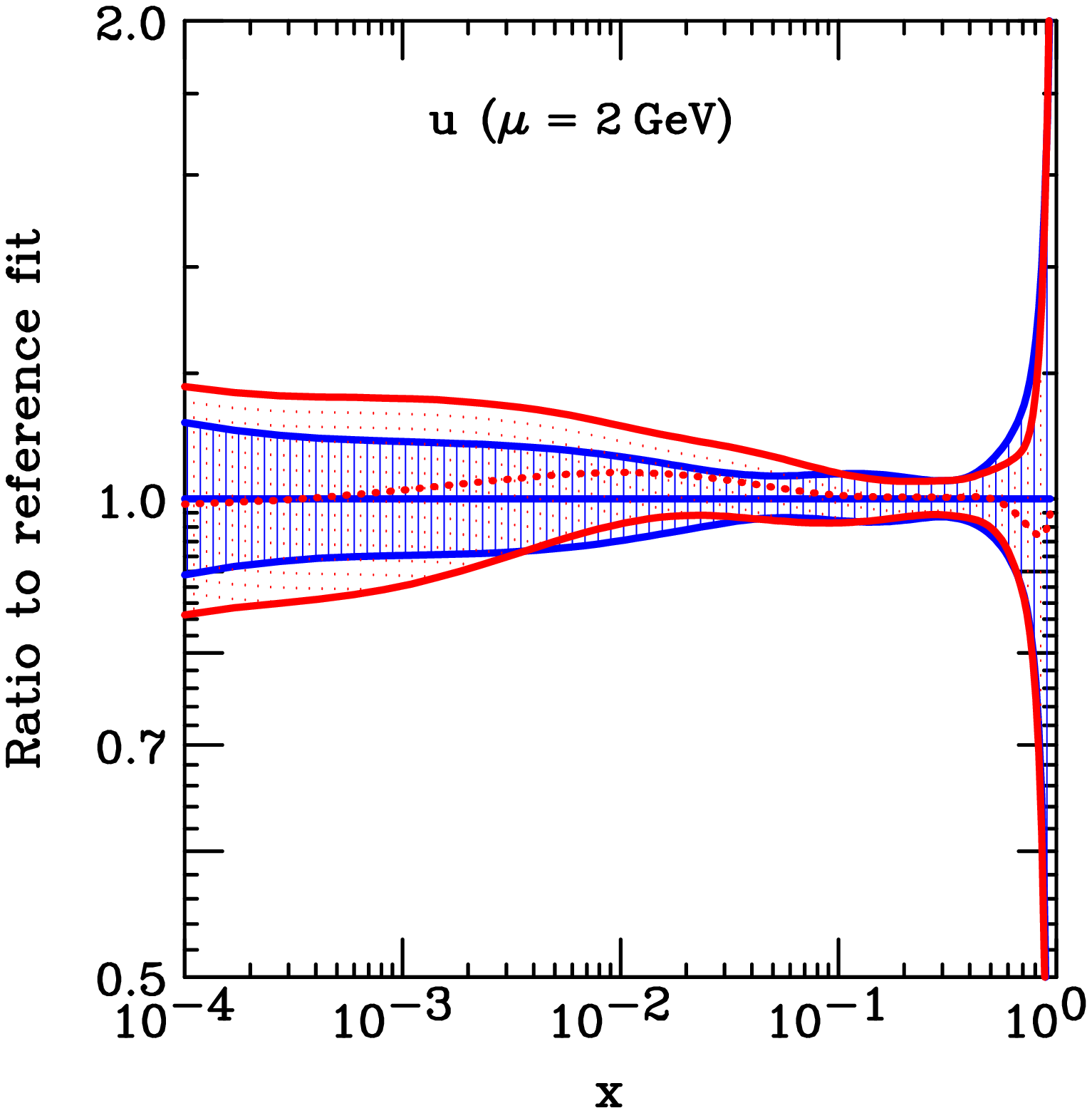}}
\hfill{}\resizebox{0.49\textwidth}{!}{ \includegraphics[clip]{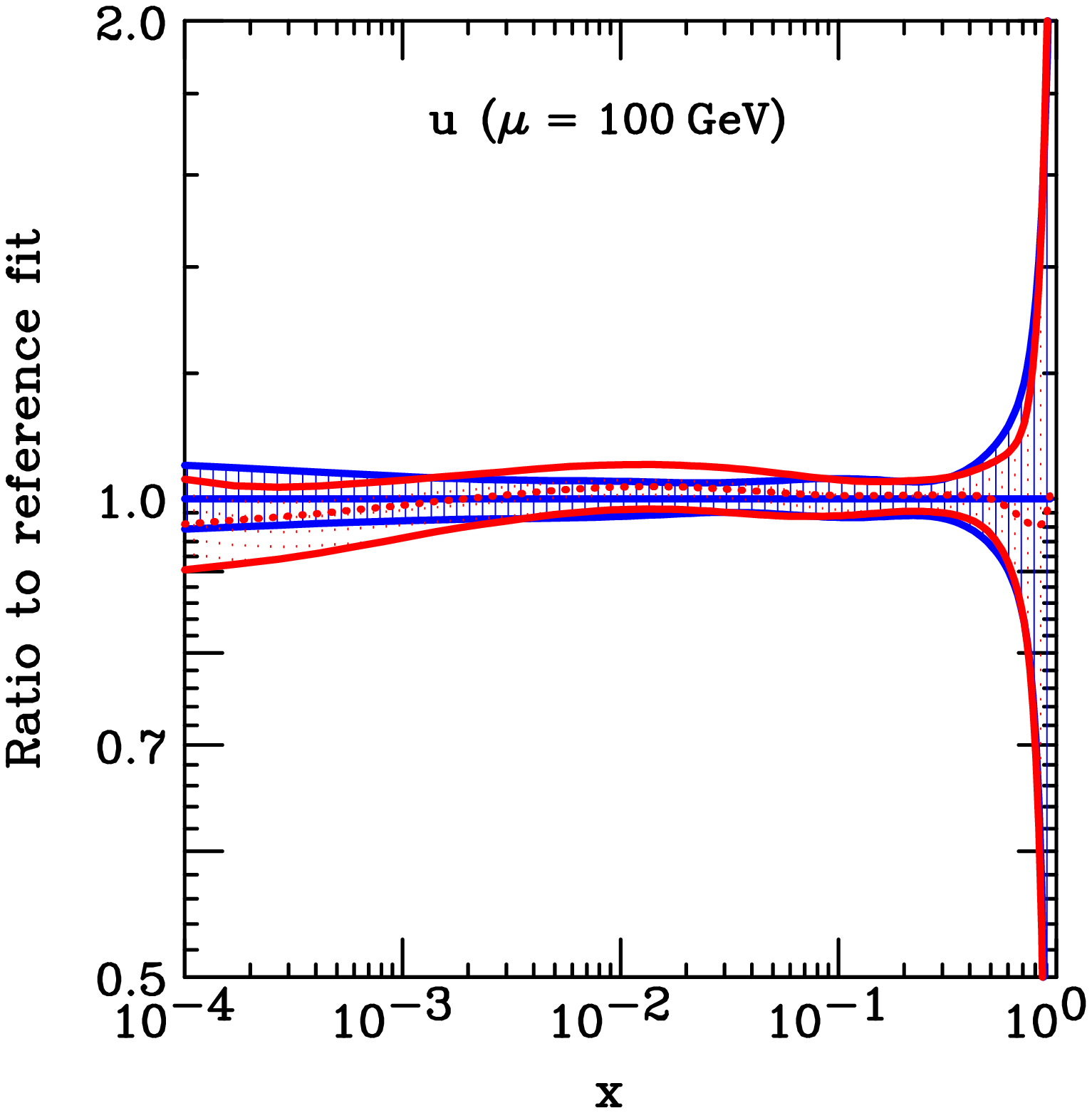}} 
\par\end{centering}
\caption{ Similar to  Fig.\ \ref{figs:ct10gluon}, but for the $u$ quark. }
\label{figs:ct10uquark} 
\end{figure}

Figure ~\ref{figs:ct10uquark} compares $u(x,\mu)$ from the CTEQ6.6 and CT10 
sets. Again, CT10 lies within the 
90\% CL range derived from CTEQ6.6.  However, $u(x,\mu)$ has increased to 
a value close to the CTEQ6.6 estimated upper limit at $x \sim 0.02$, 
even at scale  $\mu = 100 \, \mathrm{GeV}$, 
again as a result of modifications discussed in Sec.\ \ref{sec:NewTheory}.
(No such increase is observed in $d(x,\mu)$, which undergoes
qualitatively similar changes in other aspects.)

\begin{figure}[tbh]
\begin{centering}
\resizebox*{0.49\textwidth}{!}{ \includegraphics[clip]{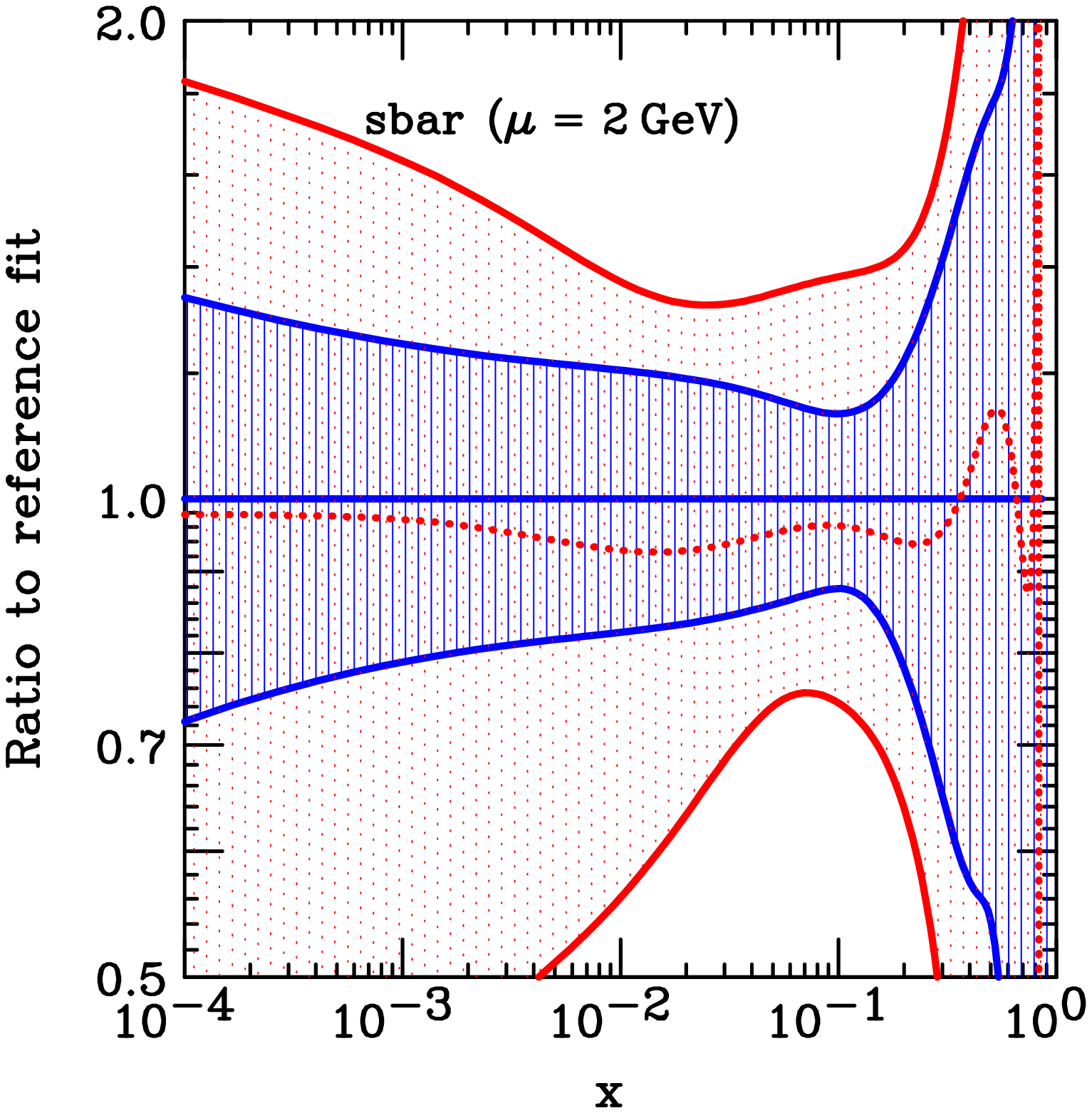}}
\hfill{}\resizebox{0.49\textwidth}{!}{ \includegraphics[clip]{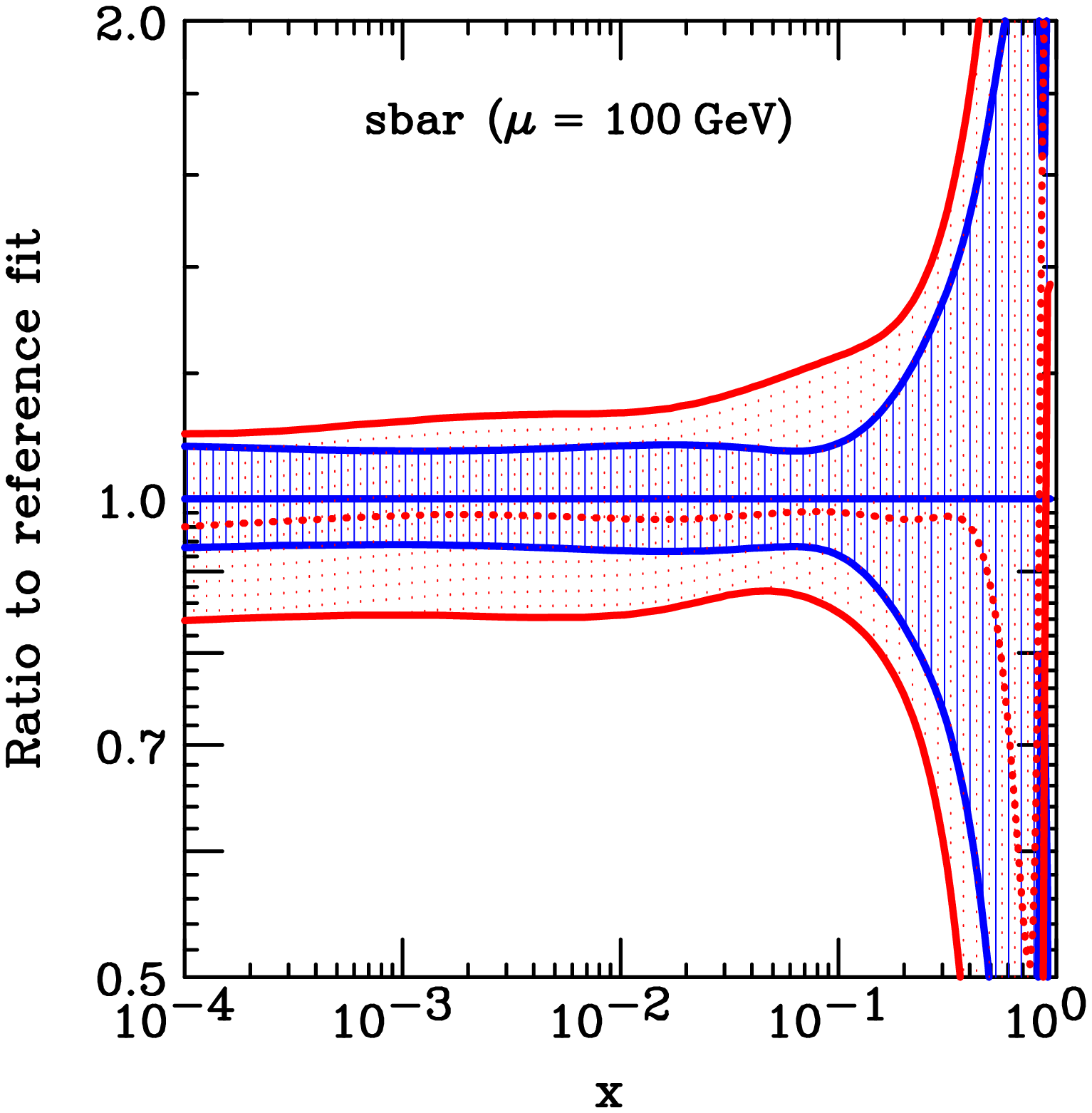}} 
\par\end{centering}
\caption{ Similar for Fig.\ \ref{figs:ct10gluon}, but for the $s=\bar{s}$ quark. }
\label{figs:ct10squark}
\end{figure}

Comparison of CT10 with CTEQ6.6 distributions for strange (anti-) quarks 
($s(x,\mu) = \bar{s}(x,\mu)$) is shown in Fig.\ \ref{figs:ct10squark}.  
Here the CT10 central fit again lies well inside 
the CTEQ6.6 uncertainty estimate; however, the CT10 
uncertainty on strangeness is much larger than in CTEQ6.6, as a result of 
the more flexible parametrization assumed in CT10.

\begin{figure}[tbh]
\begin{centering}

\hfill{}\resizebox{0.49\textwidth}{!}
{ \includegraphics[clip]{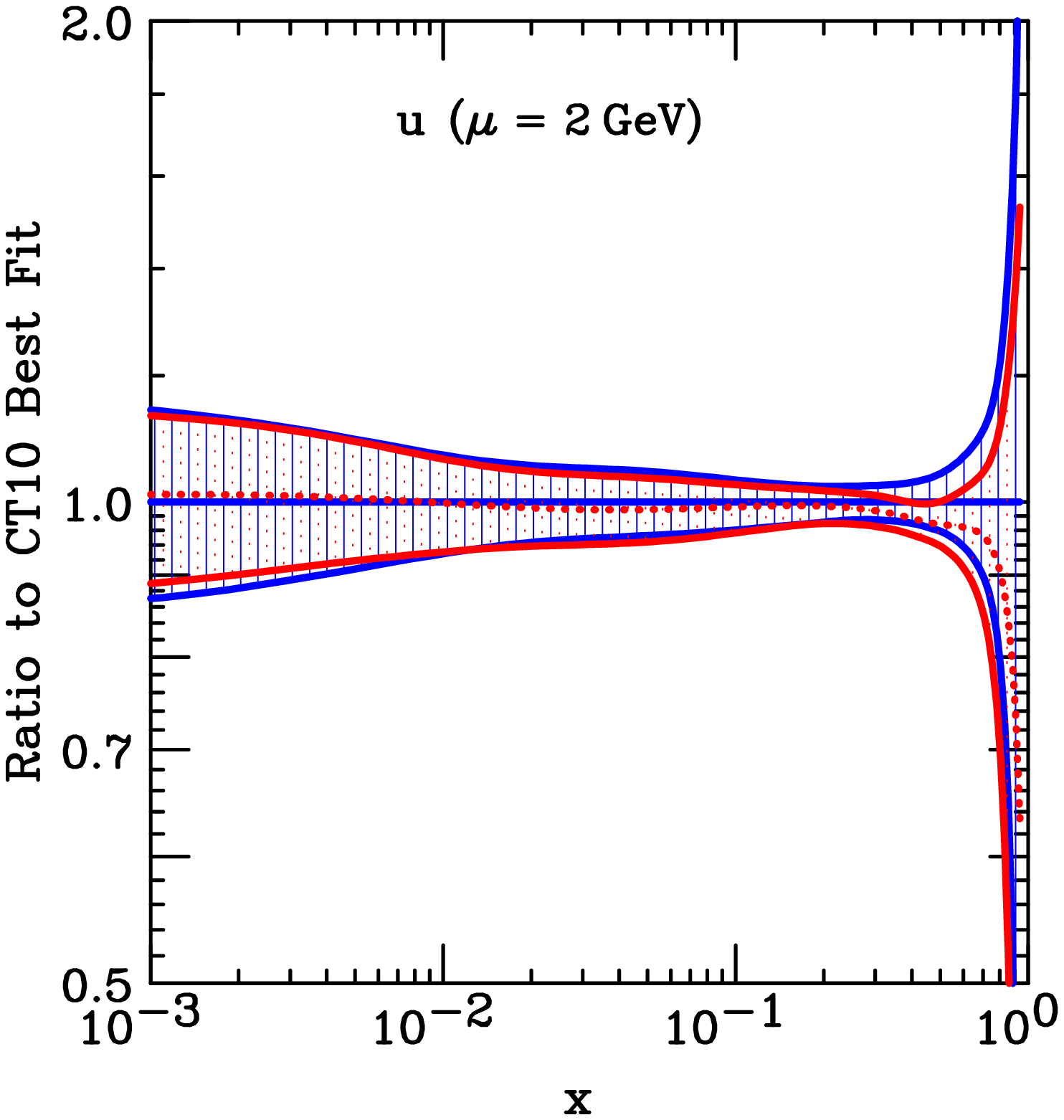}}
\hfill \resizebox*{0.49\textwidth}{!}
{ \includegraphics[clip]{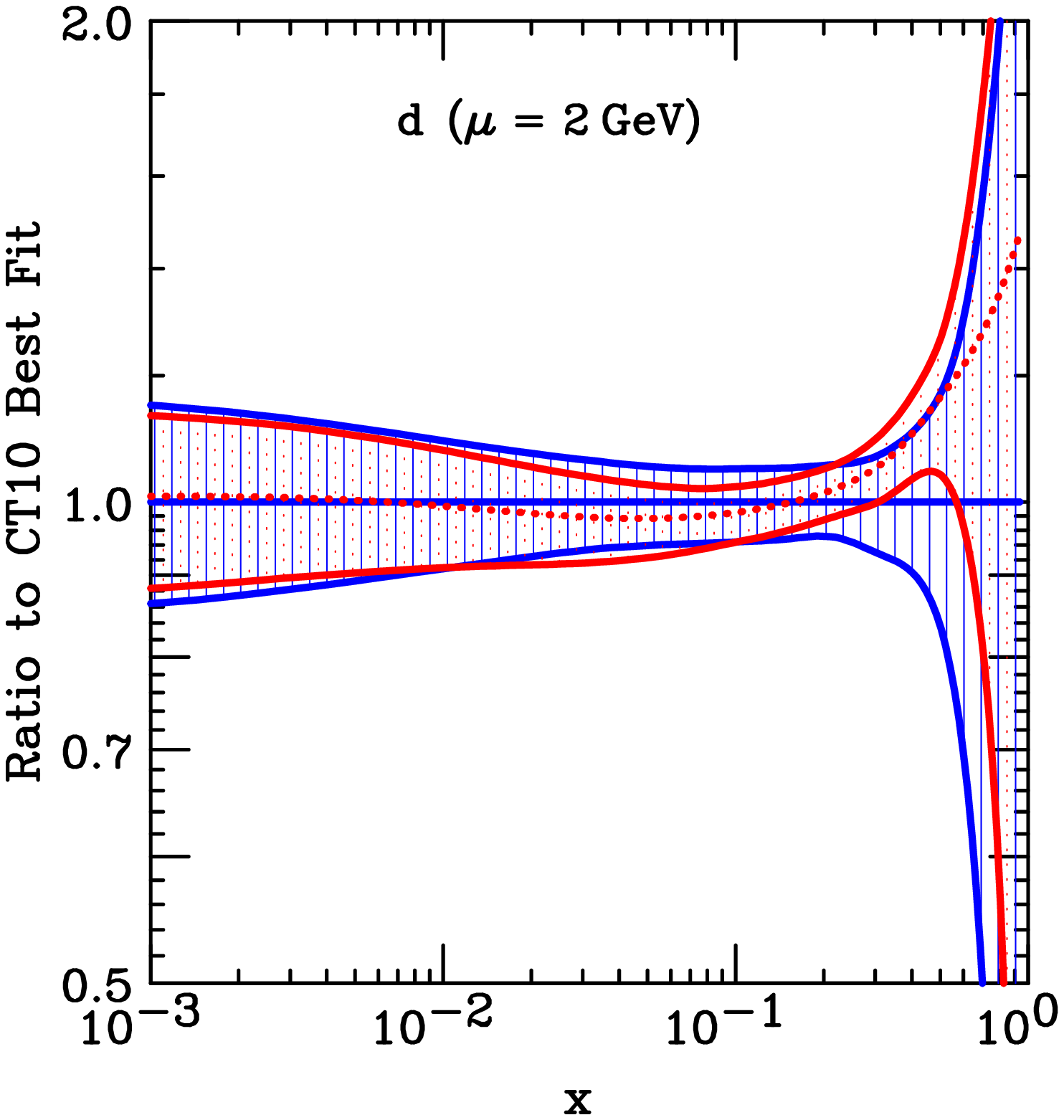}} 
\par\end{centering}
\caption{Comparisons of the CT10 and CT10W $u$-quark (left) and
  $d$-quark (right) best-fit PDFs, and their
  uncertainties, for scales of 2 GeV (left) and 100 GeV (right). The
  best-fit CT10 distributions are used as a reference. The
  CT10 best-fit PDFs and PDF uncertainties are indicated by
  solid curves and hatched bands, while those of the CT10W set are indicated by   dashed curves and dotted bands.}
\label{figs:ct10Wudquark} 
\end{figure}

\begin{figure}[tbh]
\begin{centering}
\includegraphics[scale=0.4]{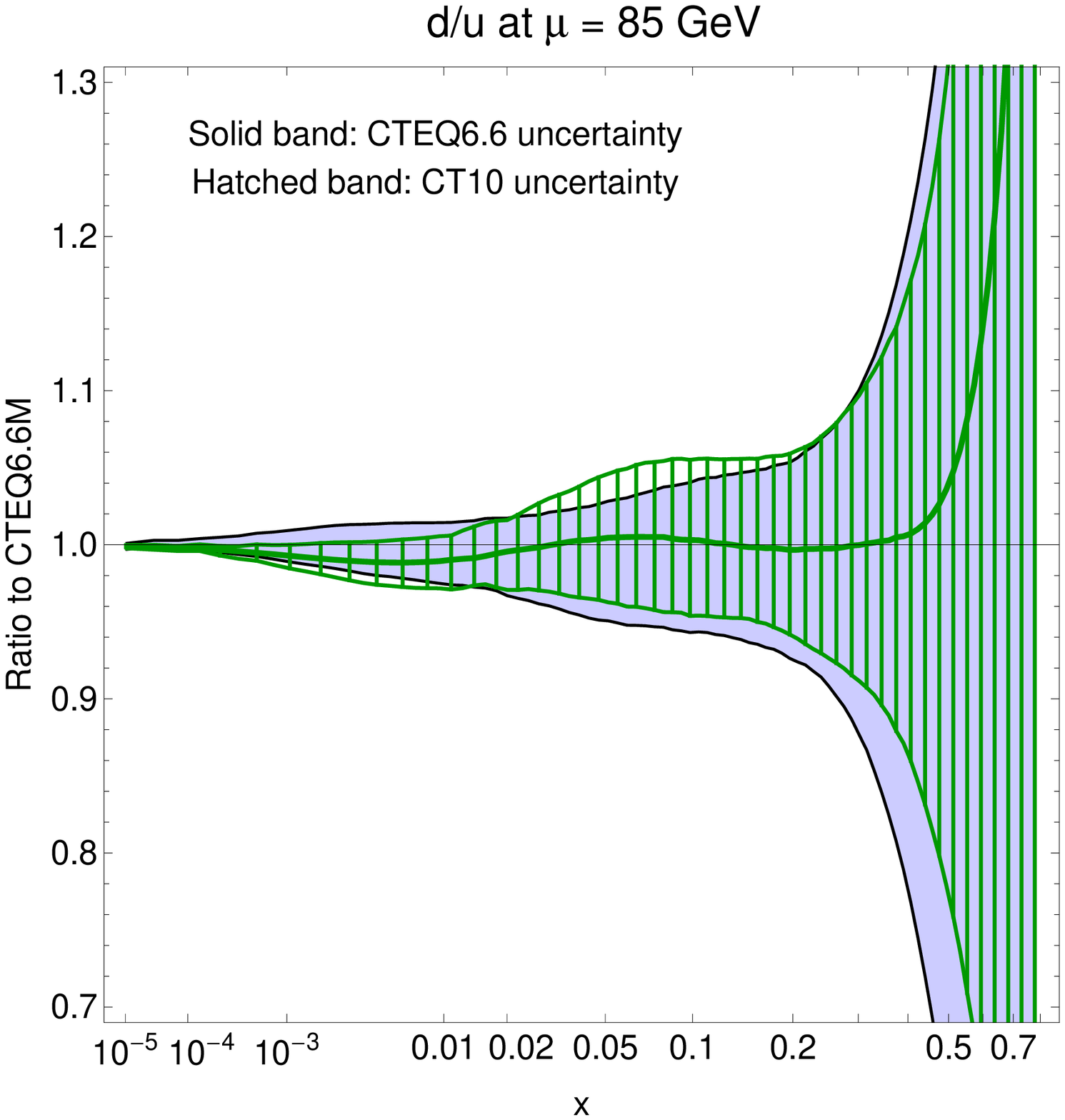}\quad\quad\quad\includegraphics[scale=0.4]{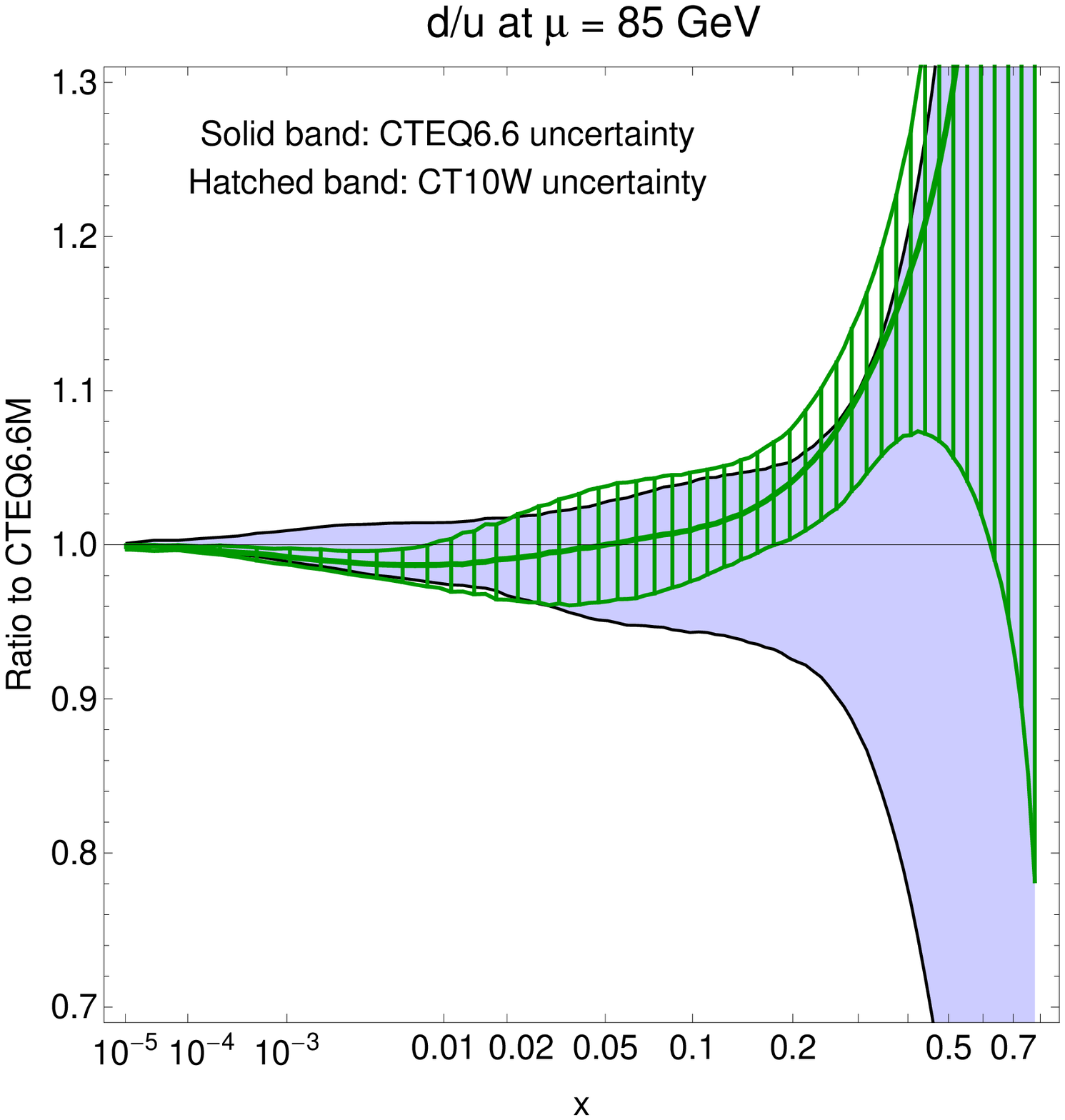}
\par\end{centering}
\caption{The  $d/u$ ratio for CT10 (left) and CT10W (right) versus that for CTEQ6.6, 
at scale $\mu=85$ GeV.}
\label{figs:ct10doveru}
\end{figure}

Figure\ \ref{figs:ct10Wudquark} compares
the best-fit PDFs and uncertainty ranges 
for the $u$ and $d$ quark PDFs in the CT10 and CT10W sets. 
(The PDFs for the gluon and sea quarks (not shown) 
are more or less the same in the 
two sets). The PDFs are compared at scale $\mu=2$ GeV, but the pattern
of their differences persists at larger scales as well. 
The up quark distribution of CT10W is smaller than that of CT10 at
$x$ of about 0.2 and above, whereas the down quark distribution is
larger in this $x$ region. 
These two changes are induced by the inclusion of
the D\O~ Run-II $A_\ell$ data. While the uncertainties on $u$ and $d$
PDFs themselves do not change much between CT10 and CT10W,
the $d/u$ ratio for CT10W, shown in Fig.\ \ref{figs:ct10doveru},
has a markedly
different slope at $x>0.01$ and reduced uncertainty, 
as compared to CT10. Clearly, the precise $A_\ell$ data has important
implications for the large-$x$ $d/u$ ratio and observables sensitive
to it.

\clearpage

\section{Quality of fits to individual data sets \label{sec:quality}}
We will now address the consistency of the CT10(W) global fits with
each of the 29 (31) data sets included in the fit. This issue can be
explored with several techniques employed by one of us (J.\ P.)
recently in Refs.~\cite{Collins:2001es,Pumplin:2009sc,Pumplin:2009nm}.
All these approaches require to redo the global fit after
introducing special features, such 
as variable $\chi^2$ weights for the individual data sets, or
a special eigenvector basis in the PDF parameter
space. 

Alternatively, one might assess the consistency between various
data sets directly from the best fit, by studying 
the $\chi^2$ values for each individual experiment. 
In a sample of $N_{exp}$ experiments with $N_n$ data points
each, $\chi^2_n$ values will be smaller 
than their most probable values, $N_n$, in some experiments, 
and larger than $N_n$ in other experiments. 
Comparison of observed frequencies of $\chi^2_n$ with
the expected probabilities would reveal how well the experiments
are fit in their ensemble; and it is more informative than just 
the global $\chi^2$ for all experiments. 
For example, the frequency distribution can help one to identify experiments 
that are fitted too well or too poorly, even 
if the global $\chi^2$ is
excellent.

Such a comparison can be done with the $\chi^2_n$ frequencies 
directly, but it requires an integration of 
several $\chi^2(N_n)$ distributions 
with non-identical degrees of freedom, $N_n$. A faster method 
uses a secondary statistical 
distribution $S$ derived from the $\chi^2$ distribution, 
such that $S$ closely resembles some standard
distribution and is maximally independent of $N_n$.
 
Several distributions of this kind are known to exist (see, e.g.,
Ref.~\cite{Lewis:1988}, and references therein), with one of the
simplest ones attributed to R.\ A.\ Fisher \cite{Fisher:1925}. 
Fisher's approximation shows that the
function $S$ in Eq.~\ref{eq:S_n} (with the subscript $n$ ignored)
closely follows the standard normal distribution even for small values of $N$.  
The theoretical 
distribution for $\chi^{2}(N)$  at $N\to \infty$ 
is approximately Gaussian with the mean and standard deviation 
of $N\pm\sqrt{2N}$,
which implies that the distribution for $S$ approaches a Gaussian one
with the mean 0 and standard deviation 1. The utility of 
$S$ comes from the fact that its Gaussian approximation 
is already quite accurate for $N$
as small as 10, and it becomes symmetric (not skewed in either direction)
faster than the $\chi^2$ distribution itself (whose skewness is not neglible 
for up to $N\approx 30$).\footnote{At $N\rightarrow \infty$, 
the skewness parameter of the $S(N)$
  distribution is asymptotically four times
  smaller than that of the $\chi^2(N)$ distribution.} The $S$ 
values can thus be used to compare the fit quality
among experiments with varying numbers of data points,
in a simple manner that avoids lengthier calculations based on the
direct analysis of $\chi^2$.

\begin{figure}[tbh]
 \vskip 10pt 
\begin{centering}
\resizebox*{0.50\textwidth}{!}{ \includegraphics[clip,scale=0.4]{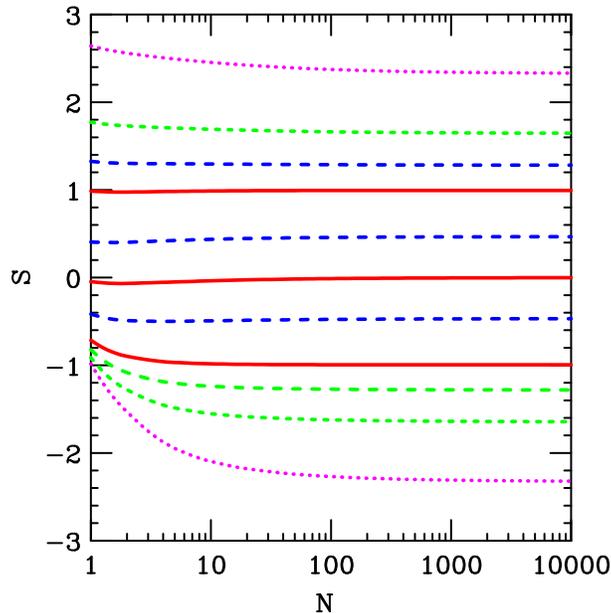}} 
\par\end{centering}
\vskip -15pt 
\caption{ Values of $S$ corresponding to cumulative probability $p=1$ (bottom),
$5$, $10$, $16$, $32$, $50$, $68$, $84$, $90$, $95$, and
$99\%$ (top). The three solid curves contain the middle $68\%$ of the 
distribution. }
\label{figs:pma} \vskip 5pt 
\end{figure}

The accuracy of the Gaussian approximation for $S$ is demonstrated by Fig.\
\ref{figs:pma}. Here we plot contours of the constant cumulative
probability in the plane of $N$ and $S$. The lines
correspond to $S$ values for the cumulative probability ranging from
1\% to 99\%, for each given $N$.
Note that the three solid curves, which contain the middle 68\% of
the distribution, lie very close to $S=-1$, $0$, and $+1$. This 
is entirely expected to happen for the Gaussian limit $N\to\infty$; 
but it is seen
here to be a good approximation even down to $N\approx10$.
For our purposes, the important curves in Fig.\ \ref{figs:pma} are
the top three, which contain cumulative probabilities of $90$, $95$,
and $99\%$---e.g., 
the chance of exceeding the value $S=1.3$, 1.6 and 2.4, are 
$10\%$, $5\%$, $1\%$, respectively, 
for the whole range of $N$ that appears in PDF fitting.

\begin{figure}[tbh]
 \vskip 10pt 
\begin{centering}
\mbox{ \resizebox{0.40\textwidth}{!}{ \includegraphics[clip,scale=0.25]{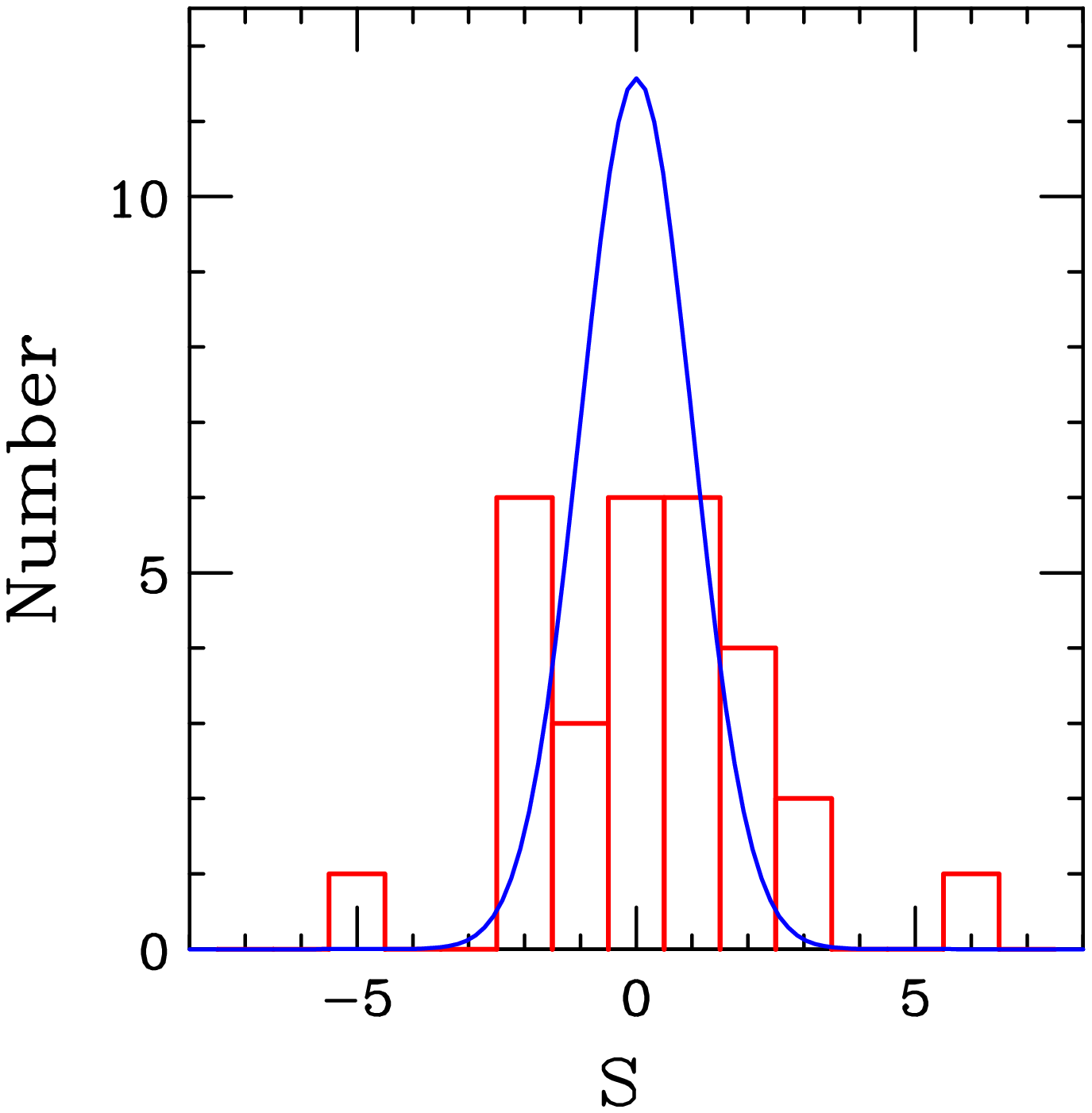}}
\hfill{}\resizebox{0.40\textwidth}{!}{ \includegraphics[clip,scale=0.25]{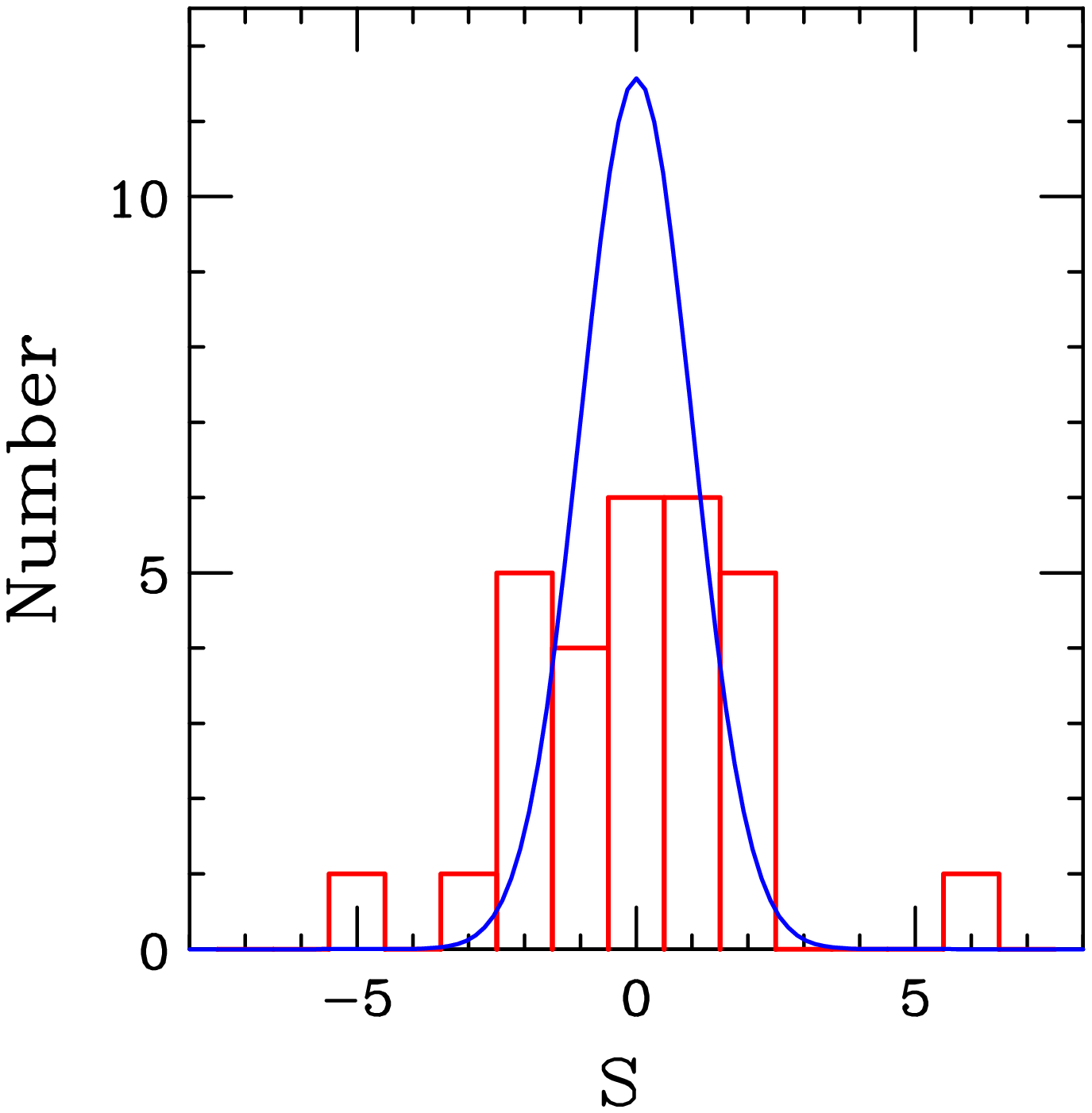}}
} 
\par\end{centering}
\vskip -15pt 
\caption{ Distribution of the $S$ parameter across 
the 29 data sets used in CT10. The left-hand side is the  CT10
fit; the right-hand side uses a more flexible parametrization with 71 free
parameters.}
\label{figs:hist} \vskip 5pt 
\end{figure}

The left side of 
Fig.~\ref{figs:hist} shows a histogram of the $S$-values for the 29 
data sets included in the CT10 best fit.
The smooth bell curve 
is a Gaussian distribution with mean 0 and variance 1. 
The observed histogram is compatible with a zero mean, but its variance
is larger than unity; it would agree better with a 
Gaussian with the standard deviation of 2 or 3. This indicates some
tension between the experiments, of the magnitude compatible with the
findings in other recent studies
\cite{Pumplin:2009sc,Pumplin:2009nm}. It has been
observed, for example, that discrepancies between contributions to
$\chi^2$ from individual experiments, which are
expected to obey the standard normal distribution, 
in fact follow a wider normal distribution, with a variance 
of about 2 \cite{Pumplin:2009sc}. It is also interesting to note that
such level of discrepancy appears to be independent of the
flexibility of the PDF parametrizations. 
The right-hand side of Fig.~\ref{figs:hist} shows 
a histogram of the $S$ parameter in a
fit with a much more flexible Chebyshev parametrization, which
triples the number of the total parameters compared to the CT10
parametrization. In this fit, the $S$ distribution still preserves 
the overall, too wide, shape and does not eliminate 
the two most-outlying points. (The outlying point on the right 
is the NMC proton DIS data.  The outlying point on the left 
is the CCFR $F_3$ data.)
The analysis of
the $S$ distribution leads us to believe that non-negligible tensions
do exist between the subsets of the current global hadronic data,
regardless of the number of free parameters in the PDFs, 
and contrary 
to the existing claims of the opposite
\cite{Ball:2010de}.$^2$\footnotetext[2]{By this measure, 
  similar tensions between the experiments appear to exist in the
  NNPDF2.0 analysis \cite{Ball:2010de}. The overall $\chi^2$=1.27 of
  that analysis is larger than our 1.1. The 
$S$ parameter
  distribution of the NNPDF2.0 fit, computed from the breakdown 
  of the $\chi^2$ values over experiments in Tables 1 and 10 
  of Ref.~\cite{Ball:2010de}, is significantly broader than the
  expected normal distribution.}

\section{Applications to Tevatron and LHC Physics \label{sec:applications}}

In this section, we examine the impact of the CT10(W) 
parton distribution functions
on the production of $W$, $Z$, top quark, Higgs boson and 
representative new physics signals at the Tevatron Run-II and the
LHC. The processes selected are important 
for benchmark measurements of the Standard Model 
parameters or illustrate typical patterns of the PDF dependence in
new physics searches, as discussed in some detail in the published CTEQ6.6 
paper \cite{Nadolsky:2008zw}.
In addition, we also comment on 
a recently published measurement of D\O~ Run-II dijet
invariant mass distribution \cite{d0-di-jet}. 

\subsection{$W$ and $Z$ Physics}

\begin{figure}[p]
\begin{centering}
\includegraphics[clip,scale=0.52]{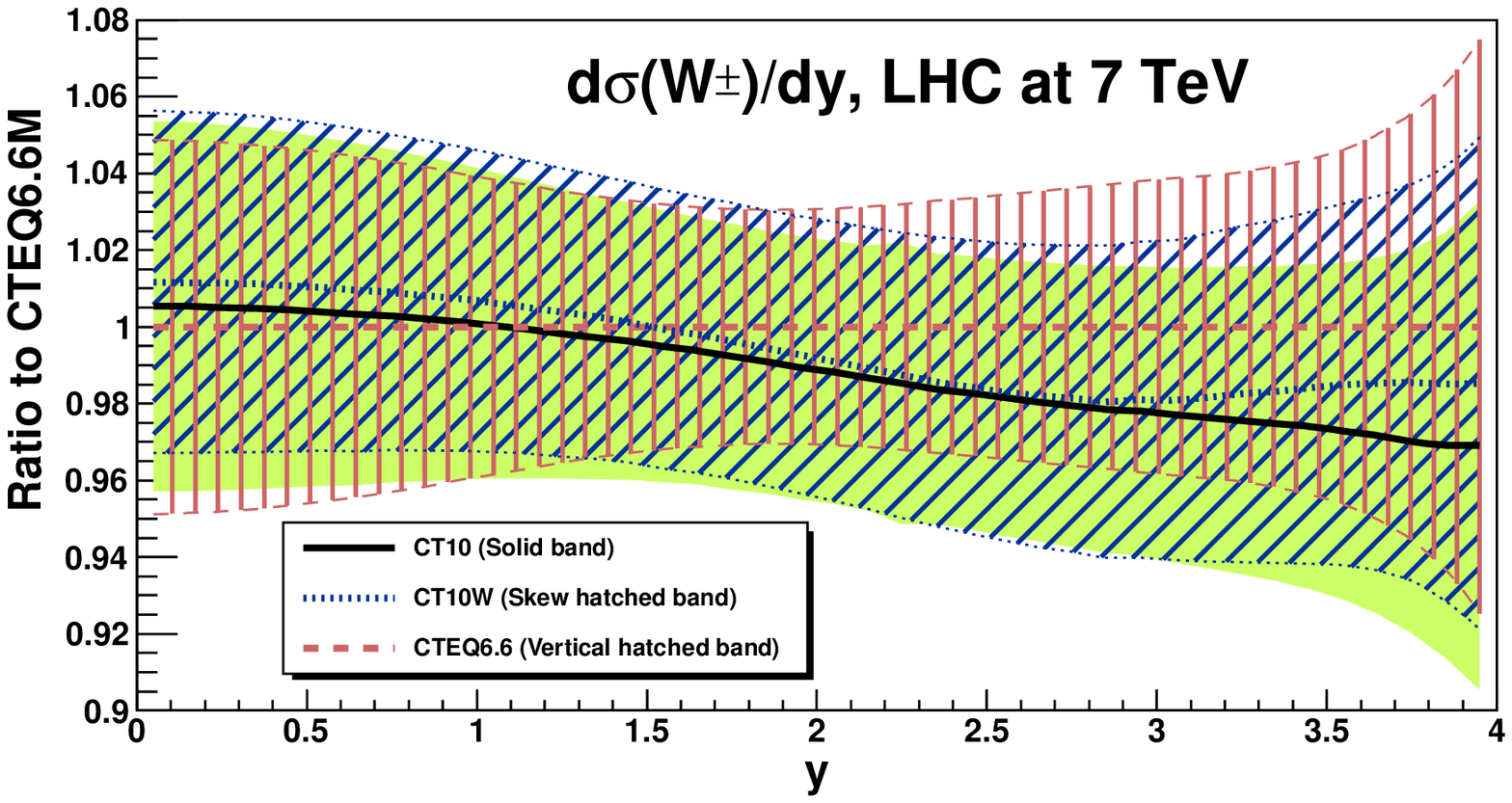} 
\includegraphics[clip,scale=0.52]{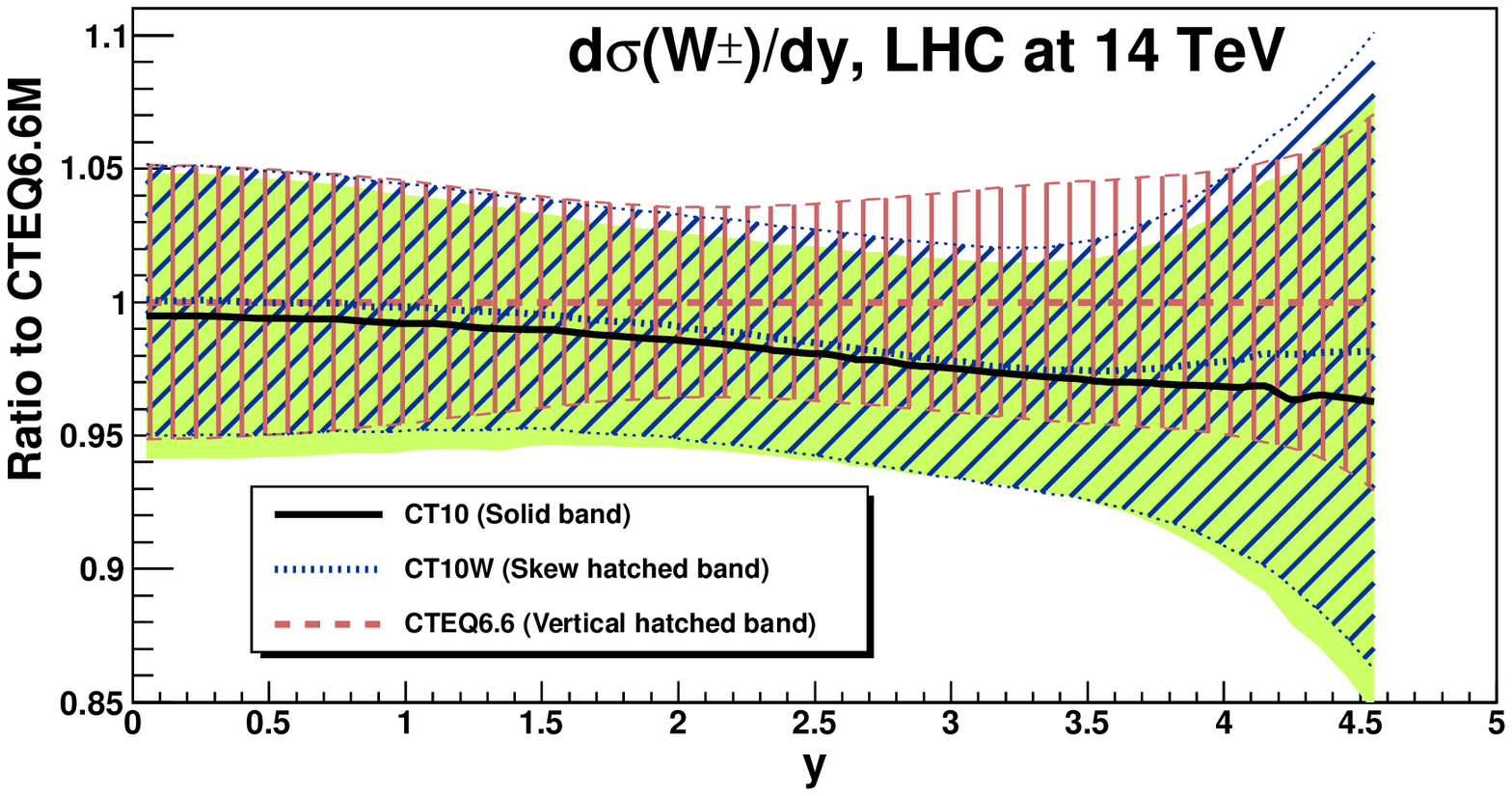}
\includegraphics[clip,scale=0.52]{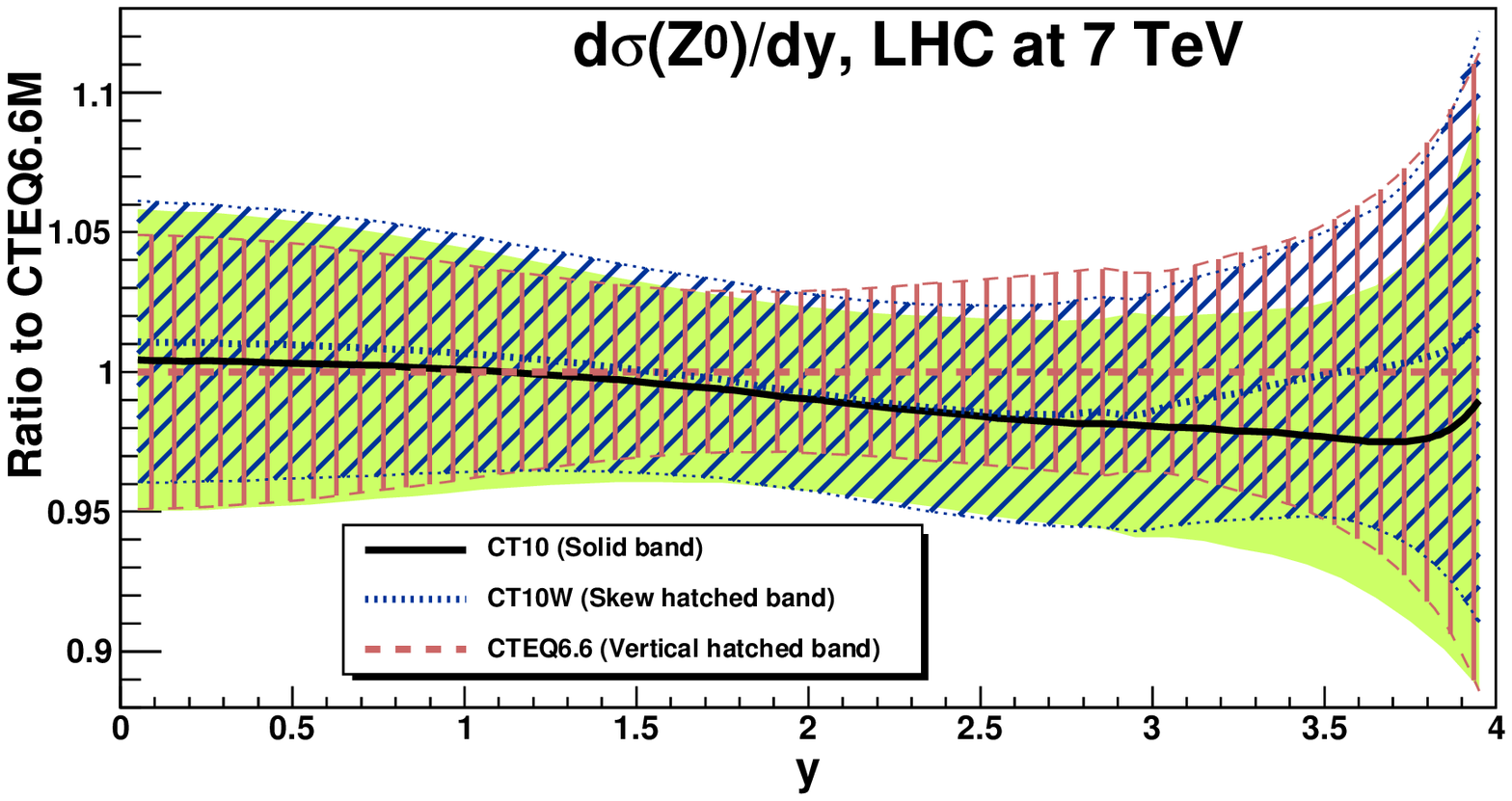} 
\includegraphics[clip,scale=0.52]{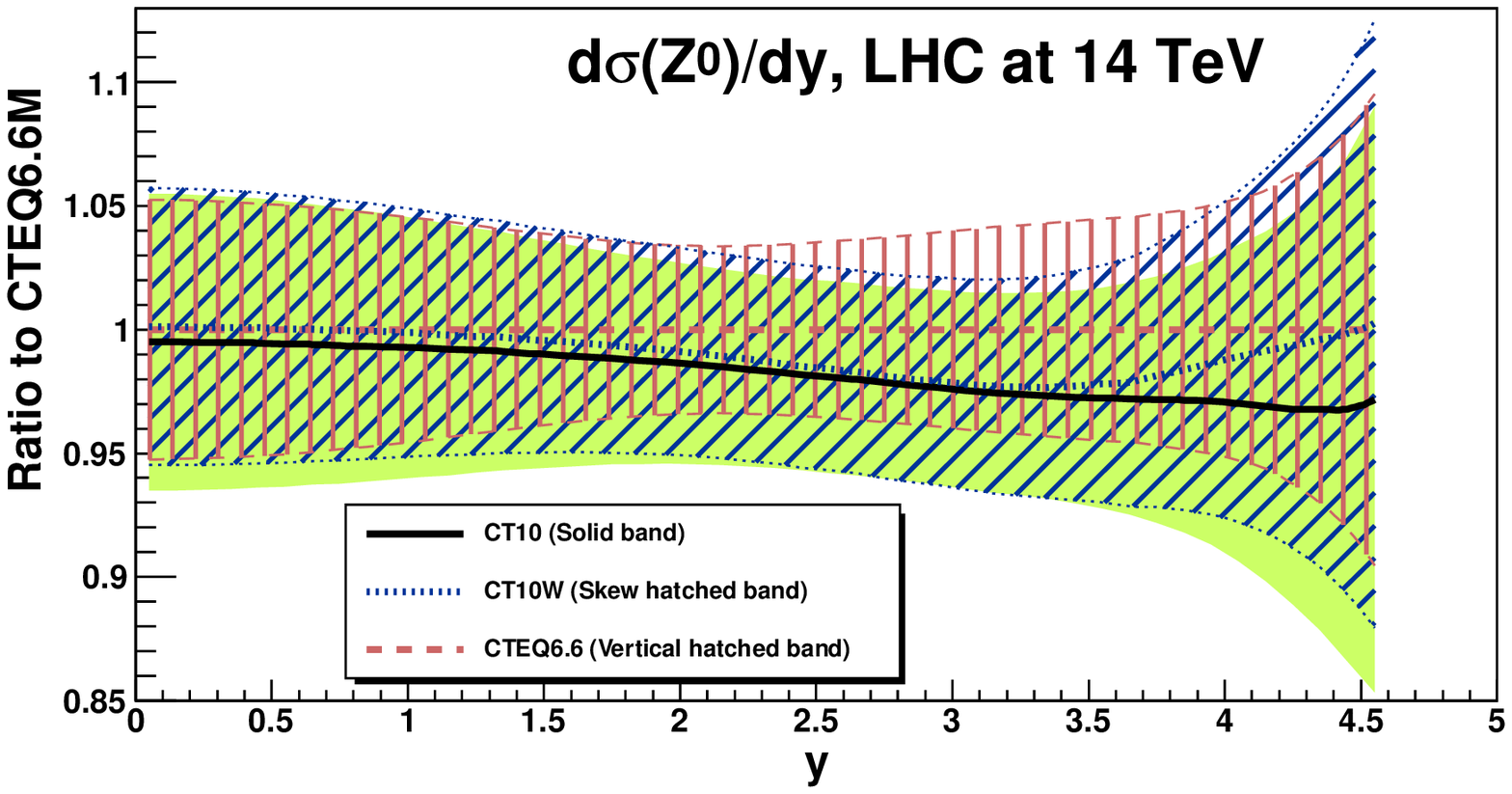} 
\par\end{centering}
\caption{Ratios of NLO rapidity distributions  
of $W$ boson production 
and of $Z$ boson production, relative to the corresponding ratios 
in the CTEQ6.6 best fit, at the LHC.}

\label{figs:DISTWZLHC} 

\end{figure}

Figure \ref{figs:DISTWZLHC} shows the PDF uncertainty bands for
the rapidity distributions $d\sigma/dy$ in inclusive 
$W^\pm$ and $Z$ boson production at
the LHC ($\sqrt{s}=7$ and 14 TeV), calculated at NNLL+NLO
using a $Q_T$ resummation program ResBos
\cite{Balazs:1995nz,Balazs:1997xd,Landry:2002ix} and 
CT10 (green solid fill), CT10W (blue skew-hatched fill), 
and CTEQ6.6 (red vertical fill) PDF eigenvector sets. 
Each cross section is normalized to the
corresponding cross section for the CTEQ6.6M PDF.
The CT10 and CT10W central predictions are similar to those of CTEQ6.6,
but have slightly larger PDF uncertainties for the reasons explained
in Sec.~\ref{sec:NewTheory}. 

\begin{figure}[p]
\begin{centering}
\includegraphics[clip,scale=0.52]{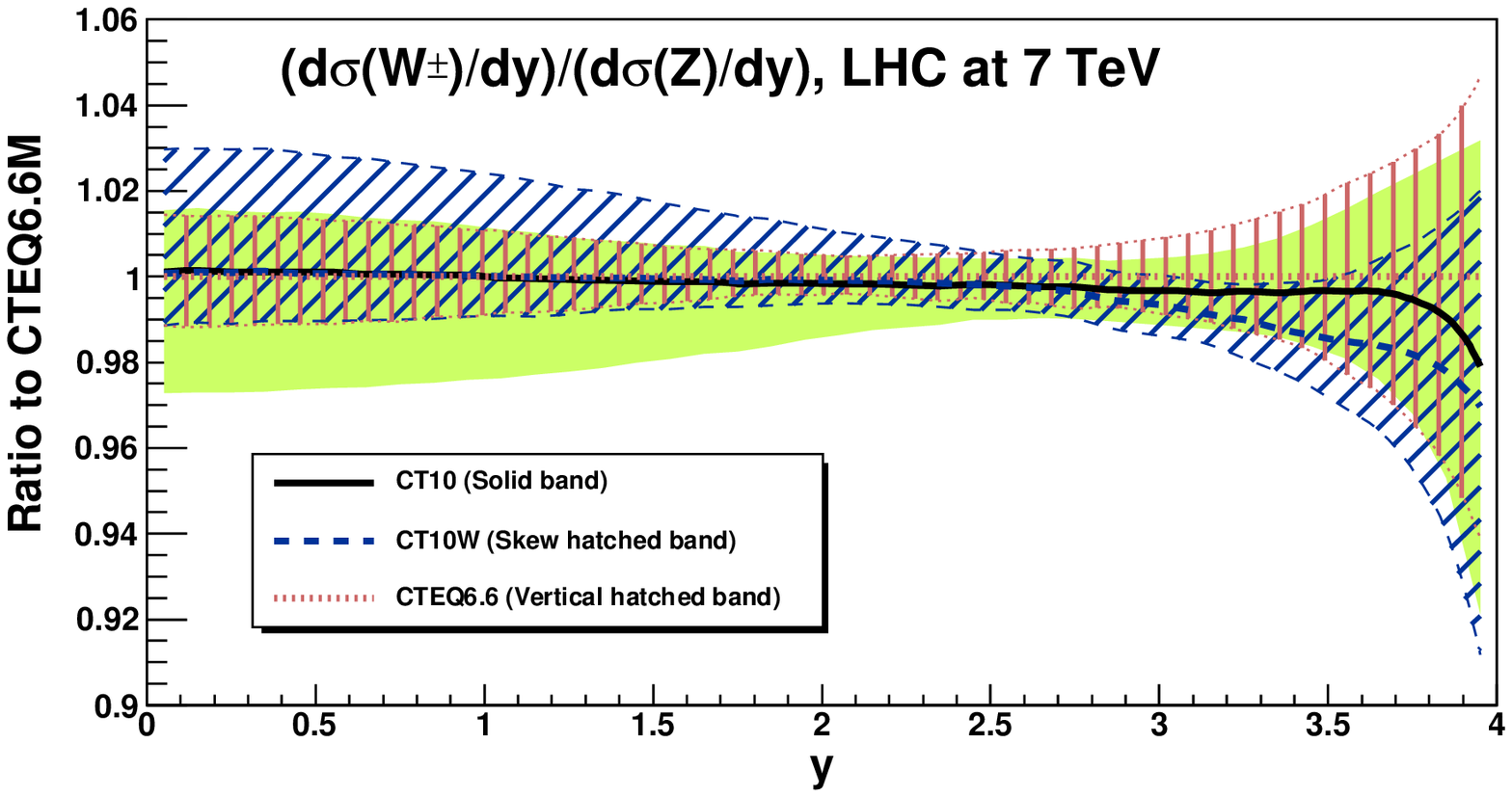}
\includegraphics[clip,scale=0.52]{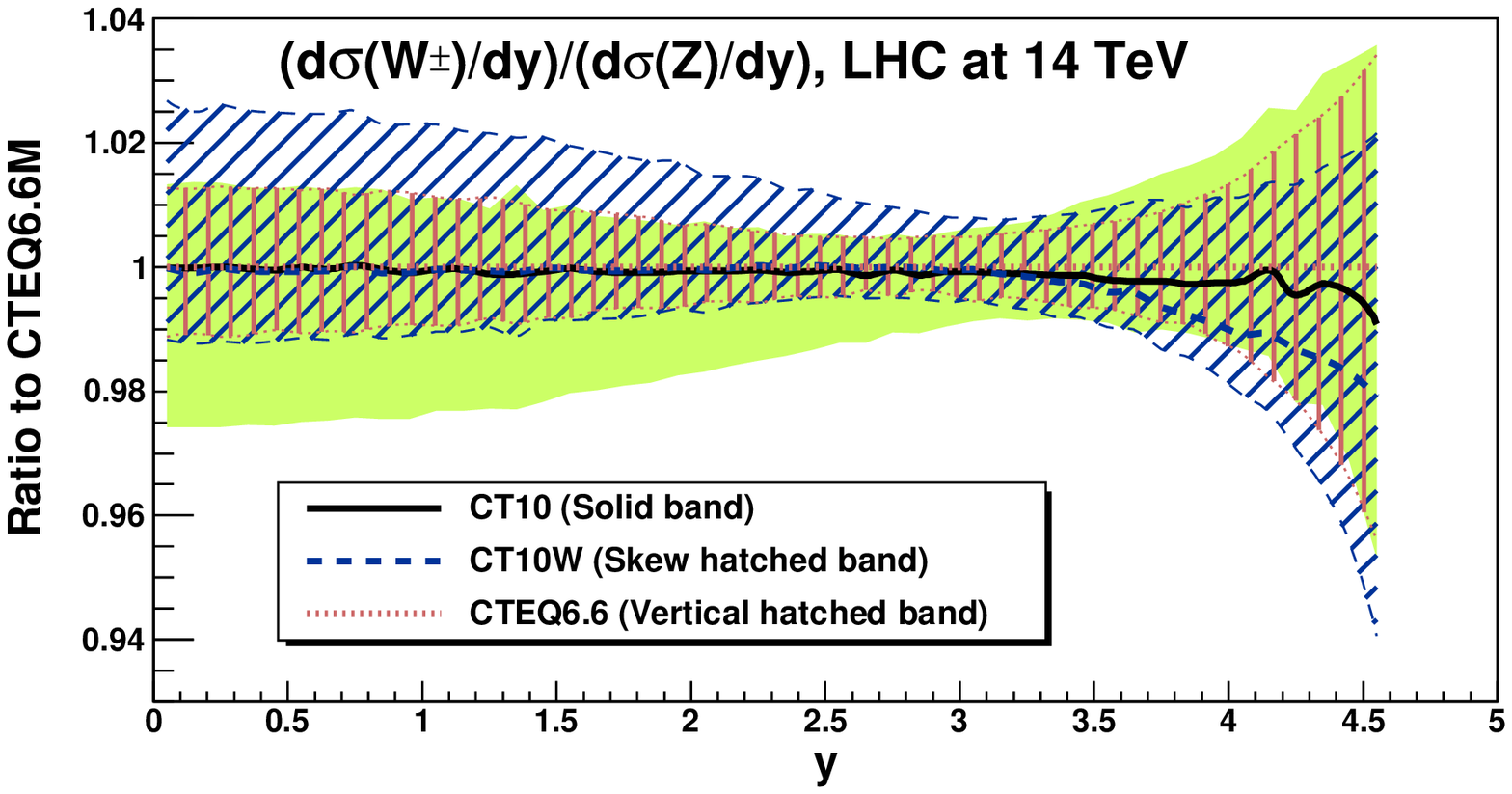}
\includegraphics[clip,scale=0.52]{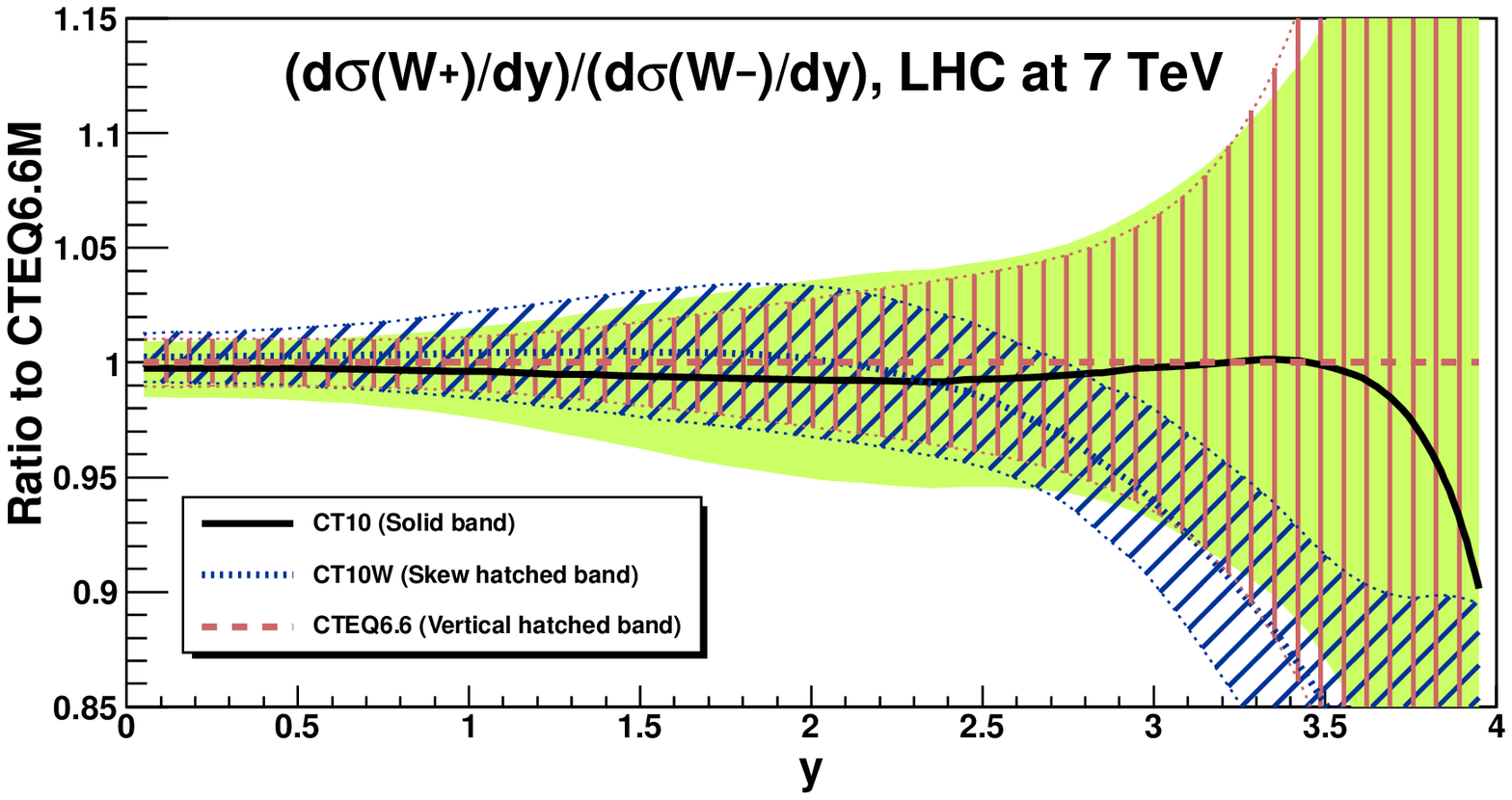}
\includegraphics[clip,scale=0.52]{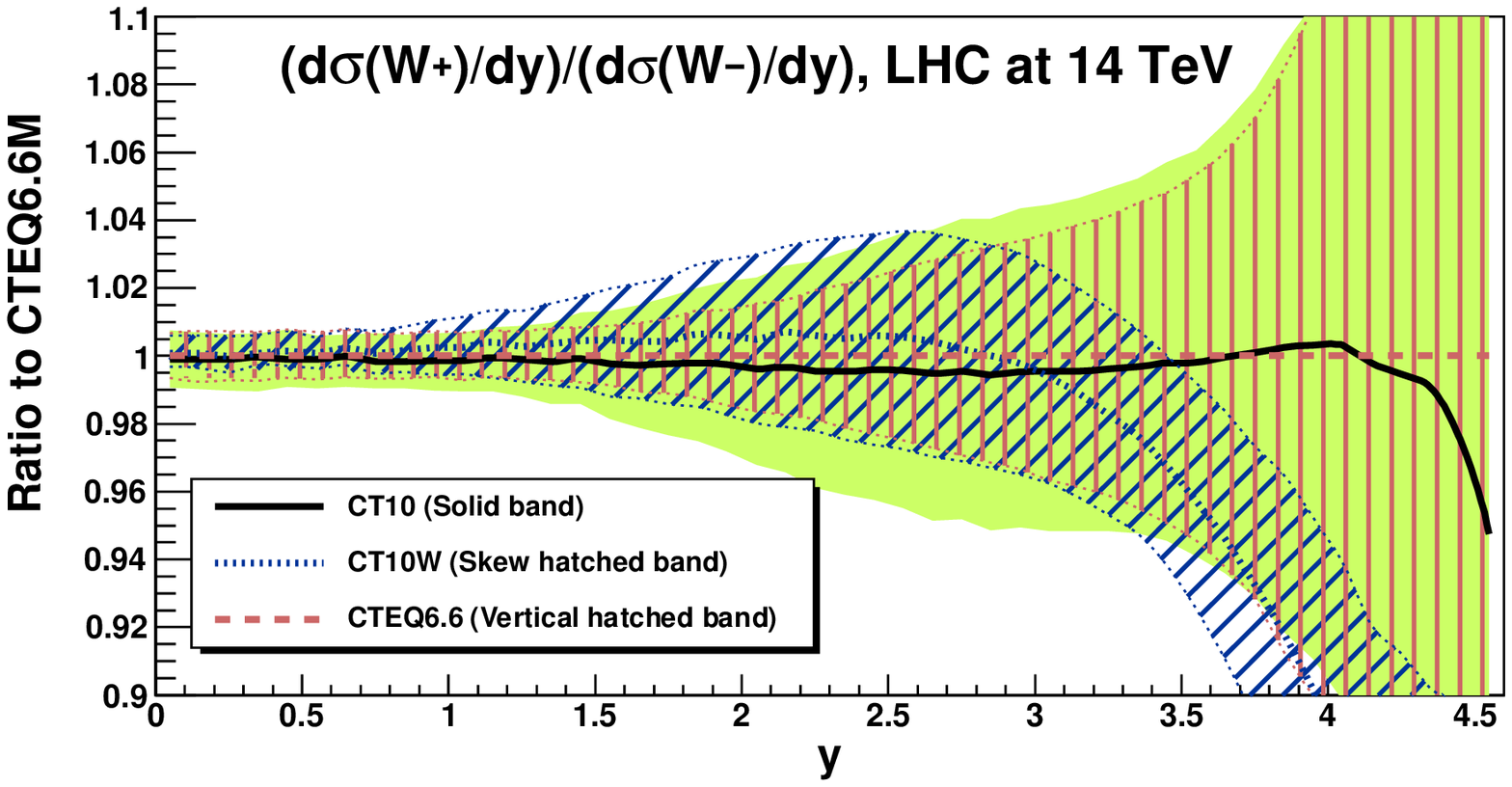} 
\par\end{centering}
\caption{CT10, CT10W, and CTEQ6.6 PDF uncertainty bands for the ratios
$(d\sigma(W^\pm)/dy)/(d\sigma(Z)/dy)$ (upper two subfigures) and 
$(d\sigma(W^+)/dy)/(d\sigma(W^-)/dy)$ (lower two subfigures), at the
  LHC energies 7 and 14 TeV.}
\label{figs:RratioZW} 
\end{figure}

Figure\ \ref{figs:RratioZW} shows the
uncertainty bands of three PDF sets for the ratio 
$(d\sigma(W^\pm)/dy)/(d\sigma(Z)/dy)$ of the $W^\pm$ and $Z$
production cross sections in the upper two subfigures, and
for the ratio $(d\sigma(W^+)/dy)/(d\sigma(W^-)/dy)$ 
of $W^+$ and $W^-$ production cross sections in the lower two
subfigures. 
The ratios obtained with CT10W are smaller than the CTEQ6.6 and CT10 ratios at
large rapidities ($y>2-3$), and they are slightly larger than the
CTEQ6.6 and CT10 ratios at small rapidities.   
For the ratio of the rapidity distributions of $W^{+}$+$W^{-}$ and $Z$, 
both CT10 and CT10W sets predict larger PDF uncertainties than 
does CTEQ6.6, in the region where 
the rapidity $y$ of the boson is less than about 3. 
This is a result of the more 
flexible parametrization of the strange (anti-strange) quark PDF 
employed in the CT10 and CT10W PDFs. 
However, for the ratio of $W^{+}$ to $W^{-}$, 
the CT10 predictions provide a slightly
smaller PDF uncertainty than does CTEQ6.6, and CT10W has an 
even smaller
uncertainty. The latter is a result of the inclusion of the D\O~ Run-II $W$ lepton asymmetry data, which reduces the uncertainty in $d/u$, 
especially in the large $x$ region. 

Finally, we examine PDF-driven 
correlations between the total cross sections for the 
$W$ boson and the $Z$ boson at the Tevatron Run-II and the LHC. Following the 
method described in Ref. \cite{Nadolsky:2008zw}, we show tolerance ellipses 
for various cross sections of $W^+$, $W^-$ and $Z$ bosons, calculated at  
NLO in QCD, unless specified otherwise. 

\begin{figure}[h]
\includegraphics[scale=0.5]{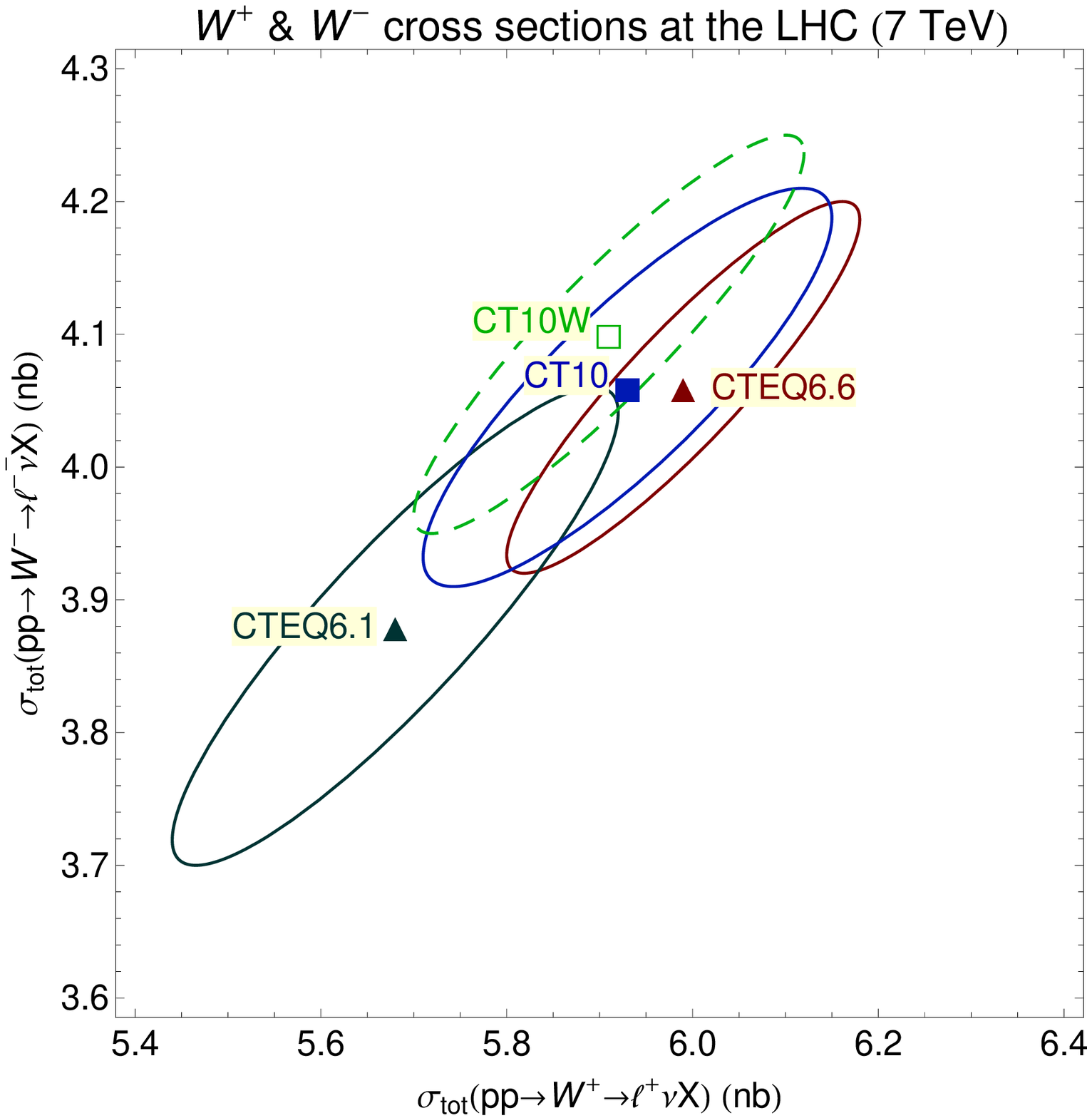}
\includegraphics[scale=0.5]{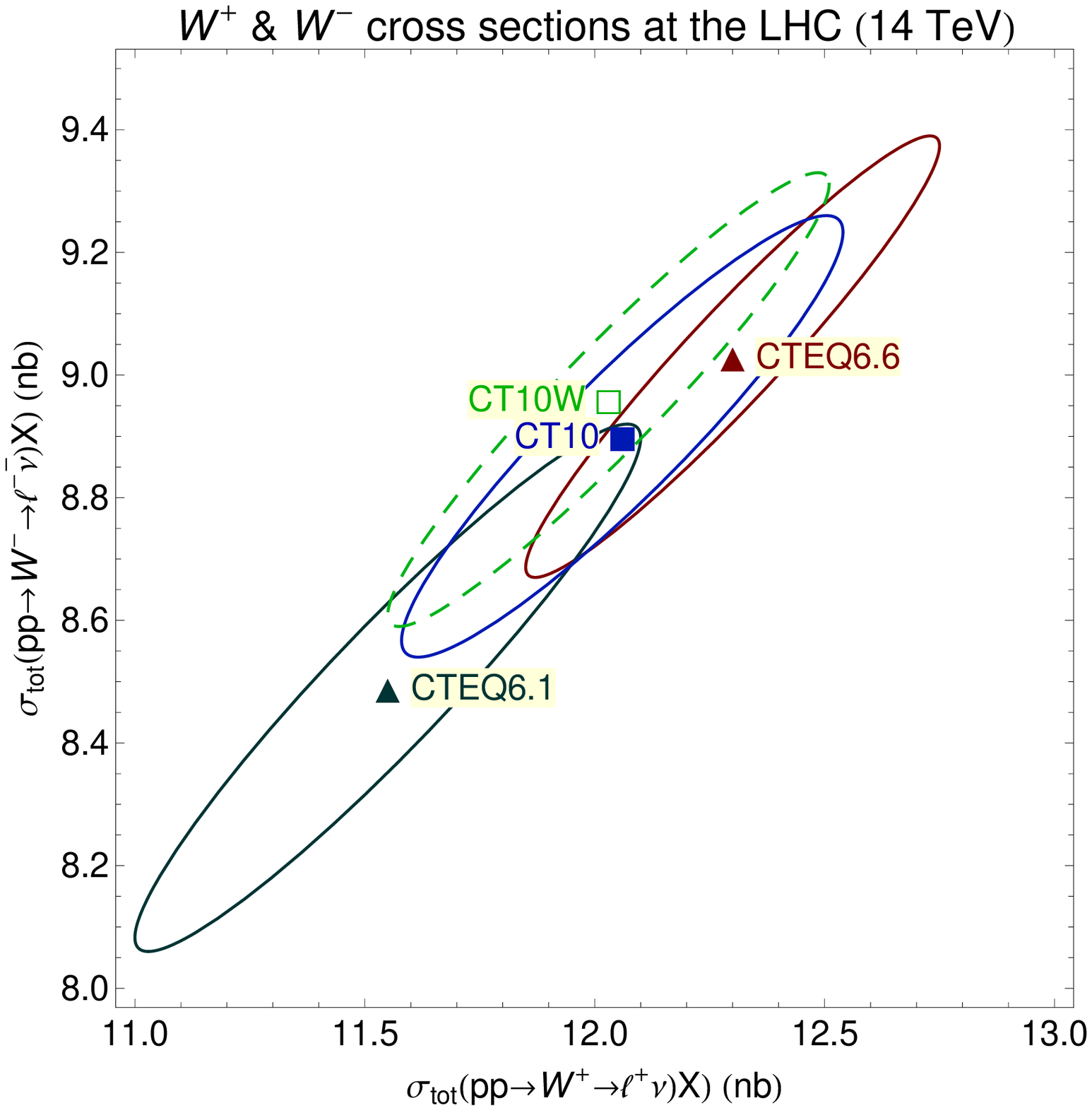}

\caption{
Total cross sections for inclusive $W^+$ and $W^-$ boson production at
the LHC, obtained with the recent CTEQ PDFs and shown with 
their PDF uncertainty ellipses.
}

\label{figs:wpwm} 

\end{figure}

Figure\ \ref{figs:wpwm} shows the comparison between $W^+$ and $W^-$
total cross sections at the LHC. Compared to CT10, the CT10W set predicts
slightly smaller $W^+$ total cross sections and larger $W^-$ cross sections
(with the latter increased by  1-2\%).
The correlation between CT10(W) $W^+$ and $W^-$ cross sections is
relaxed somewhat compared to CTEQ6.6, reflecting 
larger flexibility of the CT10(W) input parametrizations.

\begin{figure}[h]
\includegraphics[scale=0.45]{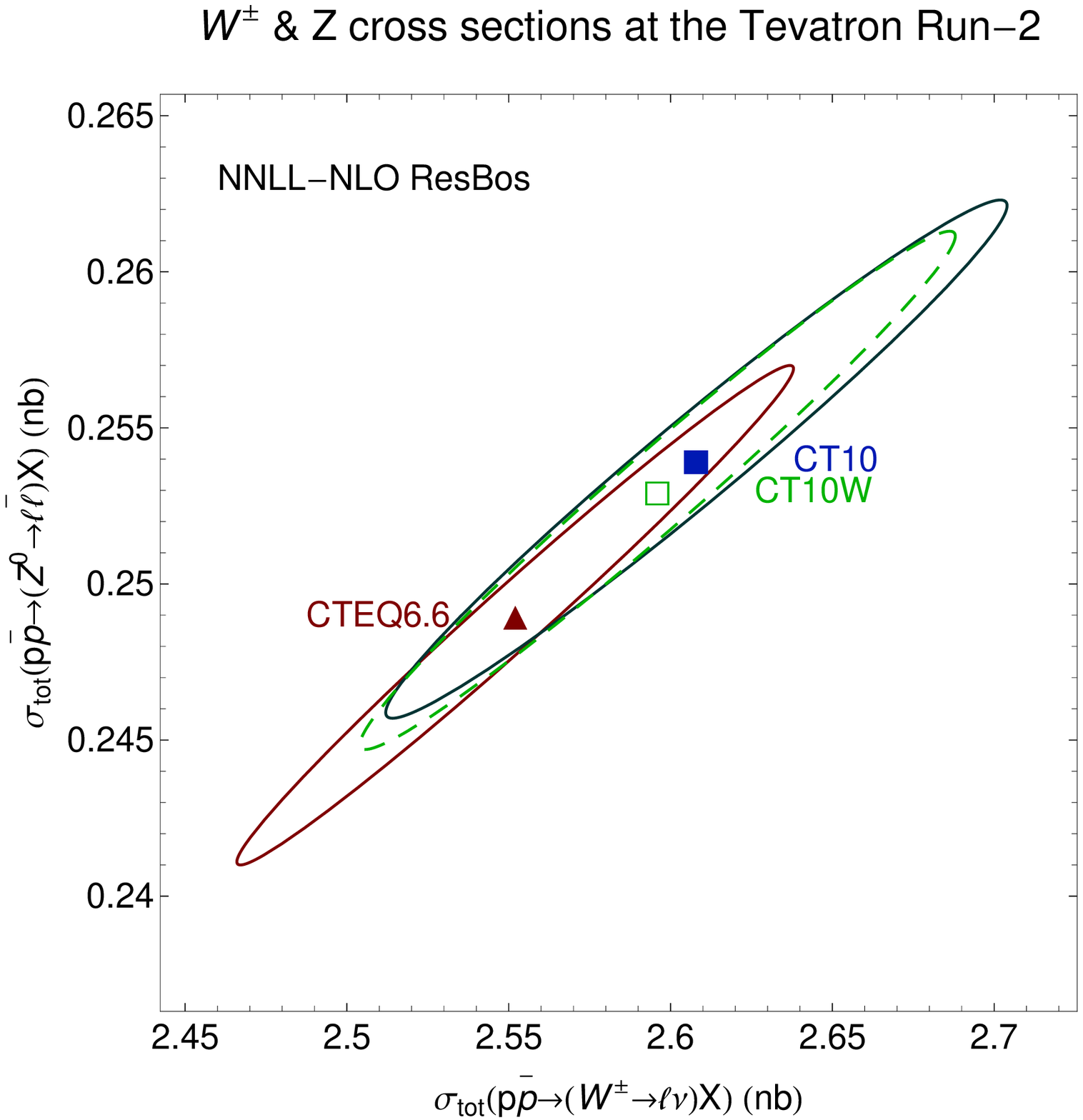}
\includegraphics[scale=0.45]{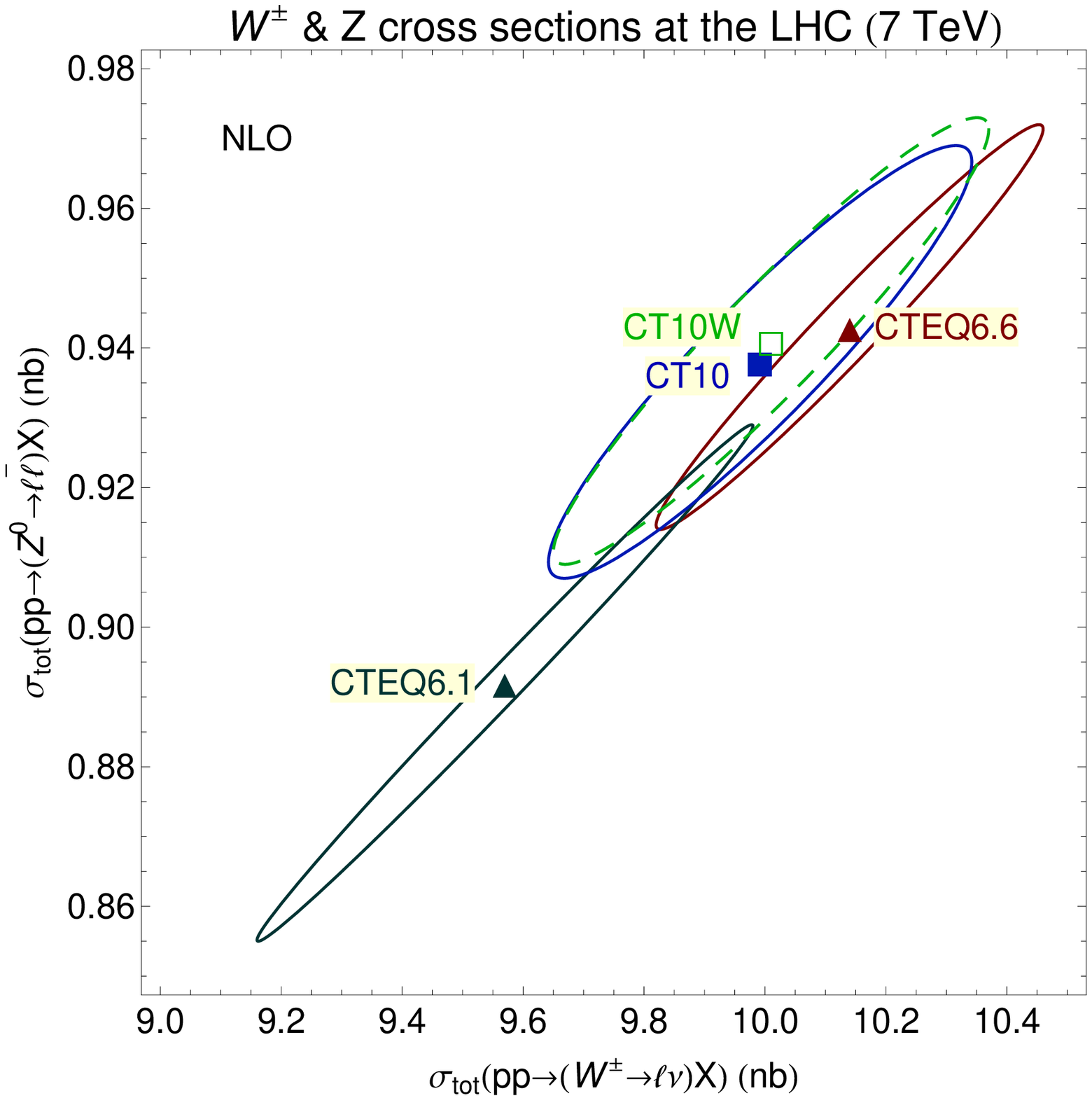}
\includegraphics[scale=0.45]{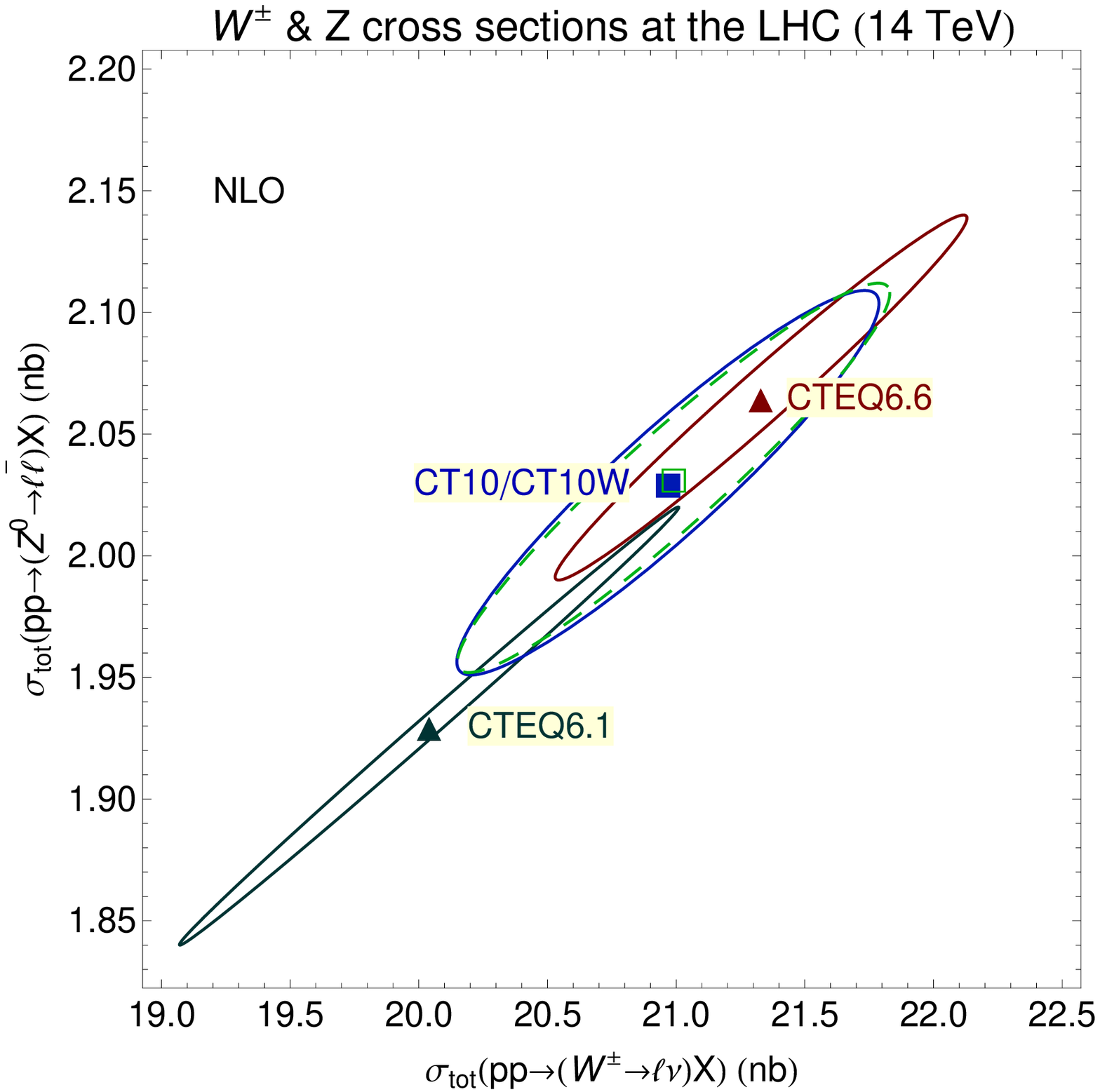}

\caption{
Total cross sections for inclusive $W^\pm$ and $Z$ production
at the Tevatron Run-II and the LHC, obtained with 
the recent CTEQ PDFs and shown with 
their PDF uncertainty ellipses.
}

\label{figs:wz}

\end{figure}

Fig.\ \ref{figs:wz} shows 
the $W^\pm$ and $Z$ total cross sections at the Tevatron Run-II and  the LHC.
At the Tevatron, the CT10 and CT10W cross sections are larger by 1\%
than the respective CTEQ6.6 cross sections for both $W^\pm$ and $Z$,
which is within the PDF uncertainty ellipse for either PDF set. Also, 
the CT10 and CT10W ratios of $W^\pm$ and $Z$ cross sections 
at the Tevatron are the same as that for CTEQ6.6. However, while the central
CT10(W) cross sections at the LHC also agree  
with their CTEQ6.6 counterparts within the PDF
uncertainties, there is a noticeable difference between the CT10(W)
and CTEQ6.6 ratios of $W^\pm$ and $Z$ cross sections. In addition, the PDF
uncertainties of the $W^\pm$ and $Z$ cross sections are less
correlated in the case of CT10(W), as a result of additional
freedom in the CT10(W) strangeness PDF.

\subsection{Other Significant Processes}

\begin{figure}[ht]

\begin{centering}
\includegraphics[clip,scale=0.41]{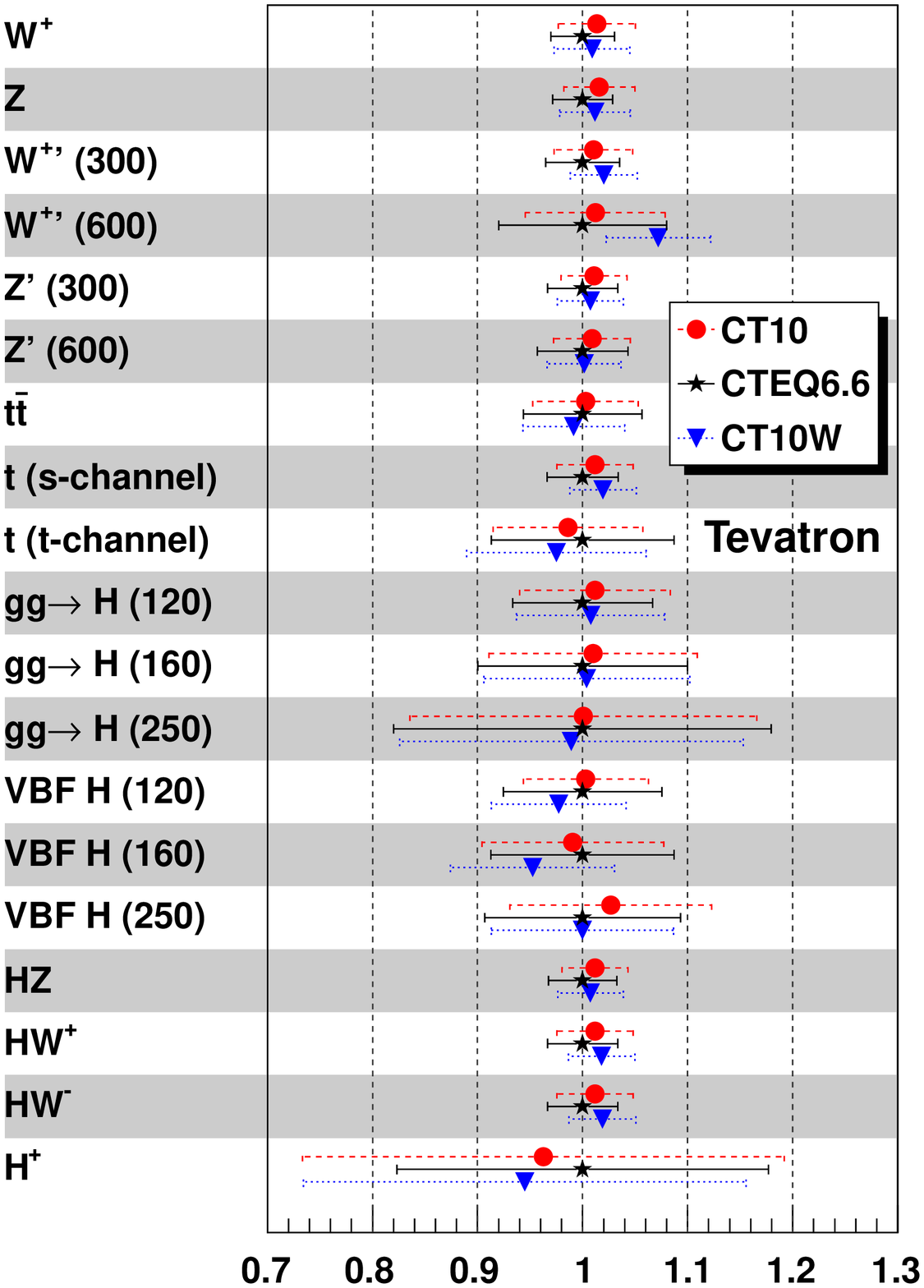}
\includegraphics[clip,scale=0.41]{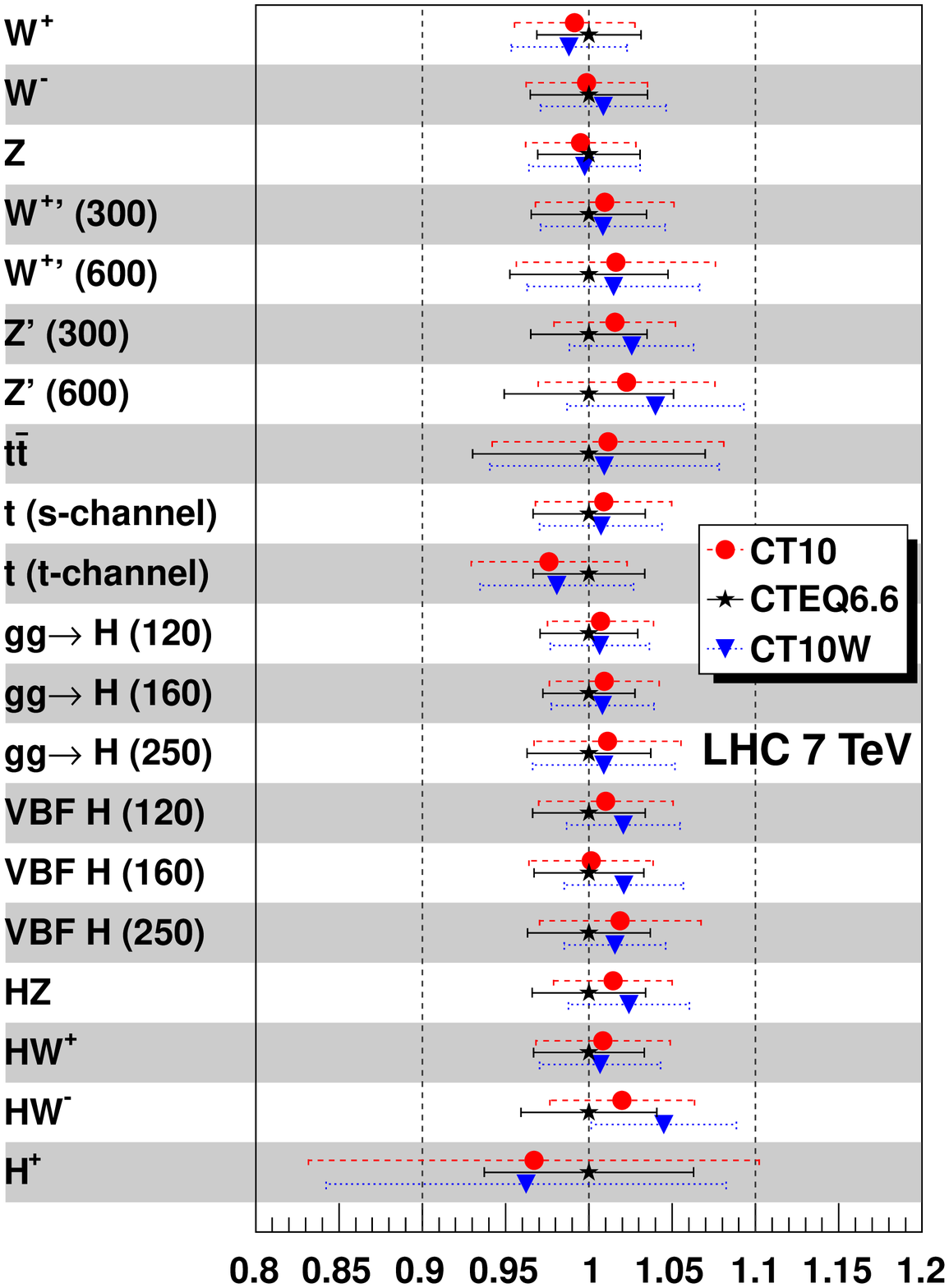}
\includegraphics[clip,scale=0.41]{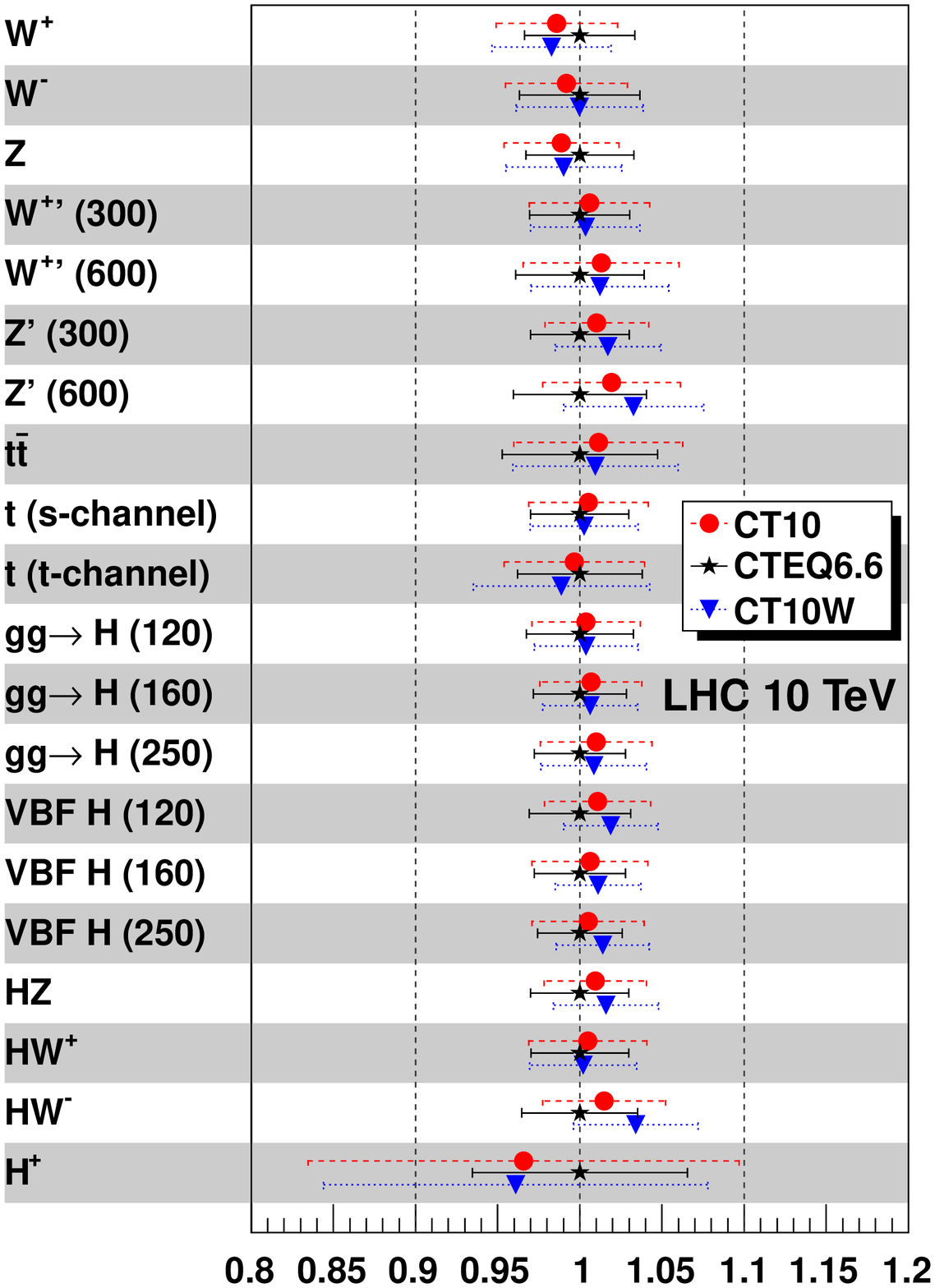}
\includegraphics[clip,scale=0.41]{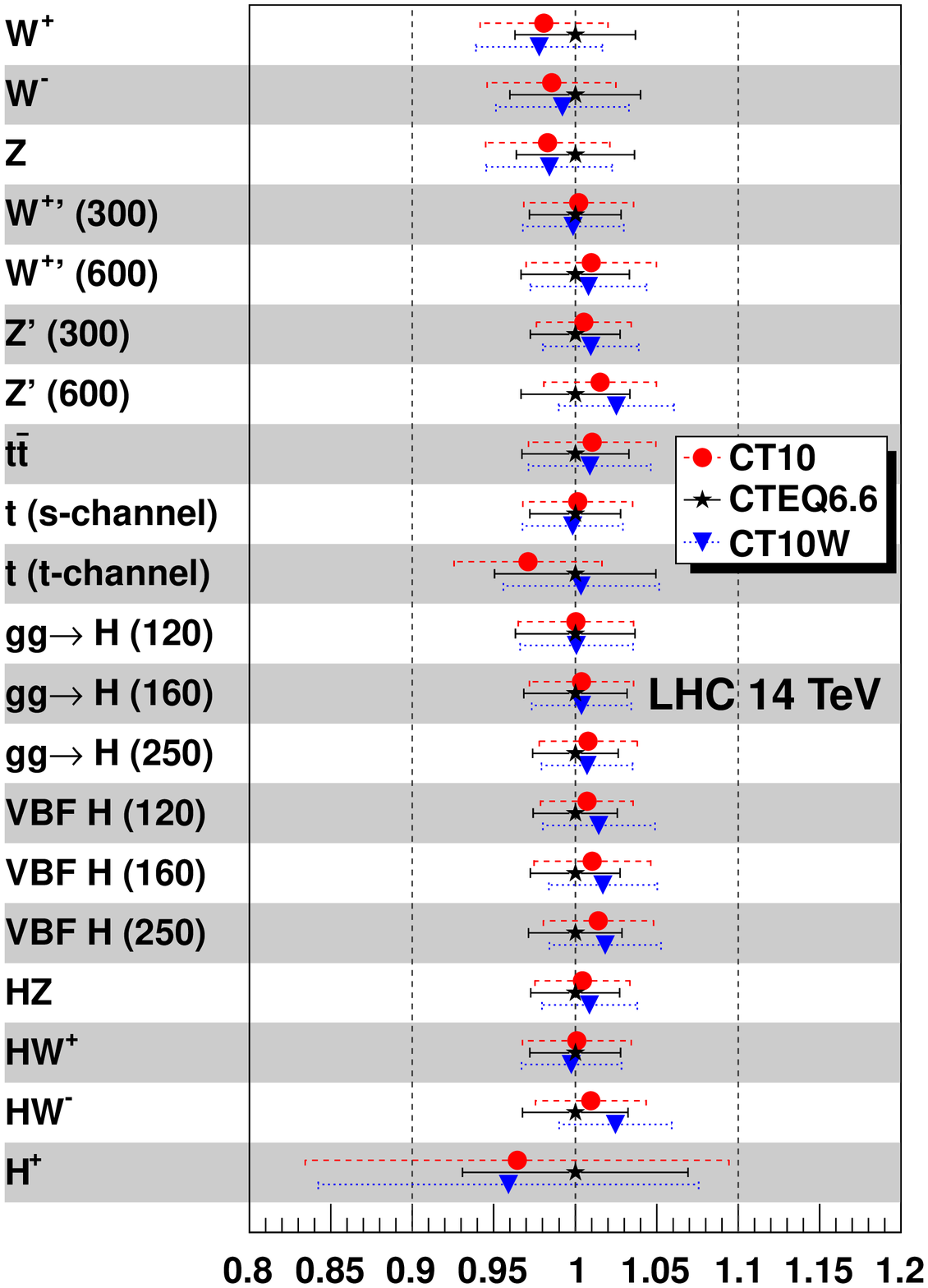} 
\par\end{centering}

\caption{Ratios of NLO total cross sections obtained using CT10 and
  CT10W to those using CTEQ6.6M PDFs, in various scattering
  processes at the Tevatron Run-II and LHC.}

\label{figs:XSECct10w}

\end{figure}

To illustrate the impact of the CT10(W) PDFs 
on hadron collider phenomenology, we 
compare the total cross sections of some selected  
processes at the Tevatron Run-II 
and the LHC (at center-of-mass energies 7 TeV, 10 TeV and 14 TeV). 
The processes include the production of 
$W^+$, $W^-$, and $Z$ bosons, also discussed above; 
top-quark ($t \bar t$) pairs; 
single top-quark in $s$ and $t$ channels; 
Standard Model (SM) Higgs boson via gluon fusion 
($gg \rightarrow H$, with Higgs boson mass being 120 GeV, 160 GeV or 250 GeV) \cite{Spira:1995mt}; 
SM Higgs boson via weak gauge boson fusion ($VV \rightarrow H$) \cite{Arnold:2008rz};
associated production of SM Higgs boson and a weak gauge boson ($HW^+$, $HW^-$ and 
$HZ$); ``sequential'' heavy weak bosons, $W^{\prime+}$ and $Z^{\prime}$, 
with masses 300 GeV or 600 GeV; and a 200 GeV charged Higgs boson via 
$c \bar s \rightarrow H^+$, as predicted by the two-Higgs-doublet model. 
(The couplings of $W^{\prime}$ and $Z^{\prime}$ bosons 
to fermions are taken to be the same as those in the Standard Model.)

Fig.\ \ref{figs:XSECct10w} 
shows the ratios of the NLO total cross sections, obtained  
using CT10 and CT10W PDFs, to those obtained using CTEQ6.6 PDFs. 
For most of the cross sections, CT10
and CT10W sets provide similar predictions and uncertainties, which 
are also in good agreement with those from CTEQ6.6 ({\it i.e.}, well within the PDF
uncertainty band). 
At the LHC, the PDF uncertainties in 
CT10 and CT10W predictions for some processes are larger than
those in CTEQ6.6 predictions, reflecting the changes in the framework
of the fit discussed in Sec.~\ref{sec:NewTheory}. At the Tevatron, 
the CT10(W) PDF uncertainties 
tend to be about the same as those for CTEQ6.6, with a notable
exception of $t\bar t$ production cross sections, which have
a smaller PDF uncertainty with the CT10W set, because of 
stricter constraints on the 
up- and down-quark PDFs at the relevant $x$ values.

Another notable change is in the 
$W^{\prime +}$ (600 GeV) production cross section 
at the Tevatron, which is enhanced  with CT10W PDFs
as a result of the increase in the large-$x$ down quark PDF
driven by the $A_\ell$ data.
At the 14 TeV LHC, 
the total cross sections of $W$ and $Z$ bosons decrease, while 
those of $Z^\prime$ and $HW^-$ increase. 
The decrease in the central value of the  
$c \bar s \rightarrow H^+$ cross section in CT10 and CT10W predictions 
 is due to the decrease in the strange 
quark PDF at the relevant $x$ values; however, its uncertainty also
increases with CT10 or CT10W, as compared to the predictions based 
on the CTEQ6.6 PDFs.

\subsection{
Dijet 
Invariant Mass Distributions}

\begin{figure}[b]
\includegraphics[width=0.49\textwidth]{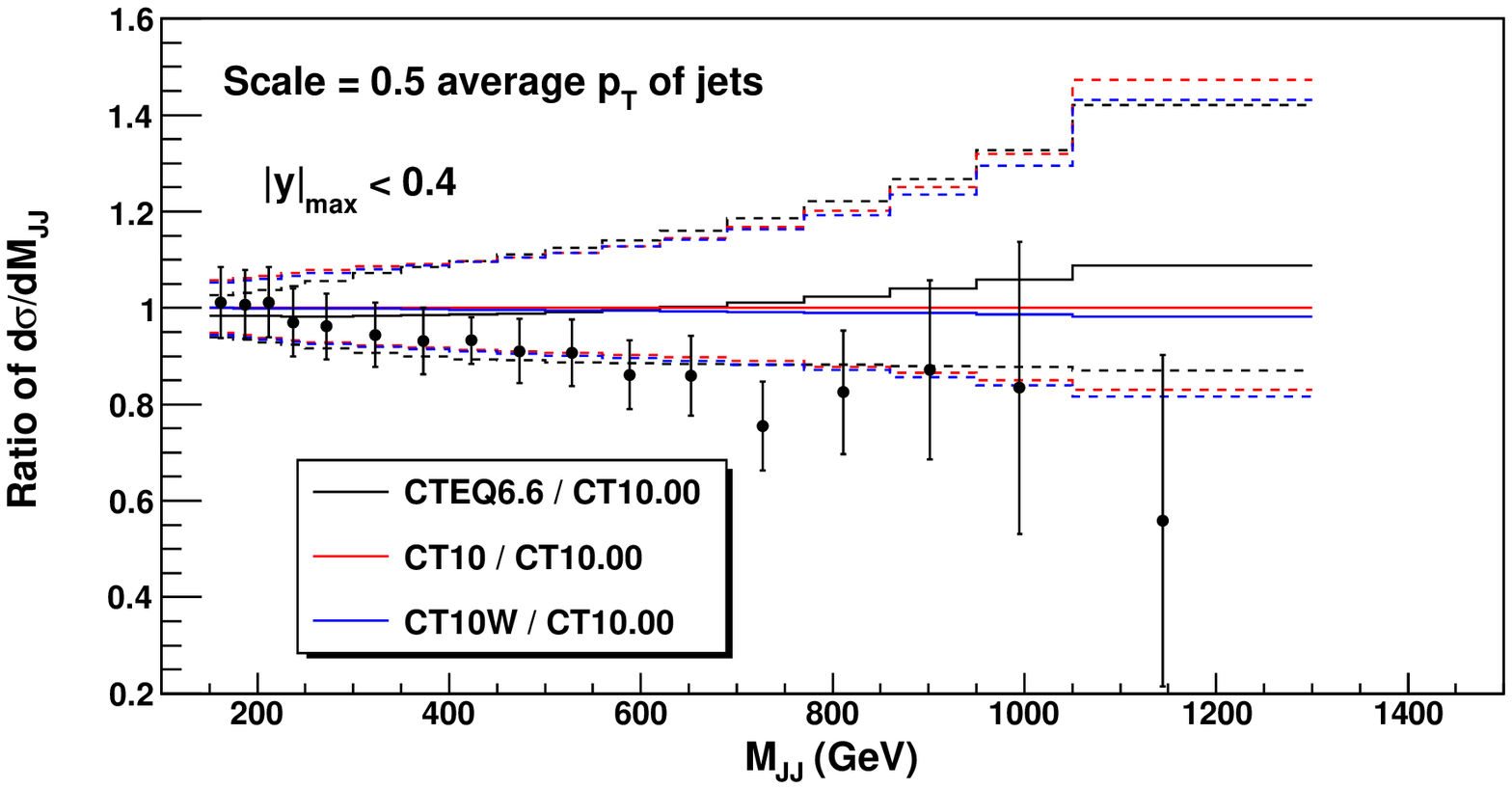}
\includegraphics[width=0.49\textwidth]{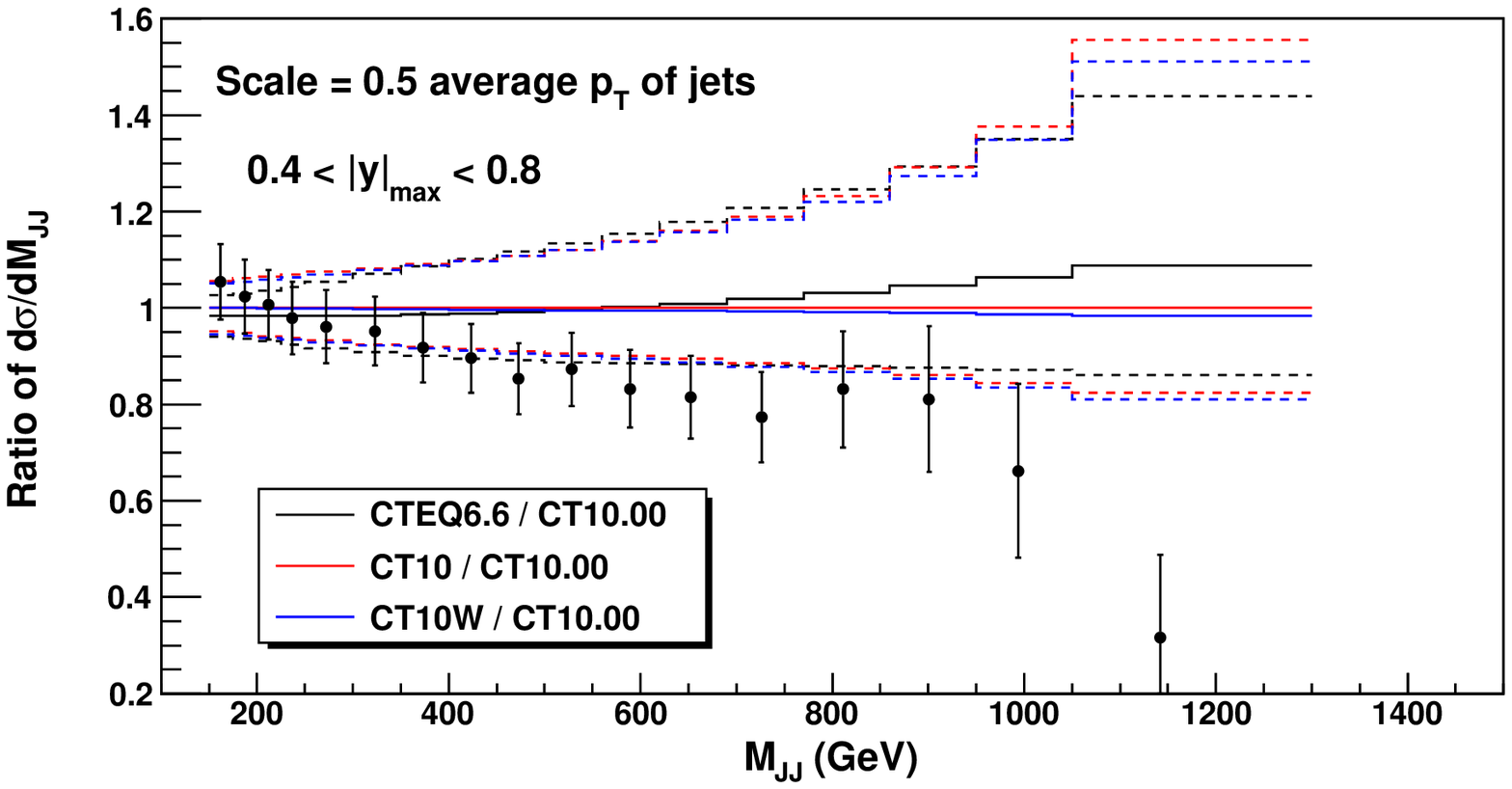}
\includegraphics[width=0.49\textwidth]{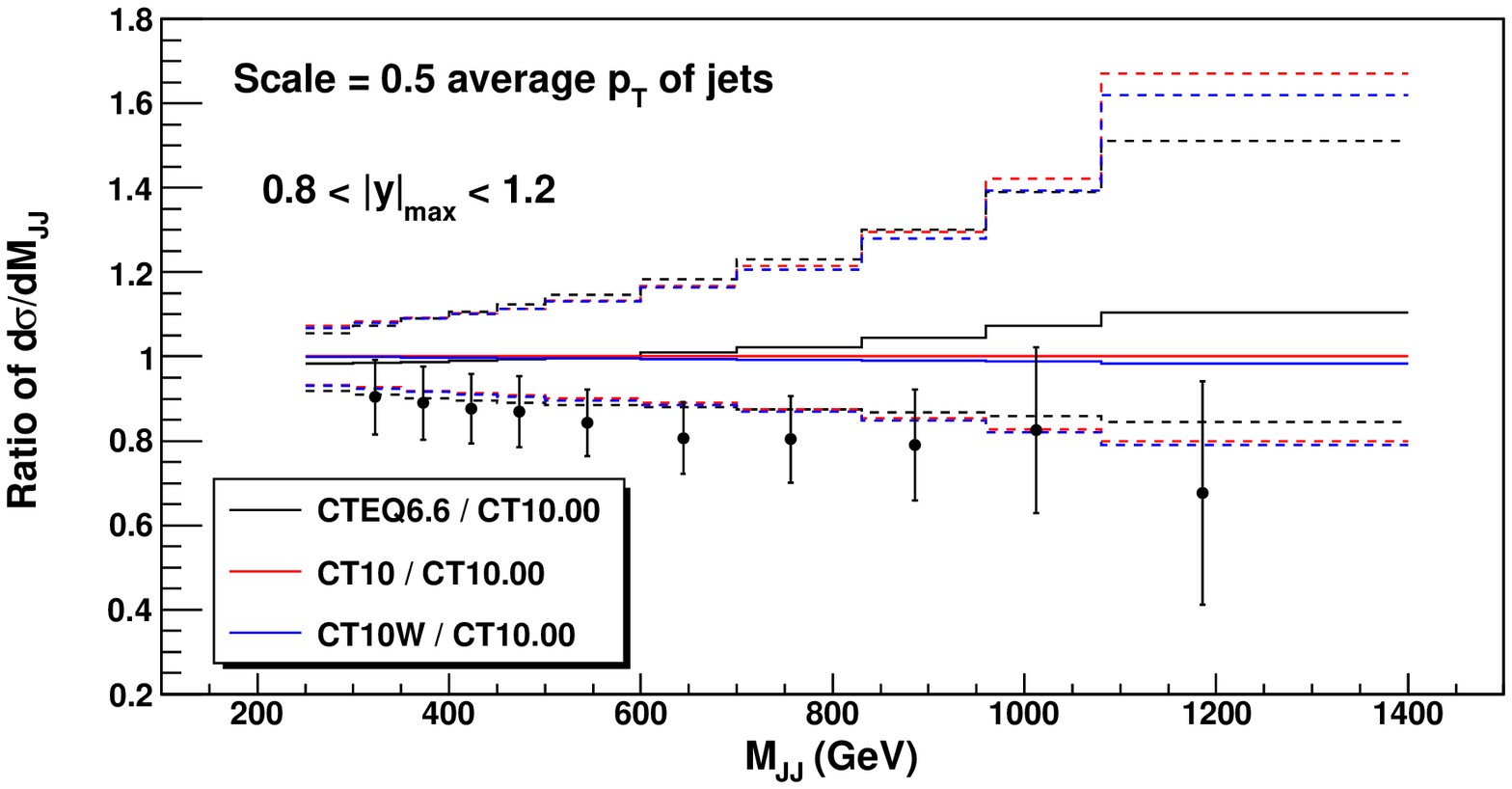}
\includegraphics[width=0.49\textwidth]{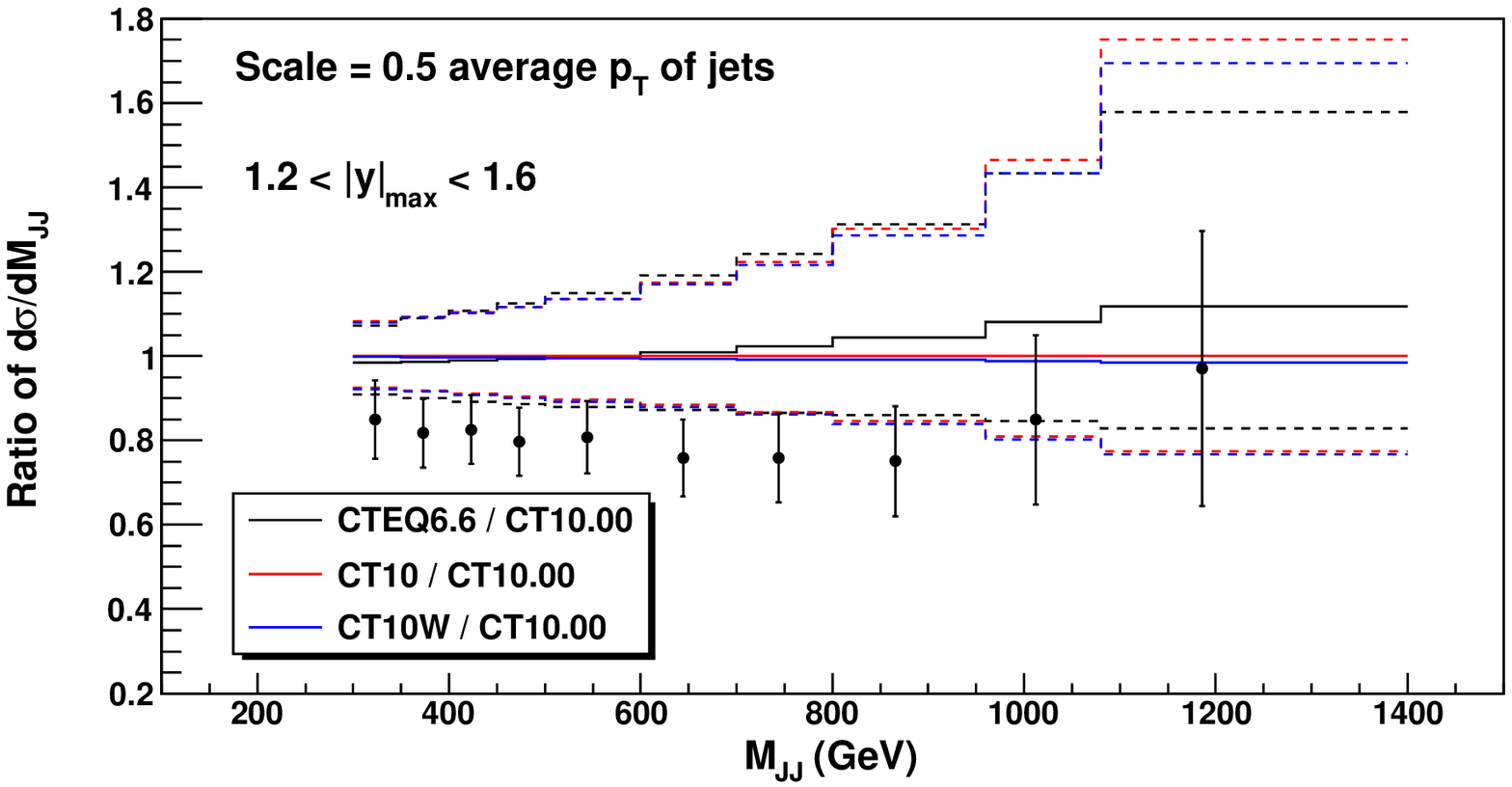}
\includegraphics[width=0.49\textwidth]{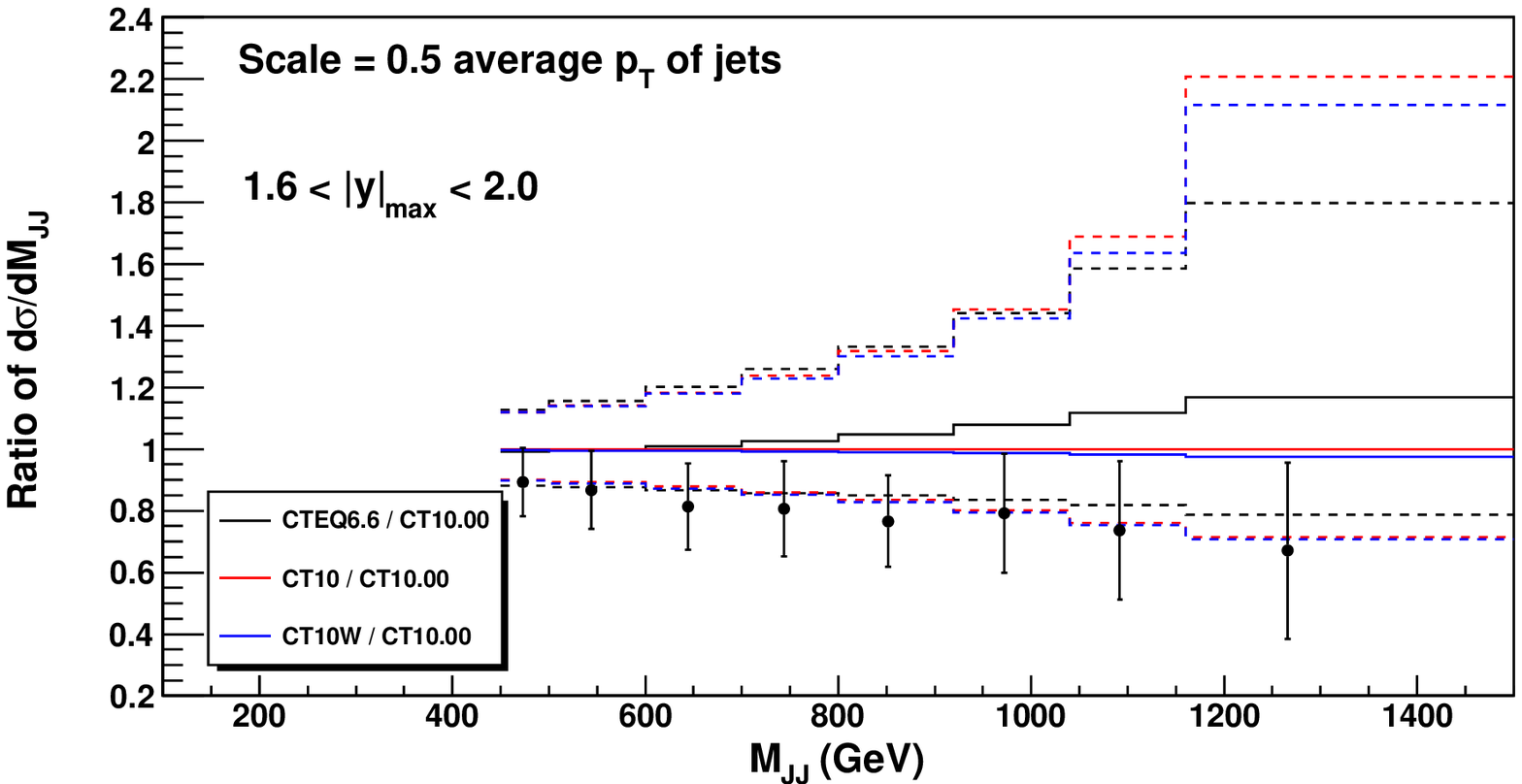}
\includegraphics[width=0.49\textwidth]{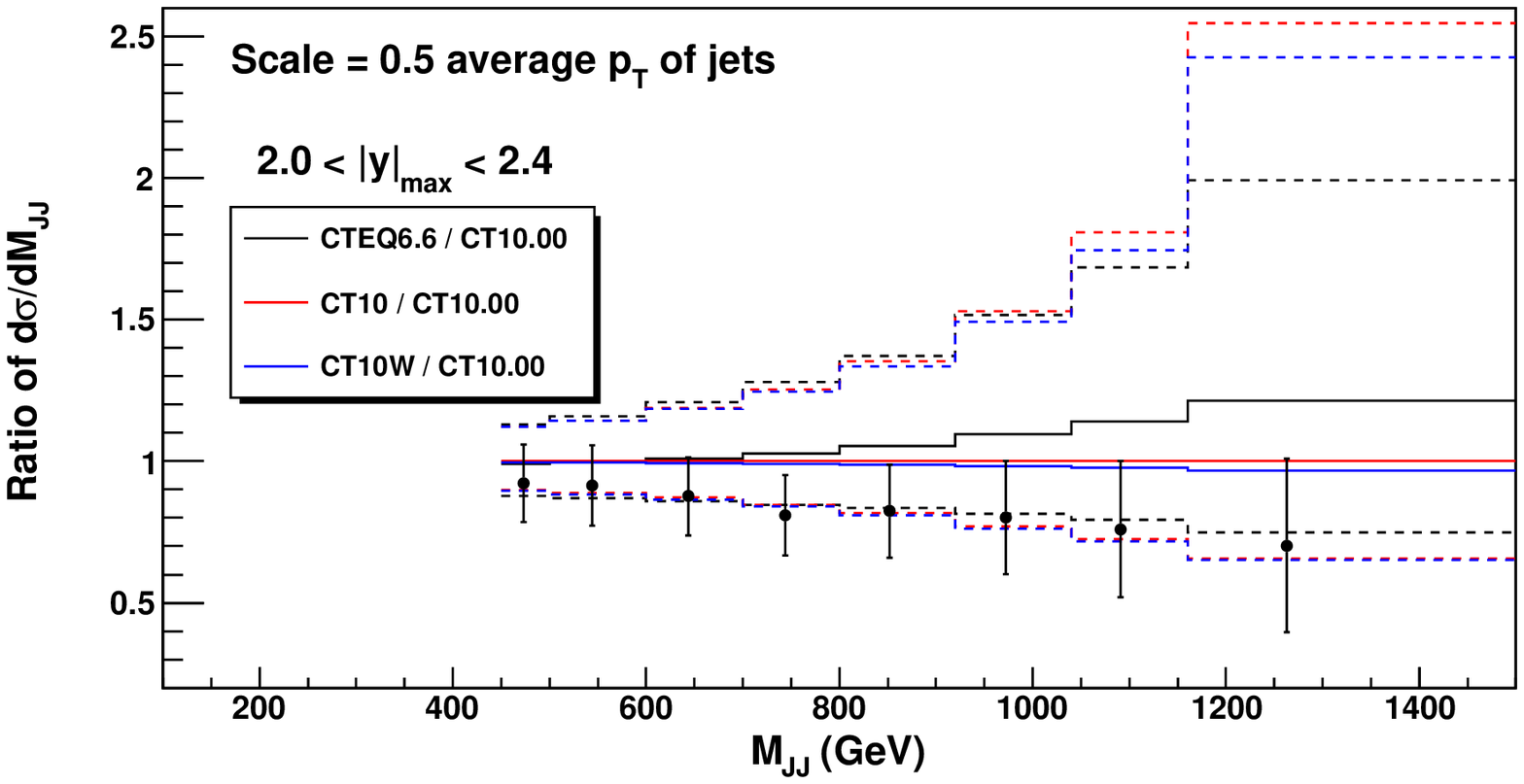}

\caption{Comparison of \protect{D\O~} Run-II data for dijet invariant mass distributions  \cite{d0-di-jet} with NLO theoretical predictions and their PDF uncertainties for CTEQ6.6 (black), CT10 (red) and CT10W (blue) PDFs. The cross sections are normalized to theoretical predictions based 
on the best-fit CT10.00 PDF set.}

\label{figs:DIJET6} 
\end{figure}

Recently, the D\O~ Collaboration reported their measurement of the 
dijet 
invariant mass 
distribution \cite{d0-di-jet}, in which a comparison was made to an NLO theory 
calculation (with FastNLO code \cite{fastnlo}) using the CTEQ6.6M PDF set, 
with both the renormalization and factorization scales set equal to the average of the 
transverse momentum of {\it the jet pair},  
$\langle p_T\rangle \equiv (p_T^{\mbox{\small jet 1}} + p_T^{\mbox{\small jet 2}})/2 $. 
In Fig.~2 of \cite{d0-di-jet}, it appears that the predictions using the CTEQ6.6M 
PDFs cannot describe the data in the large dijet invariant mass region.
Below, we shall examine the above analysis with a different choice 
of the hard scale, $\langle p_T\rangle /2$ rather than $\langle p_T\rangle$, 
which is approximately equal to the scale used 
in our theoretical cross sections for the 
Tevatron Run-I and Run-II inclusive jet data.
The reason for examining the predictions with this choice of the scale
is that the high-$x$ gluon distribution in our global fits is primarily determined by the Tevatron inclusive jet data. At NLO, the size of the predicted jet cross sections, and thus the size of the gluon distribution determined, 
depends tangibly on the assumed renormalization and factorization scales \cite{Stump:2003yu}. The gluon distribution in this $x$ region would have been different, had the average transverse momentum of the dijet pair been used in the global fit. Of course, both scales are equally valid for the dijet cross section evaluation, but it is important to understand any differences generated by the use of one scale for the the PDF determination and another for the evaluation of the dijet cross section.\footnote{The two processes are clearly related, and consist basically of the same events.} Such scale uncertainties form a part 
of theoretical uncertainties arising in  PDF determination.

Fig.\ \ref{figs:DIJET6} shows the NLO dijet invariant mass
distributions at the Tevatron Run-II, $d\sigma/dM_{jj}$, 
for CTEQ6.6 (black), CT10 (red) and CT10W (blue) PDFs, normalized to $d\sigma/dM_{jj}$ for the CT10.00 PDF, and including the PDF uncertainties. 
The cross sections are computed in bins of 
$|y|_{max}=\mbox{max}(|y_{\mbox{\small jet 1}}|,|y_{\mbox{\small jet 2}}|)$,  
with the renormalization and factorization scales 
chosen to be $\langle p_T\rangle /2$.
The D\O~ data, with statistical and total systematic errors 
added in quadrature, are also shown.
We find that with this choice of the scale,  
all three PDF sets are in better agreement with the
data than the conclusions of the D\O~ paper \cite{d0-di-jet} indicate, 
although an overall systematic shift, of order of the  systematic
shifts observed in the CT09 study of the related single-inclusive jet
distributions \cite{Pumplin:2009nk}, may further improve the agreement.
As shown in the figure, the predictions for the central fits
of CT10 and CT10W PDFs are close to each other and closer to 
the data than CTEQ6.6. 

The extent of the CTEQ6.6, CT10, and CT10W PDF uncertainty bands 
in this ratio is larger, by a factor of two, than those derived from the MSTW2008 PDFs. As a result, the MSTW2008 predictions are within our error bands,
although the reverse is not true.\footnote{This observation is consistent
with the analysis of the dijet invariant mass distribution of CDF Run-II
data \cite{d0-di-jet}, which also shows a sizable uncertainty band 
using CTEQ6.6 PDFs.}
The PDF error bands for large dijet masses are not symmetric; the upper side has more variation than the lower side. The asymmetry arises because 
dijet production at large $M_{jj}$ and $|y|_{max}$ 
is dominated by quark-quark or quark-antiquark
scatterings, with a smaller contribution from gluon-quark scattering.
Since the quark distributions are relatively better determined at 
medium to large $x$ values, the differential 
cross section of the  dijet invariant mass distribution 
cannot become too small. (The quark-gluon scattering process can only 
increase the cross sections.)

\section{Conclusions \label{sec:Conclusions}}

With the LHC is reporting its first cross sections, 
it becomes even more important to provide the best tools 
necessary for accurate predictions and comparisons to 
those cross sections. We have produced two new PDF sets, 
CT10 and CT10W, intended   for comparisons to data at 
the Tevatron and LHC. The two PDF sets include new data, 
primarily the DIS combined data sets from HERA \cite{2009wt},
the rapidity distribution of $Z^{0}$ production at the Tevatron,
 and the Tevatron Run-II $W$ lepton asymmetry data 
from the D\O~ Collaboration, 
as well as several 
improvements to the global fitting procedure. 
The latter includes more flexible PDF parametrizations, 
the treatment of experimental normalizations in the 
same manner as other systematic uncertainties, the 
removal of weights associated with the data sets (except for 
the $W$ lepton asymmetry data in the case of CT10W), and a 
more dynamical determination of the allowed tolerance 
along each eigenvector direction.

Due to the difficulty in fitting both the Tevatron 
Run-II $W$ lepton asymmetry data and the other data sets in 
the global analysis (primarily, the deuteron/proton DIS cross section 
ratio from the NMC experiment), 
we have produced two new families of PDFs, CT10 and CT10W. 
CT10 is obtained without using the D\O~ Run-II $W$ lepton 
asymmetry data, while CT10W contains those high-luminosity 
data with added weights to ensure 
reasonable agreement.  The resulting predictions for LHC 
benchmark cross sections, at  both 7 TeV and 14 TeV, are 
generally consistent with those from the older CTEQ6.6 PDFs, in some 
cases with a slightly larger uncertainty band. The latter 
is a result of the greater flexibility included in this new 
generation of global fits. Most noticeable differences in various 
cross sections, such as the charged Higgs boson 
and extra heavy gauge boson production,
are induced by changes in
the strange-quark PDF, the gluon PDF in the small-$x$ region, 
and the up-quark and down-quark PDFs in the medium to large $x$ region. 

As compared to the CTEQ6.6 prediction, both CT10 and CT10W predict a smaller 
PDF induced uncertainty in the total cross section 
for the top-quark pair production at the Tevatron Run-II.
No large differences are observed 
for LHC predictions between the CT10 and CT10W PDF sets, except 
in those observables that are sensitive to the ratio of down-quark 
to up-quark PDFs. One example is the ratio of the 
rapidity distributions of the $W^-$ and $W^+$ bosons produced at the LHC. 

In summary, the CT10 and CT10W sets are based on the most 
up-to-date information about the PDFs available 
from global hadronic experiments.   
There are 26 free parameters in both new PDF sets; thus, 
there are 26 eigenvector directions and a total of 52 error 
PDFs for both CT10 and CT10W. The CT10 and CT10W  PDF error 
sets, along with the accompanying $\alpha_s$ error sets, 
allow for a complete calculation of the combined 
PDF+$\alpha_s$ uncertainties for any observable ~\cite{Lai:2010nw}. 
To support calculations for heavy-quark production in 
the fixed-flavor-number factorization scheme, 
we provide additional PDF sets CT10(W).3F and CT10(W).4F, 
obtained from the best-fit CT10.00 and CT10W PDF sets by QCD evolution 
with three and four active quark flavors. 
All the relevant PDF sets discussed in this paper are available as a part of 
the LHAPDF library \cite{lhapdf} and from our website ~\cite{ct10website}.

\paragraph*{Acknowledgments}
We thank A. Glazov for advice 
on the implementation of HERA-1 correlated errors, 
and M. Wobisch for help with implementation of
D\O~ Run-II dijet production data.
We also thank S. Forte, J. Linnemann, L. Lyons, 
F. Olness, J. Rojo, H. Schellman, D. Stump, R. Thorne, J. Zhu, and 
CTEQ members for useful discussions.
We are grateful to  the Center for High Energy Physics 
at Peking University in China (visited by C.P.Y.), 
National Center for Theoretical Sciences 
in Taiwan (visited by C.P.Y. and H.L.L),  
and the Aspen Center for Physics in Aspen, Colorado 
(visited by P.N.), for their hospitality 
during the work on a part of this study.
This work was supported in part 
by the U.S. DOE Early Career Research Award DE-SC0003870;
by the U.S. National Science Foundation under grant PHY-0855561;
by the National Science Council of Taiwan under grants
NSC-98-2112-M-133-002-MY3 and NSC-99-2918-I-133-001;
by LHC Theory Initiative Travel Fellowship awarded by the U.S. National Science Foundation under grant PHY-0705862;
and by Lightner-Sams Foundation.

\clearpage

\appendix
\section*{Appendix: Agreement of QCD theory with the combined HERA-1 data\label{app:CombHERA}}

In this Appendix, we provide additional details on the comparison of
CT10 predictions with the combined HERA data, and the origin of the
increase in $\chi^2$ that is observed when the independent HERA data sets
are combined. The $\chi^2/N$ values in the
intervals $x<0.001$, $0.001<x<0.1$, and $x>0.1$, found in the CT10
best fit to the combined (CT10.00) and separate (CT10-like) 
HERA-1 data sets, as well as in the
CTEQ6.6M fit, are listed in Table~\ref{tab:chi2HERA1}. At $x<0.001$,
$\chi^2/N$ is about 1.19 for the combined HERA-1 set,
vs. 0.81-0.84 in the fits to the separate sets. At $x>0.1$, where
irregular scatter is obvious in the plots of both $e^+p$ and $e^-p$ NC
sets (cf. Fig.~\ref{fig:CT10redNC}),
$\chi^2/N$ is increased upon the combination of the data sets from 1.25
to 1.43. 

\begin{table}
\begin{tabular}{|c|c|c|c|c|c|c|}
\hline 
$x$ range  & \multicolumn{2}{c|}{CT10.00, comb.} & \multicolumn{2}{c|}{CT10-like, sep.} & \multicolumn{2}{c|}{CTEQ6.6M}\tabularnewline
 & $N$ & $\chi^{2}/N$ & $N$ & $\chi^{2}/N$ & $N$ & $\chi^{2}/N$\tabularnewline
\hline
\hline 
<0.001  & 63  & \textbf{1.19 } & 68 & \textbf{0.81} & 68 & \textbf{0.84}\tabularnewline
\hline 
0.001-0.1  & 298  & 0.94 & 485 & 0.92 & 485 & 0.92\tabularnewline
\hline 
>0.1  & 150 & \textbf{1.43} & 257 & \textbf{1.26} & 257 & \textbf{1.25}\tabularnewline
\hline
\end{tabular}\caption{Numbers of data points ($N$) and $\chi^{2}/N$ 
found in the CT10 best fit to the combined (CT10.00) and separate (CT10-like) 
HERA-1 data sets, as well as in the CTEQ6.6M fit.
\label{tab:chi2HERA1}}

\end{table}

\begin{figure}[p]
\includegraphics[height=0.29\textheight]{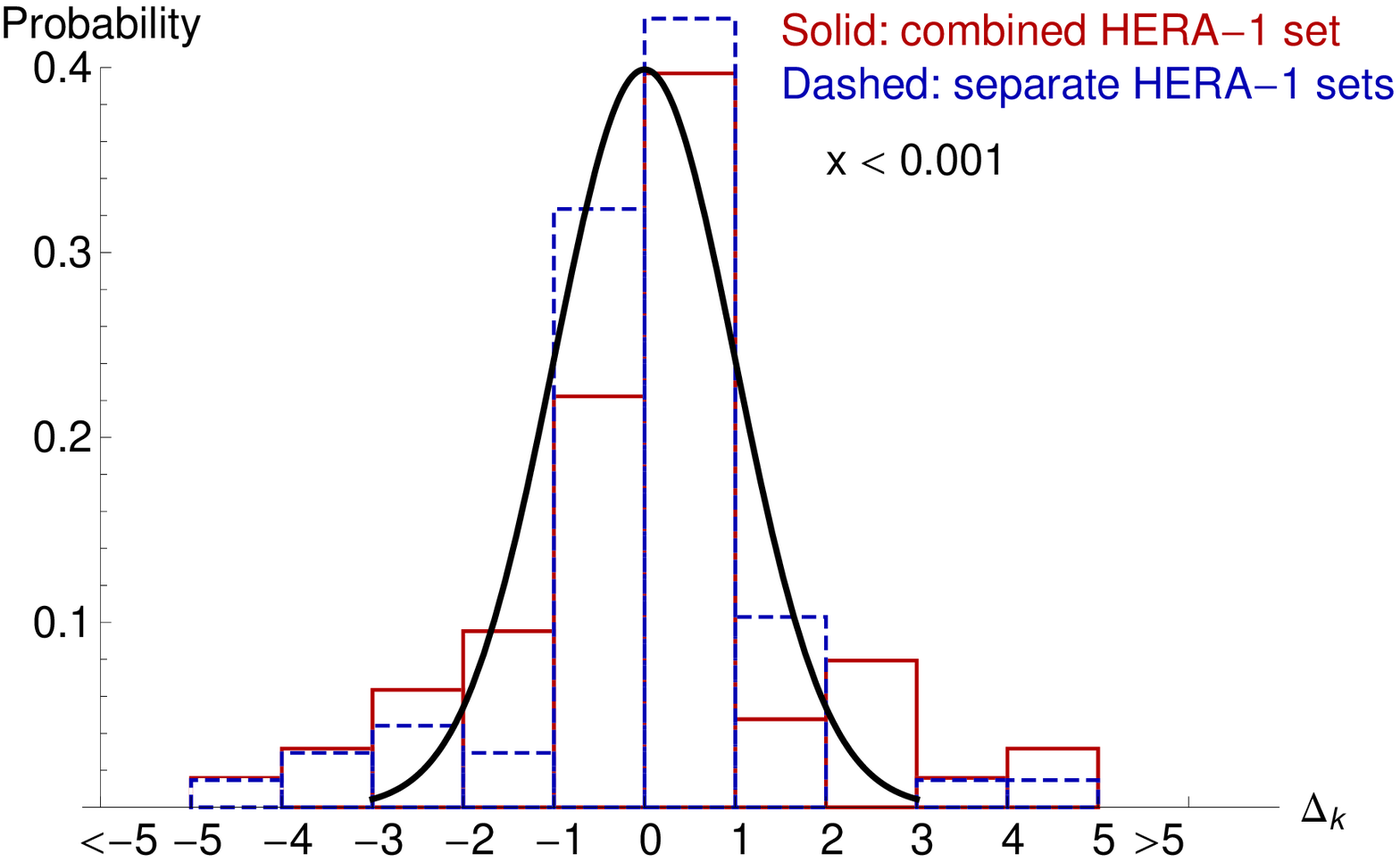}

\includegraphics[height=0.29\textheight]{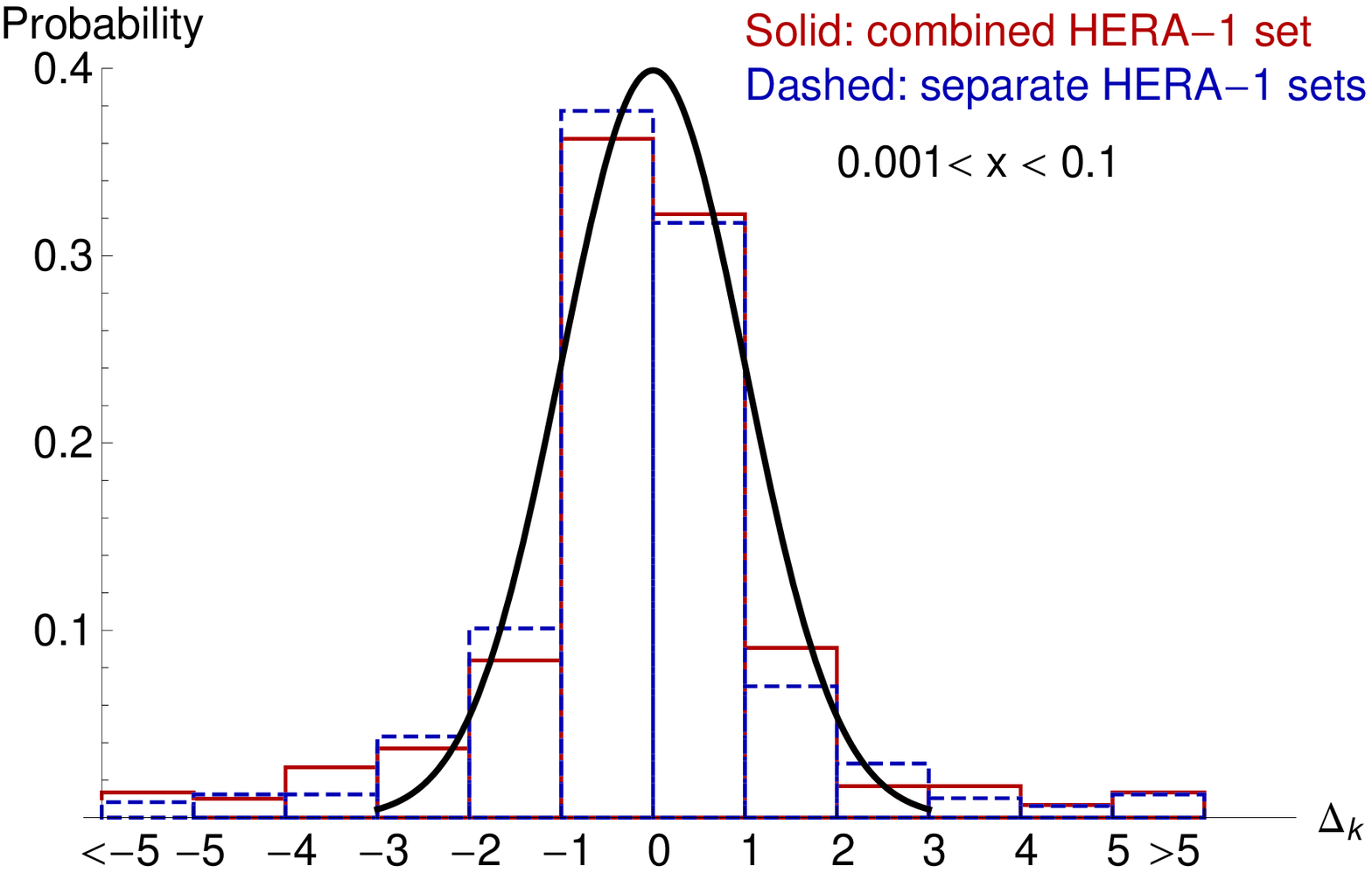}

\includegraphics[height=0.29\textheight]{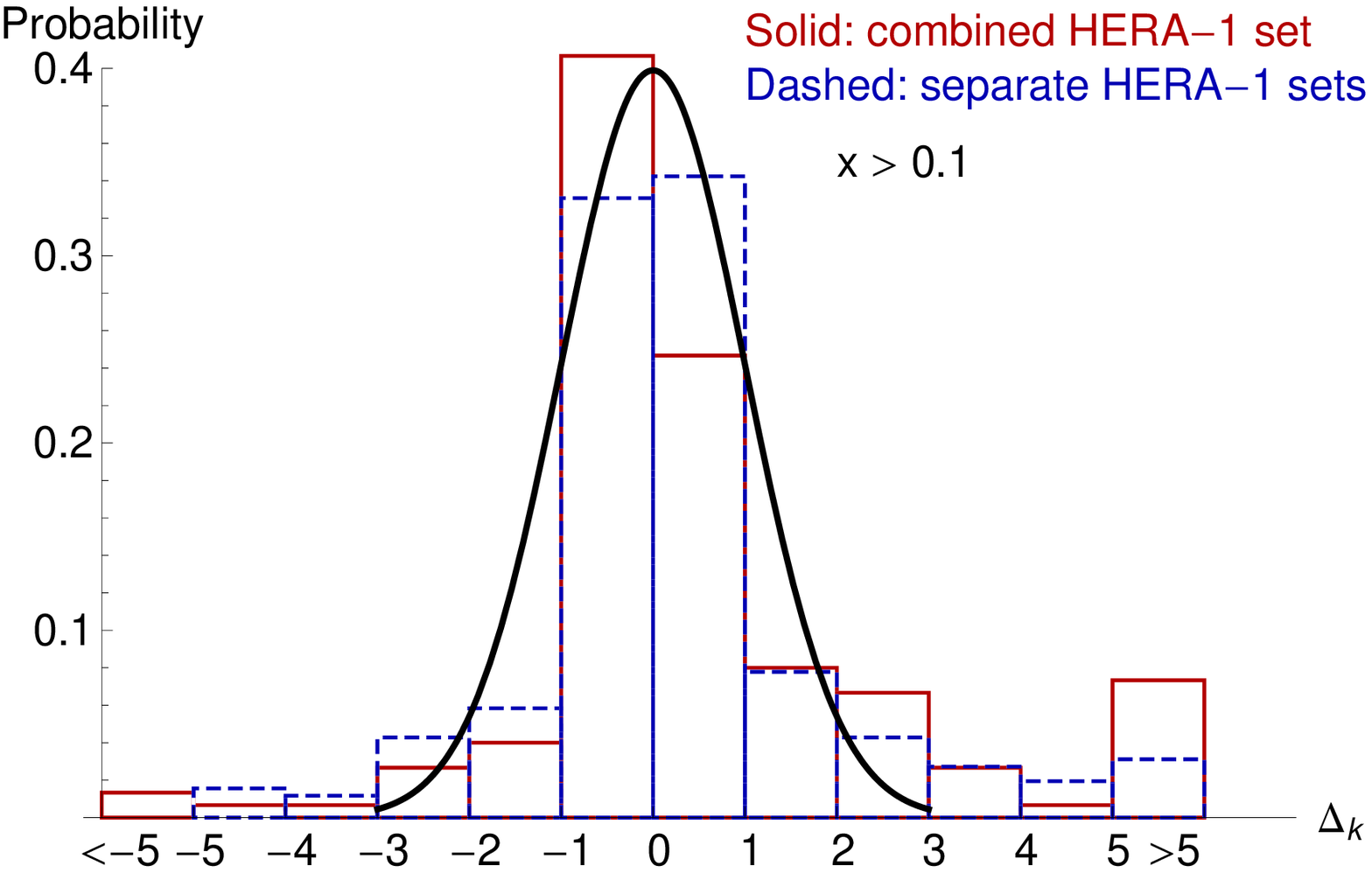}

\caption{Comparison of relative frequency 
distributions of residuals $\Delta_k$ defined in Eq.~(\ref{Deltak})
for neutral current HERA data at in the CT10 fits to the combined HERA set (solid
lines) and separate HERA data sets (dashed lines), 
at $x<0.001$ (upper figure), $0.001<x<0.1$ (middle figure)
and $x>0.1$ (lower figure).\label{fig:HERAresiduals}}

\end{figure}
To see if these increases in $\chi^2$ may be caused 
by systematic discrepancies,
we plot histograms of relative frequencies of 
$\chi^2$ residuals for each data point $k=1,...N$, 
\begin{equation}
\Delta_k = \delta_k^2\ \mbox{sign}(\delta_k),
\end{equation}
with
\begin{equation}
\delta_k = \frac{T_{k}(\{a_{\mbox{best-fit}}\}) - D_{k} + \sum_{\alpha}^{N_{\lambda}}
\lambda_{\alpha, \mbox{best-fit}}\beta_{k\alpha}}{s_k},
\label{Deltak}
\end{equation}
in each $x$ range listed in Table~\ref{tab:chi2HERA1}, and in notations of 
Sec.~\ref{sec:HERA-1}.
In an excellent
fit, the residuals $\Delta_k$ 
follow a standard normal distribution, with a mean
of zero and a unit standard deviation. A non-zero mean observed in the actual
$\Delta_k$ distribution would indicate a systematic
discrepancy affecting the whole histogrammed set of points; on the
other hand, a smaller or larger than normal width 
may be due to incorrectly estimated random effects 
(see Appendix B.2 in Ref.~\cite{Pumplin:2002vw}). 

Distributions of
the residuals for the best fits to the combined and separate HERA-1
sets are plotted in Fig.~\ref{fig:HERAresiduals}. 
At $0.001<x<0.1$ (central figure), frequencies of the residuals 
agree well with the standard distribution, 
regardless of whether the HERA-1 sets are separate or combined. 
At $x<0.001$, the mean of the residual distribution remains consistent
with zero upon the combination of
the data sets, while the width of the distribution increases. 
The residual distribution at $x>0.1$ also widens
and changes the shape, with more residuals having small
negative values or large positive (outlying) values, 
as compared to the fit to the separate
sets. Neither of these patterns indicates systematic 
deviations of the data from NLO QCD theory. On the other hand, the
histograms are suggestive of significant point-to-point random 
fluctuations in the NC DIS data
at $x<0.001$ and $x>0.1$, which appear to be exacerbated when the systematic
uncertainties are reduced through the combination of the data sets. 

\begin{figure}[tb]
\includegraphics[width=0.7\textwidth,height=0.8\textheight,keepaspectratio]{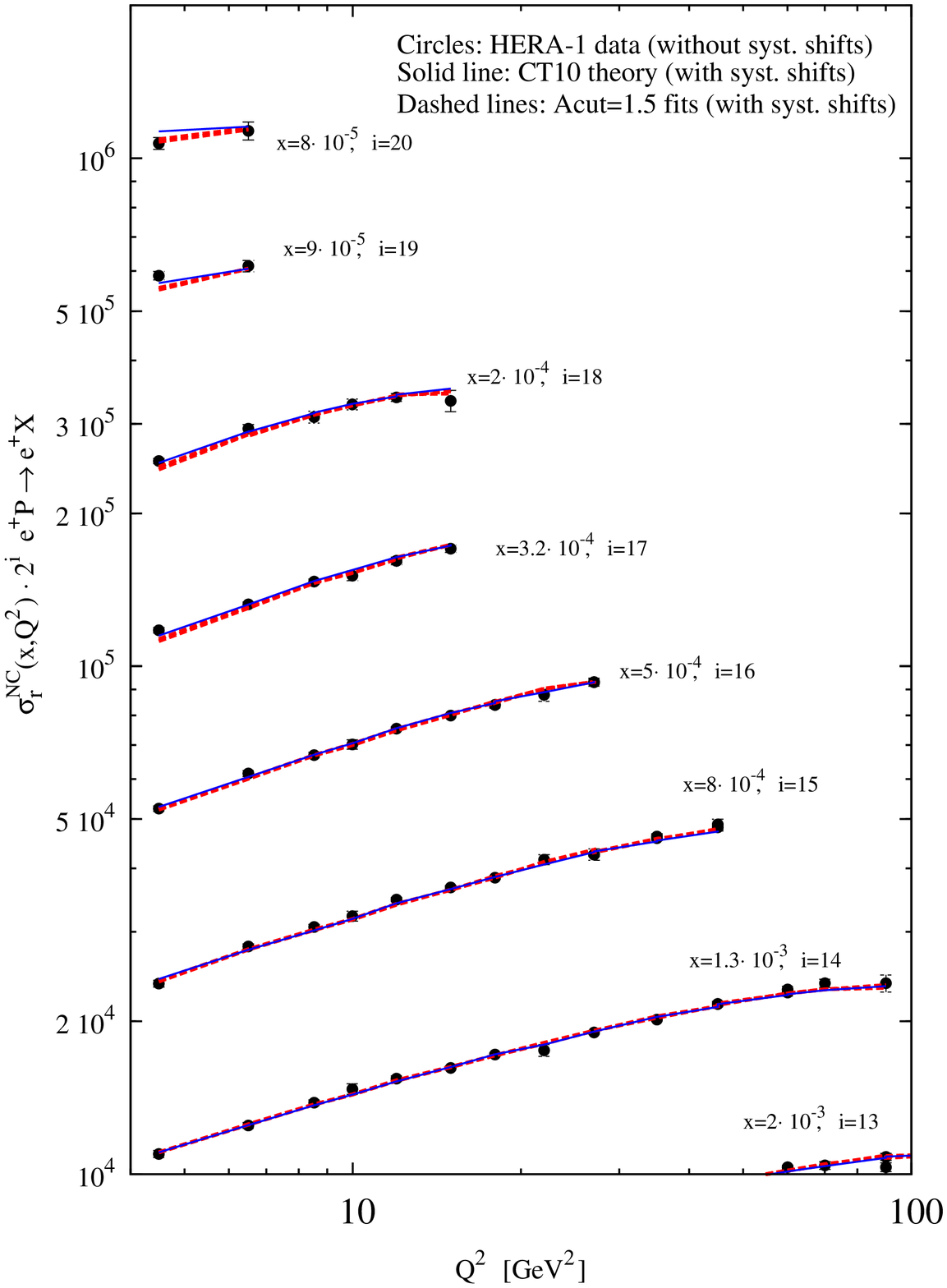}
\caption{Comparison of the HERA data for reduced DIS cross sections 
  at small $A_{gs}$
  values with the CT10 fit (blue) and two fits with $A_{cut}=1.5$
  (red). \label{fig:Acut15NC}}
\end{figure}

\begin{figure}[tb]
\includegraphics[width=0.7\textwidth,height=0.7\textheight,keepaspectratio]{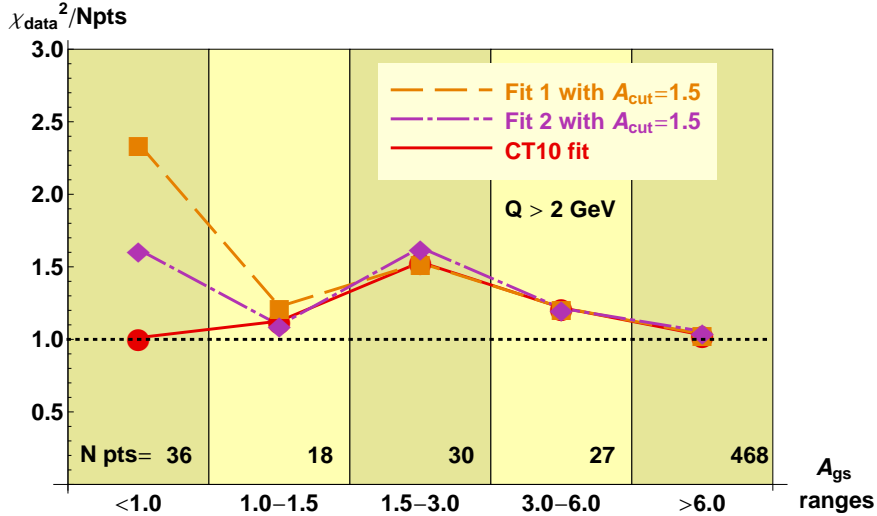}
\caption{The breakdown of $\chi^2_{data}$ values for the combined HERA
  data over the $A_{gs}$
  ranges in the CT10 fit and two fits with $A_{cut}=1.5$.\label{fig:chi2vsAcut}}
\end{figure}
An alternative perspective is provided by dependence on a 
``geometric scaling'' 
variable $A_{gs}\equiv Q^2 x^\lambda$ (with
$\lambda=0.3$), which may reveal disagreements with the NLO QCD
framework in the region of small $x$ and $Q$.\footnote{
$A_{gs}$ is proportional to the variable $\tau = Q^2/Q^2(x)$ 
arising in some saturation models \cite{Stasto:2000er,Caola:2008xr}.}  
The $A_{gs}$ parameter has been studied in recent
NNPDF1.2 and 2.0 global analyses to seek possible deviations from NLO DGLAP
factorization due to saturation or related small-$x$ phenomena
\cite{Caola:2009iy,Caola:2010cy}. In the region $A_{gs} < A_{cut}=0.5-1.5$,
 Refs.~\cite{Caola:2009iy,Caola:2010cy} found
a systematic disagreement between the $Q$ dependence of the
measured DIS cross sections and the prediction based on 
the NLO DGLAP evolution of their PDFs, according to a pattern 
consistent with saturation effects. 
This discrepancy is not expected to be remedied by NNLO corrections,
as those include large logarithms 
requiring all-order summation in the small-$x$ region. 
If confirmed, it will profoundly affect our
understanding of high-energy QCD and various phenomenological applications.

The disagreement stated by NNPDF is not significant (below $1\sigma$) if
the data at small $A_{gs}$ are included in the fit. However, it becomes
significant at the level of $2\sigma$ or more if the small-$A_{gs}$ data
are excluded while determining the PDFs (so that the PDFs are
fitted only to the large-$A_{gs}$ data, for which 
the DGLAP factorization is presumably valid),
but included at the end, when comparing 
the full data sample to the resulting theoretical cross sections. 

We repeated a part of the NNPDF study 
in the region $Q\geq 2$ GeV, where our data
are selected. Our goal is to find out if any deviations 
exist in the included $Q$ region,
 where higher-order corrections are known to be mild,
and  with the full general-mass treatment of heavy quarks. 
(The NNPDF analysis is realized 
in the zero-mass approximation and also includes DIS data in the 
less safe region
$\sqrt{2}\mbox{ GeV} < Q < 2$ GeV.) Besides the CT10 fit, 
several additional fits were
performed only to the data at $A_{gs} > A_{cut}=0.5-1.5$, 
and using several parametrizations 
of the gluon PDF at $x<10^{-3}$ to estimate the sensitivity 
to the initial parametrization choice.
\footnote{In
this exercise, we did not estimate the full uncertainty due
to the parametrization dependence. Obviously it
is larger than the (already significant) 
differences between the $A_{cut}$ fits that are explicitly presented.}  
While the outcomes of these fits bear some similarity to 
those by NNPDF, the spread of the outcomes appears to be too wide 
to corroborate the existence of the deviations. 

In more detail, some fits with the imposed $A_{cut}$ constraints 
produce systematic deficits in theoretical cross sections 
at $A_{gs}$ below 1.0, in a pattern that is similar
to that observed by NNPDF. Since the largest discrepancies are
observed in the fits to the data above $A_{cut}=1.5$,  
we focus on two representative fits with this $A_{cut}$ value 
for the rest of the discussion. Fig.~\ref{fig:Acut15NC} compares 
the CT10 fit and two fits with $A_{cut}=1.5$ to a
subset of HERA data at small $x$ and $Q$. The theoretical predictions
in this figure are shifted toward the data by the amounts found from
the correlation matrix for experimental systematic errors. All three
fits agree well with the data at large $x$ and $Q$, but downward
deviations of the $A_{cut}$ fits emerge at $A_{gs} < 0.5$,
corresponding to the lowest $Q$ values in 
the upper four $x$ bins in Fig.~\ref{fig:Acut15NC}.

Fig.~\ref{fig:chi2vsAcut} shows the breakdown of $\chi^2$
contributions from the data points, 
given by the first term (without $\sum
\lambda_\alpha^2$) in Eq.~(\ref{Chi2sys}), by various ranges of $A_{gs}$. 
In the fitted region $A_{gs}\geq 1.5$, 
the $A_{cut}$ fits result in $\chi^2$ that 
is the same or slightly better (by no more than 10-15 units) 
than the total $\chi^2$ observed in the CT10 fit, 
$\chi^2=608$ for 525 data points and 114 systematic error parameters. 
In the interval $1.0< A_{cut} < 1.5$, the $A_{cut}$ fits agree closely
with the data, as well as with the CT10 fit. At $A_{gs} < 1.0$, 
the CT10 fit results in an essentially ideal value of $\chi^2/N\approx
1$, while the deficit in the predictions of the $A_{cut}$ fits
increases their $\chi^2$ considerably. The magnitude 
of the deficits varies by large amounts between the $A_{cut}$ fits, 
with their $\chi^2$ taking any values between 1 and 2.5 in the $A_{gs} <
1.0$ region. Similar distributions of $\chi^2$ vs. $A_{gs}$ are obtained if only the data that ``causally connected'' by the DGLAP evolution \cite{Caola:2009iy} are included; see the equivalent of Fig.~\ref{fig:chi2vsAcut}
for this case on  the CT10 website \cite{ct10website}.

It is interesting to compare the breakdown of our $\chi^2$ values 
in Fig.~\ref{fig:chi2vsAcut}  with that in two NNPDF2.0 fits 
without and with the $A_{gs} > 1.5$ cut, taken from 
Fig.~8 in Ref.~\cite{Caola:2010cy}. Note again that the $Q$ cuts assumed by 
CT10 and NNPDF2.0 (and the data samples included) 
are slightly different. The quality of the fits 
obtained by the two groups is comparable, with
$\chi^2/N=1.18$ (1.14) for the combined HERA-1 data 
in the CT10 fit (NNPDF2.0 fit \cite{Ball:2010de}). 
In the CT10 fit, both the small-$A_{gs}$
and large-$A_{gs}$ ranges,  $A_{gs} < 1.5$ and $A_{gs} > 3.0$, are
fitted very well ($\chi^2/N\approx 1$), while somewhat
higher-than-ideal $\chi^2/N\approx 1.5$ is observed at $1.5 < A_{gs} < 3.0$.
In the NNPDF2.0 fit, the region $A_{gs} > 6.0$ has a lower
$\chi^2/N\approx 0.9$ than in the CT10 fit, but the quality of the fit
progressively deteriorates, as $A_{gs} $ decreases, and gets worse
than that in the CT10 fit at $A_{gs} < 1.5$. 
With the $A_{gs}$ cut placed at 1.5, the
NNPDF fit significantly disagrees with the data in the whole
excluded region $A_{gs} < 1.5$, with $\chi^2/N > 1.7$;
some deterioration of $\chi^2$ is also observed in the borderline
region of the fitted data, $1.5 < A_{gs} < 3.0$. In our analysis, 
the CT10 fit and $A_{cut} =1.5$ fits are very close for all
$A_{gs}$ above 1.0, with more pronounced 
differences showing up only at $A_{cut} < 1.0$. 

Taken together, the results of the two groups suggest {\em
instability} of the outcomes of the $A_{cut}$ fits outside of the
fitted region of the DIS data. Indeed, all examined fits, without or
with the cuts, produce close results when describing the fitted
data; but their small differences in the fitted region cause  
significant differences outside of it. 

Several features of the $A_{cut}$ fits may
contribute to the instability. Backward DGLAP evolution 
from a high $\mu$ scale to lower scales requires to know accurately the $x$
and $Q$ derivatives of the PDFs, given that very distinct shapes of the
PDFs at the low scale may correspond to close shapes of the PDFs at
the high scale. With the data at the smallest $x$
and $Q$ excluded, the $A_{cut}$ fit loses sensitivity to the 
derivatives in the $x$ region where the PDFs are varying rapidly. 
Extrapolation from the fitted region, 
with only a limited lever arm in $x$ and $Q$ available for it, 
may be inaccurate at the smallest $A_{gs}$ values considered. 

The $A_{cut}$ fits do not fully evaluate the experimental systematic
parameters $\lambda_\alpha$,  some of which affect mostly
small $x$ and $Q$ values and are excluded from the fit by  the
$A_{cut}$ condition. While wrong estimation of
experimental systematics may not explain all observed
discrepancies, the systematic effects shift the
data (or theory) predictions at small $A_{gs}$ 
in approximately the same way as the $A_{cut}$ fits do and, hence,
require careful consideration. 
\clearpage

\bibliographystyle{apsrev}
\bibliography{ct10-main}

\begin{thebibliography}{64}
\expandafter\ifx\csname natexlab\endcsname\relax\def\natexlab#1{#1}\fi
\expandafter\ifx\csname bibnamefont\endcsname\relax
  \def\bibnamefont#1{#1}\fi
\expandafter\ifx\csname bibfnamefont\endcsname\relax
  \def\bibfnamefont#1{#1}\fi
\expandafter\ifx\csname citenamefont\endcsname\relax
  \def\citenamefont#1{#1}\fi
\expandafter\ifx\csname url\endcsname\relax
  \def\url#1{\texttt{#1}}\fi
\expandafter\ifx\csname urlprefix\endcsname\relax\def\urlprefix{URL }\fi
\providecommand{\bibinfo}[2]{#2}
\providecommand{\eprint}[2][]{\url{#2}}

\bibitem[{\citenamefont{Alekhin et~al.}(2010)\citenamefont{Alekhin, Blumlein,
  Klein, and Moch}}]{Alekhin:2009ni}
\bibinfo{author}{\bibfnamefont{S.}~\bibnamefont{Alekhin}},
  \bibinfo{author}{\bibfnamefont{J.}~\bibnamefont{Blumlein}},
  \bibinfo{author}{\bibfnamefont{S.}~\bibnamefont{Klein}}, \bibnamefont{and}
  \bibinfo{author}{\bibfnamefont{S.}~\bibnamefont{Moch}},
  \bibinfo{journal}{Phys. Rev.} \textbf{\bibinfo{volume}{D81}},
  \bibinfo{pages}{014032} (\bibinfo{year}{2010}).

\bibitem[{\citenamefont{Ball et~al.}(2010)}]{Ball:2010de}
\bibinfo{author}{\bibfnamefont{R.~D.} \bibnamefont{Ball}} \bibnamefont{et~al.}
  (\bibinfo{collaboration}{Neural Network PDF Collaboration}),
  \bibinfo{journal}{Nucl. Phys.} \textbf{\bibinfo{volume}{B838}},
  \bibinfo{pages}{136} (\bibinfo{year}{2010}).

\bibitem[{\citenamefont{Pumplin et~al.}(2009)}]{Pumplin:2009nk}
\bibinfo{author}{\bibfnamefont{J.}~\bibnamefont{Pumplin}} \bibnamefont{et~al.},
  \bibinfo{journal}{Phys. Rev.} \textbf{\bibinfo{volume}{D80}},
  \bibinfo{pages}{014019} (\bibinfo{year}{2009}).

\bibitem[{\citenamefont{Aaron et~al.}(2010)}]{2009wt}
\bibinfo{author}{\bibfnamefont{F.~D.} \bibnamefont{Aaron}} \bibnamefont{et~al.}
  (\bibinfo{collaboration}{H1 Collaboration}), \bibinfo{journal}{JHEP}
  \textbf{\bibinfo{volume}{01}}, \bibinfo{pages}{109} (\bibinfo{year}{2010}).

\bibitem[{\citenamefont{Martin et~al.}(2009)\citenamefont{Martin, Stirling,
  Thorne, and Watt}}]{Martin:2009iq}
\bibinfo{author}{\bibfnamefont{A.~D.} \bibnamefont{Martin}},
  \bibinfo{author}{\bibfnamefont{W.~J.} \bibnamefont{Stirling}},
  \bibinfo{author}{\bibfnamefont{R.~S.} \bibnamefont{Thorne}},
  \bibnamefont{and} \bibinfo{author}{\bibfnamefont{G.}~\bibnamefont{Watt}},
  \bibinfo{journal}{Eur. Phys. J.} \textbf{\bibinfo{volume}{C63}},
  \bibinfo{pages}{189} (\bibinfo{year}{2009}).

\bibitem[{\citenamefont{Nadolsky et~al.}(2008)}]{Nadolsky:2008zw}
\bibinfo{author}{\bibfnamefont{P.~M.} \bibnamefont{Nadolsky}}
  \bibnamefont{et~al.}, \bibinfo{journal}{Phys. Rev.}
  \textbf{\bibinfo{volume}{D78}}, \bibinfo{pages}{013004}
  (\bibinfo{year}{2008}).

\bibitem[{\citenamefont{Aaltonen et~al.}(2008)}]{Aaltonen:2008eq}
\bibinfo{author}{\bibfnamefont{T.}~\bibnamefont{Aaltonen}} \bibnamefont{et~al.}
  (\bibinfo{collaboration}{CDF Collaboration}), \bibinfo{journal}{Phys. Rev.}
  \textbf{\bibinfo{volume}{D78}}, \bibinfo{pages}{052006}
  (\bibinfo{year}{2008}).

\bibitem[{\citenamefont{Abazov et~al.}(2008{\natexlab{a}})}]{:2008hua}
\bibinfo{author}{\bibfnamefont{V.~M.} \bibnamefont{Abazov}}
  \bibnamefont{et~al.} (\bibinfo{collaboration}{\protect{D\O~ Collaboration}}),
  \bibinfo{journal}{Phys. Rev. Lett.} \textbf{\bibinfo{volume}{101}},
  \bibinfo{pages}{062001} (\bibinfo{year}{2008}{\natexlab{a}}).

\bibitem[{\citenamefont{Affolder et~al.}(2001)}]{Affolder:2001fa}
\bibinfo{author}{\bibfnamefont{A.~A.} \bibnamefont{Affolder}}
  \bibnamefont{et~al.} (\bibinfo{collaboration}{CDF Collaboration}),
  \bibinfo{journal}{Phys. Rev.} \textbf{\bibinfo{volume}{D64}},
  \bibinfo{pages}{032001} (\bibinfo{year}{2001}).

\bibitem[{\citenamefont{Abbott et~al.}(2001)}]{Abbott:2000kp}
\bibinfo{author}{\bibfnamefont{B.}~\bibnamefont{Abbott}} \bibnamefont{et~al.}
  (\bibinfo{collaboration}{\protect{D\O~ Collaboration}}),
  \bibinfo{journal}{Phys. Rev.} \textbf{\bibinfo{volume}{D64}},
  \bibinfo{pages}{032003} (\bibinfo{year}{2001}).

\bibitem[{\citenamefont{Aaltonen et~al.}(2010)}]{Aaltonen:2010zza}
\bibinfo{author}{\bibfnamefont{T.~A.} \bibnamefont{Aaltonen}}
  \bibnamefont{et~al.} (\bibinfo{collaboration}{CDF Collaboration})
  (\bibinfo{year}{2010}), \eprint{0908.3914}.

\bibitem[{\citenamefont{Abazov et~al.}(2007)}]{Abazov:2007jy}
\bibinfo{author}{\bibfnamefont{V.~M.} \bibnamefont{Abazov}}
  \bibnamefont{et~al.} (\bibinfo{collaboration}{\protect{D\O~ Collaboration}}),
  \bibinfo{journal}{Phys. Rev.} \textbf{\bibinfo{volume}{D76}},
  \bibinfo{pages}{012003} (\bibinfo{year}{2007}).

\bibitem[{\citenamefont{Acosta et~al.}(2005)}]{Acosta:2005ud}
\bibinfo{author}{\bibfnamefont{D.}~\bibnamefont{Acosta}} \bibnamefont{et~al.}
  (\bibinfo{collaboration}{CDF Collaboration}), \bibinfo{journal}{Phys. Rev.}
  \textbf{\bibinfo{volume}{D71}}, \bibinfo{pages}{051104}
  (\bibinfo{year}{2005}).

\bibitem[{\citenamefont{Abazov et~al.}(2008{\natexlab{b}})}]{d0_e_asy}
\bibinfo{author}{\bibfnamefont{V.~M.} \bibnamefont{Abazov}}
  \bibnamefont{et~al.} (\bibinfo{collaboration}{\protect{D\O~ Collaboration}}),
  \bibinfo{journal}{Phys. Rev. Lett.} \textbf{\bibinfo{volume}{101}},
  \bibinfo{pages}{211801} (\bibinfo{year}{2008}{\natexlab{b}}).

\bibitem[{\citenamefont{Abazov et~al.}(2008{\natexlab{c}})}]{d0_mu_asy}
\bibinfo{author}{\bibfnamefont{V.~M.} \bibnamefont{Abazov}}
  \bibnamefont{et~al.} (\bibinfo{collaboration}{\protect{D\O~ Collaboration}}),
  \bibinfo{journal}{Phys. Rev.} \textbf{\bibinfo{volume}{D77}},
  \bibinfo{pages}{011106} (\bibinfo{year}{2008}{\natexlab{c}}).

\bibitem[{\citenamefont{Pumplin}(2010)}]{Pumplin:2009sc}
\bibinfo{author}{\bibfnamefont{J.}~\bibnamefont{Pumplin}},
  \bibinfo{journal}{Phys. Rev.} \textbf{\bibinfo{volume}{D81}},
  \bibinfo{pages}{074010} (\bibinfo{year}{2010}).

\bibitem[{\citenamefont{Collins and Pumplin}(2001)}]{Collins:2001es}
\bibinfo{author}{\bibfnamefont{J.~C.} \bibnamefont{Collins}} \bibnamefont{and}
  \bibinfo{author}{\bibfnamefont{J.}~\bibnamefont{Pumplin}}
  (\bibinfo{year}{2001}), \eprint{hep-ph/0105207}.

\bibitem[{\citenamefont{Pumplin}(2009)}]{Pumplin:2009nm}
\bibinfo{author}{\bibfnamefont{J.}~\bibnamefont{Pumplin}},
  \bibinfo{journal}{Phys. Rev.} \textbf{\bibinfo{volume}{D80}},
  \bibinfo{pages}{034002} (\bibinfo{year}{2009}).

\bibitem[{\citenamefont{Stump et~al.}(2001)}]{LM}
\bibinfo{author}{\bibfnamefont{D.}~\bibnamefont{Stump}} \bibnamefont{et~al.},
  \bibinfo{journal}{Phys. Rev.} \textbf{\bibinfo{volume}{D65}},
  \bibinfo{pages}{014012} (\bibinfo{year}{2001}).

\bibitem[{\citenamefont{Pumplin et~al.}(2002)}]{Pumplin:2002vw}
\bibinfo{author}{\bibfnamefont{J.}~\bibnamefont{Pumplin}} \bibnamefont{et~al.},
  \bibinfo{journal}{JHEP} \textbf{\bibinfo{volume}{07}}, \bibinfo{pages}{012}
  (\bibinfo{year}{2002}).

\bibitem[{\citenamefont{James and Roos}(1975)}]{MINUIT}
\bibinfo{author}{\bibfnamefont{F.}~\bibnamefont{James}} \bibnamefont{and}
  \bibinfo{author}{\bibfnamefont{M.}~\bibnamefont{Roos}},
  \bibinfo{journal}{Comput. Phys. Commun.} \textbf{\bibinfo{volume}{10}},
  \bibinfo{pages}{343} (\bibinfo{year}{1975}).

\bibitem[{\citenamefont{Gribov and Lipatov}(1972)}]{Gribov:1972ri}
\bibinfo{author}{\bibfnamefont{V.~N.} \bibnamefont{Gribov}} \bibnamefont{and}
  \bibinfo{author}{\bibfnamefont{L.~N.} \bibnamefont{Lipatov}},
  \bibinfo{journal}{Yad. Fiz.} \textbf{\bibinfo{volume}{15}},
  \bibinfo{pages}{781} (\bibinfo{year}{1972}), \bibinfo{note}{\protect{[Sov. J.
  Nucl. Phys. {\bf 15} 438 (1972)]}}.

\bibitem[{\citenamefont{Altarelli and Parisi}(1977)}]{Altarelli:1977zs}
\bibinfo{author}{\bibfnamefont{G.}~\bibnamefont{Altarelli}} \bibnamefont{and}
  \bibinfo{author}{\bibfnamefont{G.}~\bibnamefont{Parisi}},
  \bibinfo{journal}{Nucl. Phys.} \textbf{\bibinfo{volume}{B126}},
  \bibinfo{pages}{298} (\bibinfo{year}{1977}).

\bibitem[{\citenamefont{Dokshitzer}(1977)}]{Dokshitzer:1977sg}
\bibinfo{author}{\bibfnamefont{Y.~L.} \bibnamefont{Dokshitzer}},
  \bibinfo{journal}{Sov. Phys. JETP} \textbf{\bibinfo{volume}{46}},
  \bibinfo{pages}{641} (\bibinfo{year}{1977}).

\bibitem[{\citenamefont{Pumplin et~al.}(2001)}]{Hessian}
\bibinfo{author}{\bibfnamefont{J.}~\bibnamefont{Pumplin}} \bibnamefont{et~al.},
  \bibinfo{journal}{Phys. Rev.} \textbf{\bibinfo{volume}{D65}},
  \bibinfo{pages}{014013} (\bibinfo{year}{2001}).

\bibitem[{\citenamefont{Salam and Rojo}(2009)}]{Salam:2008qg}
\bibinfo{author}{\bibfnamefont{G.~P.} \bibnamefont{Salam}} \bibnamefont{and}
  \bibinfo{author}{\bibfnamefont{J.}~\bibnamefont{Rojo}},
  \bibinfo{journal}{Comput. Phys. Commun.} \textbf{\bibinfo{volume}{180}},
  \bibinfo{pages}{120} (\bibinfo{year}{2009}).

\bibitem[{ct1()}]{ct10website}
\bibinfo{howpublished}{\protect{http://hep.pa.msu.edu/cteq/public/ct10.html}}.

\bibitem[{\citenamefont{Lai et~al.}(2010)}]{Lai:2010nw}
\bibinfo{author}{\bibfnamefont{H.-L.} \bibnamefont{Lai}} \bibnamefont{et~al.}
  (\bibinfo{year}{2010}), \eprint{arXiv:1004.4624}.

\bibitem[{\citenamefont{Abe et~al.}(1995)}]{Abe:1994rj}
\bibinfo{author}{\bibfnamefont{F.}~\bibnamefont{Abe}} \bibnamefont{et~al.}
  (\bibinfo{collaboration}{CDF}), \bibinfo{journal}{Phys. Rev. Lett.}
  \textbf{\bibinfo{volume}{74}}, \bibinfo{pages}{850} (\bibinfo{year}{1995}).

\bibitem[{HER()}]{HERAcorrmat}
\bibinfo{howpublished}{\protect{https://www.desy.de/h1zeus/combined\_results/}%
}.

\bibitem[{\citenamefont{Nadolsky and Sullivan}(2001)}]{Nadolsky:2001yg}
\bibinfo{author}{\bibfnamefont{P.~M.} \bibnamefont{Nadolsky}} \bibnamefont{and}
  \bibinfo{author}{\bibfnamefont{Z.}~\bibnamefont{Sullivan}}
  (\bibinfo{year}{2001}), \eprint{hep-ph/0110378}.

\bibitem[{\citenamefont{Berger et~al.}(1989)\citenamefont{Berger, Halzen, Kim,
  and Willenbrock}}]{Berger:1988tu}
\bibinfo{author}{\bibfnamefont{E.~L.} \bibnamefont{Berger}},
  \bibinfo{author}{\bibfnamefont{F.}~\bibnamefont{Halzen}},
  \bibinfo{author}{\bibfnamefont{C.~S.} \bibnamefont{Kim}}, \bibnamefont{and}
  \bibinfo{author}{\bibfnamefont{S.}~\bibnamefont{Willenbrock}},
  \bibinfo{journal}{Phys. Rev.} \textbf{\bibinfo{volume}{D40}},
  \bibinfo{pages}{83} (\bibinfo{year}{1989}).

\bibitem[{\citenamefont{Martin et~al.}(1989)\citenamefont{Martin, Roberts, and
  Stirling}}]{Martin:1988aj}
\bibinfo{author}{\bibfnamefont{A.~D.} \bibnamefont{Martin}},
  \bibinfo{author}{\bibfnamefont{R.~G.} \bibnamefont{Roberts}},
  \bibnamefont{and} \bibinfo{author}{\bibfnamefont{W.~J.}
  \bibnamefont{Stirling}}, \bibinfo{journal}{Mod. Phys. Lett.}
  \textbf{\bibinfo{volume}{A4}}, \bibinfo{pages}{1135} (\bibinfo{year}{1989}).

\bibitem[{\citenamefont{Benvenuti et~al.}(1989)}]{Benvenuti:1989rh}
\bibinfo{author}{\bibfnamefont{A.~C.} \bibnamefont{Benvenuti}}
  \bibnamefont{et~al.} (\bibinfo{collaboration}{BCDMS Collaboration}),
  \bibinfo{journal}{Phys. Lett.} \textbf{\bibinfo{volume}{B223}},
  \bibinfo{pages}{485} (\bibinfo{year}{1989}).

\bibitem[{\citenamefont{Benvenuti et~al.}(1990)}]{Benvenuti:1989fm}
\bibinfo{author}{\bibfnamefont{A.~C.} \bibnamefont{Benvenuti}}
  \bibnamefont{et~al.} (\bibinfo{collaboration}{BCDMS Collaboration}),
  \bibinfo{journal}{Phys. Lett.} \textbf{\bibinfo{volume}{B237}},
  \bibinfo{pages}{592} (\bibinfo{year}{1990}).

\bibitem[{\citenamefont{Aubert et~al.}(1987)}]{Aubert:1987da}
\bibinfo{author}{\bibfnamefont{J.~J.} \bibnamefont{Aubert}}
  \bibnamefont{et~al.} (\bibinfo{collaboration}{European Muon Collaboration}),
  \bibinfo{journal}{Nucl. Phys.} \textbf{\bibinfo{volume}{B293}},
  \bibinfo{pages}{740} (\bibinfo{year}{1987}).

\bibitem[{\citenamefont{Whitlow}(1990)}]{Whitlow:1990dr}
\bibinfo{author}{\bibfnamefont{L.~W.} \bibnamefont{Whitlow}}, Ph.D. thesis,
  \bibinfo{school}{Stanford University} (\bibinfo{year}{1990}),
  \bibinfo{note}{report SLAC-0357, and references therein}.

\bibitem[{\citenamefont{Amaudruz et~al.}(1992)}]{Amaudruz:1992bf}
\bibinfo{author}{\bibfnamefont{P.}~\bibnamefont{Amaudruz}} \bibnamefont{et~al.}
  (\bibinfo{collaboration}{New Muon Collaboration}), \bibinfo{journal}{Phys.
  Lett.} \textbf{\bibinfo{volume}{B295}}, \bibinfo{pages}{159}
  (\bibinfo{year}{1992}).

\bibitem[{\citenamefont{Badelek and Kwiecinski}(1992)}]{Badelek:1991qa}
\bibinfo{author}{\bibfnamefont{B.~M.} \bibnamefont{Badelek}} \bibnamefont{and}
  \bibinfo{author}{\bibfnamefont{J.}~\bibnamefont{Kwiecinski}},
  \bibinfo{journal}{Nucl. Phys.} \textbf{\bibinfo{volume}{B370}},
  \bibinfo{pages}{278} (\bibinfo{year}{1992}).

\bibitem[{\citenamefont{Virchaux and Milsztajn}(1992)}]{Virchaux:1991jc}
\bibinfo{author}{\bibfnamefont{M.}~\bibnamefont{Virchaux}} \bibnamefont{and}
  \bibinfo{author}{\bibfnamefont{A.}~\bibnamefont{Milsztajn}},
  \bibinfo{journal}{Phys. Lett.} \textbf{\bibinfo{volume}{B274}},
  \bibinfo{pages}{221} (\bibinfo{year}{1992}).

\bibitem[{\citenamefont{Bazizi and Wimpenny}(1991)}]{Bazizi:1991mq}
\bibinfo{author}{\bibfnamefont{K.}~\bibnamefont{Bazizi}} \bibnamefont{and}
  \bibinfo{author}{\bibfnamefont{S.~J.} \bibnamefont{Wimpenny}}
  (\bibinfo{year}{1991}), \bibinfo{note}{preprint UCR-DIS-91-02}.

\bibitem[{\citenamefont{Milsztajn et~al.}(1991)\citenamefont{Milsztajn, Staude,
  Teichert, Virchaux, and Voss}}]{Milsztajn:1990cc}
\bibinfo{author}{\bibfnamefont{A.}~\bibnamefont{Milsztajn}},
  \bibinfo{author}{\bibfnamefont{A.}~\bibnamefont{Staude}},
  \bibinfo{author}{\bibfnamefont{K.~M.} \bibnamefont{Teichert}},
  \bibinfo{author}{\bibfnamefont{M.}~\bibnamefont{Virchaux}}, \bibnamefont{and}
  \bibinfo{author}{\bibfnamefont{R.}~\bibnamefont{Voss}}, \bibinfo{journal}{Z.
  Phys.} \textbf{\bibinfo{volume}{C49}}, \bibinfo{pages}{527}
  (\bibinfo{year}{1991}).

\bibitem[{\citenamefont{Martin et~al.}(1994)\citenamefont{Martin, Stirling, and
  Roberts}}]{Martin:1994kn}
\bibinfo{author}{\bibfnamefont{A.~D.} \bibnamefont{Martin}},
  \bibinfo{author}{\bibfnamefont{W.~J.} \bibnamefont{Stirling}},
  \bibnamefont{and} \bibinfo{author}{\bibfnamefont{R.~G.}
  \bibnamefont{Roberts}}, \bibinfo{journal}{Phys. Rev.}
  \textbf{\bibinfo{volume}{D50}}, \bibinfo{pages}{6734} (\bibinfo{year}{1994}).

\bibitem[{\citenamefont{Lai et~al.}(1995)}]{Lai:1994bb}
\bibinfo{author}{\bibfnamefont{H.-L.} \bibnamefont{Lai}} \bibnamefont{et~al.},
  \bibinfo{journal}{Phys. Rev.} \textbf{\bibinfo{volume}{D51}},
  \bibinfo{pages}{4763} (\bibinfo{year}{1995}).

\bibitem[{\citenamefont{Catani et~al.}(2010)\citenamefont{Catani, Ferrera, and
  Grazzini}}]{Catani:2010en}
\bibinfo{author}{\bibfnamefont{S.}~\bibnamefont{Catani}},
  \bibinfo{author}{\bibfnamefont{G.}~\bibnamefont{Ferrera}}, \bibnamefont{and}
  \bibinfo{author}{\bibfnamefont{M.}~\bibnamefont{Grazzini}},
  \bibinfo{journal}{JHEP} \textbf{\bibinfo{volume}{05}}, \bibinfo{pages}{006}
  (\bibinfo{year}{2010}).

\bibitem[{\citenamefont{Accardi et~al.}(2010)}]{OwensLargeX}
\bibinfo{author}{\bibfnamefont{A.}~\bibnamefont{Accardi}} \bibnamefont{et~al.},
  \bibinfo{journal}{Phys. Rev.} \textbf{\bibinfo{volume}{D81}},
  \bibinfo{pages}{034016} (\bibinfo{year}{2010}).

\bibitem[{\citenamefont{Thorne et~al.}(2010)\citenamefont{Thorne, Martin,
  Stirling, and Watt}}]{Thorne:2010kj}
\bibinfo{author}{\bibfnamefont{R.~S.} \bibnamefont{Thorne}},
  \bibinfo{author}{\bibfnamefont{A.~D.} \bibnamefont{Martin}},
  \bibinfo{author}{\bibfnamefont{W.~J.} \bibnamefont{Stirling}},
  \bibnamefont{and} \bibinfo{author}{\bibfnamefont{G.}~\bibnamefont{Watt}}
  (\bibinfo{year}{2010}), \eprint{arXiv:1006.2753}.

\bibitem[{\citenamefont{Balazs et~al.}(1995)\citenamefont{Balazs, Qiu, and
  Yuan}}]{Balazs:1995nz}
\bibinfo{author}{\bibfnamefont{C.}~\bibnamefont{Balazs}},
  \bibinfo{author}{\bibfnamefont{J.}~\bibnamefont{Qiu}}, \bibnamefont{and}
  \bibinfo{author}{\bibfnamefont{C.-P.} \bibnamefont{Yuan}},
  \bibinfo{journal}{Phys. Lett.} \textbf{\bibinfo{volume}{B355}},
  \bibinfo{pages}{548} (\bibinfo{year}{1995}).

\bibitem[{\citenamefont{Balazs and Yuan}(1997)}]{Balazs:1997xd}
\bibinfo{author}{\bibfnamefont{C.}~\bibnamefont{Balazs}} \bibnamefont{and}
  \bibinfo{author}{\bibfnamefont{C.-P.} \bibnamefont{Yuan}},
  \bibinfo{journal}{Phys. Rev.} \textbf{\bibinfo{volume}{D56}},
  \bibinfo{pages}{5558} (\bibinfo{year}{1997}).

\bibitem[{\citenamefont{Landry et~al.}(2003)\citenamefont{Landry, Brock,
  Nadolsky, and Yuan}}]{Landry:2002ix}
\bibinfo{author}{\bibfnamefont{F.}~\bibnamefont{Landry}},
  \bibinfo{author}{\bibfnamefont{R.}~\bibnamefont{Brock}},
  \bibinfo{author}{\bibfnamefont{P.~M.} \bibnamefont{Nadolsky}},
  \bibnamefont{and} \bibinfo{author}{\bibfnamefont{C.-P.} \bibnamefont{Yuan}},
  \bibinfo{journal}{Phys. Rev.} \textbf{\bibinfo{volume}{D67}},
  \bibinfo{pages}{073016} (\bibinfo{year}{2003}).

\bibitem[{\citenamefont{Lai et~al.}(1997)}]{cteq4}
\bibinfo{author}{\bibfnamefont{H.-L.} \bibnamefont{Lai}} \bibnamefont{et~al.},
  \bibinfo{journal}{Phys. Rev.} \textbf{\bibinfo{volume}{D55}},
  \bibinfo{pages}{1280} (\bibinfo{year}{1997}).

\bibitem[{\citenamefont{Arneodo et~al.}(1995)}]{NMC}
\bibinfo{author}{\bibfnamefont{M.}~\bibnamefont{Arneodo}} \bibnamefont{et~al.}
  (\bibinfo{collaboration}{New Muon Collaboration}), \bibinfo{journal}{Phys.
  Lett.} \textbf{\bibinfo{volume}{B364}}, \bibinfo{pages}{107}
  (\bibinfo{year}{1995}).

\bibitem[{\citenamefont{Lewis}(1988)}]{Lewis:1988}
\bibinfo{author}{\bibfnamefont{T.}~\bibnamefont{Lewis}},
  \bibinfo{journal}{Austral. J. Statist.} \textbf{\bibinfo{volume}{30A}},
  \bibinfo{pages}{160} (\bibinfo{year}{1988}).

\bibitem[{\citenamefont{Fisher}(1925)}]{Fisher:1925}
\bibinfo{author}{\bibfnamefont{R.~A.} \bibnamefont{Fisher}},
  \emph{\bibinfo{title}{Statistical methods for research workers}}
  (\bibinfo{publisher}{Oliver and Boyd, Edinburgh}, \bibinfo{year}{1925}),
  chap.~\bibinfo{chapter}{4}, \bibinfo{note}{\protect{an Internet version of
  the 1st edition at http://psychclassics.yorku.ca/Fisher/Methods/}}.

\bibitem[{\citenamefont{Abazov et~al.}(2010)}]{d0-di-jet}
\bibinfo{author}{\bibfnamefont{V.~M.} \bibnamefont{Abazov}}
  \bibnamefont{et~al.} (\bibinfo{collaboration}{\protect{D\O~ Collaboration}})
  (\bibinfo{year}{2010}), \eprint{arXiv:1002.4594}.

\bibitem[{\citenamefont{Spira}(1995)}]{Spira:1995mt}
\bibinfo{author}{\bibfnamefont{M.}~\bibnamefont{Spira}} (\bibinfo{year}{1995}),
  \eprint{hep-ph/9510347}.

\bibitem[{\citenamefont{Arnold et~al.}(2009)}]{Arnold:2008rz}
\bibinfo{author}{\bibfnamefont{K.}~\bibnamefont{Arnold}} \bibnamefont{et~al.},
  \bibinfo{journal}{Comput. Phys. Commun.} \textbf{\bibinfo{volume}{180}},
  \bibinfo{pages}{1661} (\bibinfo{year}{2009}).

\bibitem[{\citenamefont{Kluge et~al.}(2006)\citenamefont{Kluge, Rabbertz, and
  Wobisch}}]{fastnlo}
\bibinfo{author}{\bibfnamefont{T.}~\bibnamefont{Kluge}},
  \bibinfo{author}{\bibfnamefont{K.}~\bibnamefont{Rabbertz}}, \bibnamefont{and}
  \bibinfo{author}{\bibfnamefont{M.}~\bibnamefont{Wobisch}}
  (\bibinfo{year}{2006}), \eprint{hep-ph/0609285}.

\bibitem[{\citenamefont{Stump et~al.}(2003)}]{Stump:2003yu}
\bibinfo{author}{\bibfnamefont{D.}~\bibnamefont{Stump}} \bibnamefont{et~al.},
  \bibinfo{journal}{JHEP} \textbf{\bibinfo{volume}{10}}, \bibinfo{pages}{046}
  (\bibinfo{year}{2003}).

\bibitem[{lha()}]{lhapdf}
\bibinfo{howpublished}{\protect{http://projects.hepforge.org/lhapdf/}}.

\bibitem[{\citenamefont{Stasto et~al.}(2001)\citenamefont{Stasto,
  Golec-Biernat, and Kwiecinski}}]{Stasto:2000er}
\bibinfo{author}{\bibfnamefont{A.~M.} \bibnamefont{Stasto}},
  \bibinfo{author}{\bibfnamefont{K.~J.} \bibnamefont{Golec-Biernat}},
  \bibnamefont{and}
  \bibinfo{author}{\bibfnamefont{J.}~\bibnamefont{Kwiecinski}},
  \bibinfo{journal}{Phys. Rev. Lett.} \textbf{\bibinfo{volume}{86}},
  \bibinfo{pages}{596} (\bibinfo{year}{2001}).

\bibitem[{\citenamefont{Caola and Forte}(2008)}]{Caola:2008xr}
\bibinfo{author}{\bibfnamefont{F.}~\bibnamefont{Caola}} \bibnamefont{and}
  \bibinfo{author}{\bibfnamefont{S.}~\bibnamefont{Forte}},
  \bibinfo{journal}{Phys. Rev. Lett.} \textbf{\bibinfo{volume}{101}},
  \bibinfo{pages}{022001} (\bibinfo{year}{2008}).

\bibitem[{\citenamefont{Caola et~al.}(2010{\natexlab{a}})\citenamefont{Caola,
  Forte, and Rojo}}]{Caola:2009iy}
\bibinfo{author}{\bibfnamefont{F.}~\bibnamefont{Caola}},
  \bibinfo{author}{\bibfnamefont{S.}~\bibnamefont{Forte}}, \bibnamefont{and}
  \bibinfo{author}{\bibfnamefont{J.}~\bibnamefont{Rojo}},
  \bibinfo{journal}{Phys. Lett.} \textbf{\bibinfo{volume}{B686}},
  \bibinfo{pages}{127} (\bibinfo{year}{2010}{\natexlab{a}}).

\bibitem[{\citenamefont{Caola et~al.}(2010{\natexlab{b}})\citenamefont{Caola,
  Forte, and Rojo}}]{Caola:2010cy}
\bibinfo{author}{\bibfnamefont{F.}~\bibnamefont{Caola}},
  \bibinfo{author}{\bibfnamefont{S.}~\bibnamefont{Forte}}, \bibnamefont{and}
  \bibinfo{author}{\bibfnamefont{J.}~\bibnamefont{Rojo}}
  (\bibinfo{year}{2010}{\natexlab{b}}), \eprint{arXiv:1007.5405}.

\end{thebibliography}

\end{document}